\documentclass{jfm}

\usepackage{graphicx}
\usepackage{newtxtext}
\usepackage{newtxmath}
\usepackage{natbib}
\usepackage{hyperref}
\hypersetup{
    colorlinks = true,
    urlcolor   = blue,
    citecolor  = black,
}

\newcommand{\RomanNumeralCaps}[1]
\linenumbers

\title{Clustering and chaotic motion of heavy inertial particles in an isolated non-axisymmetric vortex}

\author{Anu V. S. Nath\aff{1}
 \and Anubhab Roy\aff{1}   \corresp{\email{anubhab@iitm.ac.in}}}

\affiliation{\aff{1}Department of Applied Mechanics,
Indian Institute of Technology Madras, Chennai 600036}

\begin{document}
\maketitle

\begin{abstract}

We investigate the dynamics of heavy inertial particles in a flow field due to an isolated, non-axisymmetric vortex. For our study, we consider a canonical elliptical vortex - the Kirchhoff vortex and its strained variant, the Kida vortex. Contrary to the anticipated centrifugal dispersion of inertial particles, which is typical in open vortical flows, we observe the clustering of particles around co-rotating attractors near the Kirchhoff vortex due to its non-axisymmetric nature. We analyse the inertia-modified stability characteristics of the fixed points, highlighting how some of the fixed points migrate in physical space, collide and then annihilate with increasing particle inertia. The introduction of external straining, the Kida vortex being an example, introduces chaotic tracer transport. Using a Melnikov analysis, we show that particle inertia and external straining can compete, where chaotic transport can be suppressed beyond a critical value of particle inertia.

\end{abstract}

% \begin{keywords}
% Authors should not enter keywords on the manuscript, as these must be chosen by the author during the online submission process and will then be added during the typesetting process (see \href{https://www.cambridge.org/core/journals/journal-of-fluid-mechanics/information/list-of-keywords}{Keyword PDF} for the full list).  Other classifications will be added at the same time.
% \end{keywords}

% {\bf MSC Codes }  {\it(Optional)} Please enter your MSC Codes here
%%%%%%%%%%%%%%%%%%%%%%%%%%%%%%%%%%%%%%%%%%%%%%%%%%%%%%%%%%%%%%%%%%%%%%%%%%%%%%%%%%%%%%%%%%%%%%%%%%%%%%%%%%%%%%%%%%%%%%5
\section{Introduction}
\label{sec1}
Coherent vortical structures are ubiquitous in nature. In the planetary context, the flows are predominantly two-dimensional due to the strong influence of rotation. These turbulent flows naturally evolve into long-lived isolated eddies/vortices \citep{mcwilliams1984emergence}, like in oceans and planetary atmospheres — for example, the Great Red Spot on Jupiter, tropical cyclones, and Gulf Stream rings. The inverse cascade of energy and planetary rotation plays a prominent role in forming these coherent vortices \citep{vallis2017atmospheric}.%For a detailed review of the topic, see \citet{provenzale1999transport}. 
The two-dimensional (2D) turbulent flows in a rotating background will eventually self-organize into concentrated vortical lumps as observed in various simulations and observations \citep{fornberg1977numerical,basdevant1981study,babiano1987vorticity,benzi1988self}. In three-dimensional turbulent flows as well, the formation of vortical filaments has been similarly identified \citep{siggia1981numerical,vincent1991spatial,bartello1994coherent}. Thus, the later evolution of turbulent flows is strongly influenced by the underlying vorticity dynamics of these coherent eddies.

Turbulent flows in geophysical and astrophysical contexts are often dispersed with particulate matter - water droplets and ice crystals in clouds \citep{shaw2003particle}, pyroclastic flows \citep{dufek2016fluid}, wind-sand interactions in aeolian processes \citep{kok2012physics}, and dust in protoplanetary disks \citep{armitage2020astrophysics}. The coherent vortices embedded in these flows greatly influence particle transport. Tracer particles can get trapped by the vortices for a long time, much larger than the eddy turnover time, and get transported across the distances over which the eddy travels \citep{elhmaidi1993elementary}. The trapped particles can be released only after the disruption of the vortex itself. However, the particulate matter does not necessarily have negligible inertia; the finite inertia aspect of particles can make the dynamics of the suspended phase more complex with aspects of clustering \citep{bec2003fractal,bec2005multifractal,sapsis2010clustering}, and caustics \citep{crisanti1992dynamics,falkovich2002acceleration,wilkinson2005caustics}.
%,wilkinson2006caustic,gustavsson2012inertial,ravichandran2015caustics
%,deepu2017caustics}. 
Heavy inertial particles are centrifuged away by the vortex cores and get accumulated in the straining regions of the flow \citep{maxey1987gravitational}. However, in a rotating background, the heavy inertial particles get pushed by the Coriolis force into the cores of anticyclonic vortices, which is hypothesized to trigger the formation of planetesimals in the astrophysical context \citep{barge1995did,tanga1996forming,chavanis1999trapping}.

The transport of particles by various vortical structures has been extensively modelled and studied in the past few decades. \cite{batchelor1994expulsion} investigated the expulsion of heavy inertial particles from a rising bubble, considering the combined effects of gravitational sedimentation and the toroidal circulation of gas inside. \cite{marcu1995dynamics} explored the transport of inertial particles near a Burger's vortex with and without the influence of gravity. In the absence of gravity, particles with sufficiently small inertia were captured by the vortex centre, while those with large inertia exhibited stable limit cycle dynamics. The inclusion of gravity altered the dynamics, generating additional fixed points that could capture the particles. \cite{raju1997dynamics} have studied the transport of inertial particles with varying density ratios in three model flows: a solid-body vortex, a point vortex, and a stagnation point flow. \cite{eames2004settling} considered the sedimentation and dispersion of inertial particles past an isolated spherical vortex and a random distribution of spherical vortices, revealing that the interplay between particle inertia and stagnation points significantly increased the vertical dispersivity of dense particles compared to tracers. \cite{hunt2007vortices} investigated inertial particle transport near a vortex tube and steadily propagating vortex rings, providing analytical treatment and experimental comparisons. The clustering of heavy inertial particles in a pair of co-rotating vortices was explored by \cite{angilella2010dust, ravichandran2014attracting}. \cite{ravichandran2015caustics} have studied the clustering of inertial particles and the subsequent emergence of caustics in a point vortex and a system of point vortices. In the context of airborne pathogen transport through the atmosphere, recent studies by \cite{dagan2021settling} and \cite{avni2022dynamics} have modelled evaporating droplets as advected by a Lamb-Chaplygin vortex dipole, revealing that the interaction with the vortex enhanced droplets' settling time and transported them over large distances in the air. Motivated by the dispersion of droplets in warm cumulus clouds during the condensational phase, \cite{nath2022transport} investigated the dispersion of condensing droplets in a background flow modelled as an array of Taylor-Green vortices; they demonstrated a significant enhancement in droplet dispersion as they acquire more inertia by condensation.

Neighbouring coherent vortices in a turbulent flow can interact with each other and induce shearing, which could disrupt them \citep{legras2001erosion}. According to \citet{reinaud2003shape}, vortices that can withstand the highest levels of strain are those most likely to be found in an actual turbulent flow.
%They show that the shape of vortices in quasi-geostrophic turbulence is oblate spheroid of potential vorticity with an aspect ratio of $0.8$.
An elliptic vortex patch of constant vorticity is an exact solution of the incompressible 2D Euler equation \citep[see][]{kirchhoff1876mechanik,lamb1924hydrodynamics,saffman1995vortex}; below aspect ratio of 3 the vortices are both linearly \citep{love1893stability} and nonlinearly \citep{wan1986stability,tang1987nonlinear} stable. Elliptic vortex patches and their interactions have been extensively studied to understand better the stability and evolution of vortices in ideal fluid \citep[see][]{moore1971structure,kida1981motion,dritschel1990stability,legras1991elliptical,dritschel1996instability,mitchell2008evolution}, with motivations from geophysical turbulent flows \citep{dritschel1995general}. Due to the non-axisymmetric vorticity distribution, an elliptic patch of uniform vorticity in an irrotational background will rotate with a constant angular velocity while preserving its size and shape. This configuration, widely known as the Kirchhoff vortex, is well-suited for studying isolated non-axisymmetric coherent vortices in 2D flows. The tracer transport in the Kirchhoff vortex is non-chaotic as it is an integrable Hamiltonian system.

To study vortex interactions \citet{kida1981motion} proposed a model of a vortex tube in a uniform shear flow; the effects of the other vortices on a certain vortex tube may be replaced, in the first approximation, by a linear flow. \citet{moore1971structure} had earlier studied steady elliptic vortex patches in uniform shear; \citet{kida1981motion} generalized the solutions to include exact unsteady elliptic vortices. However, we would like to mention that the first study of an elliptic vortex in a specific linear flow, a simple shear flow, was carried out by Chaplygin \citep[see][]{chaplygin1899pulsating,meleshko1994chaplygin}. 
% %%%%%%%%%%%%%%%%%%%%%%%%%%%%%%%%%%%%%%%%%%%%%%%%%%%%%%%%%%%%%%%%%%%%
% \begin{figure}
%     \centering
%     \includegraphics[width=0.8\linewidth]{Fig008.pdf}
%     \caption{Schematic showing three major kinds of dynamics exhibited by a sheared elliptical vortex: (a) initial vortex patch of anti-clockwise vorticity content sheared by an external shear flow of opposite sense of vorticity content, (b-d) rotation: full rotation and periodic straining due to weak shear compared to vorticity (e-g) nutation: back and forth oscillation and periodic straining due to medium shear rate and (h-j) elongation: the dominant shear flow stretches the vortex patch indefinitely and destroys it.}
%     \label{Fig008}
% \end{figure}
% %%%%%%%%%%%%%%%%%%%%%%%%%%%%%%%%%%%%%%%%%%%%%%%%%%%%%%%%%%%%%%%%%%%%%%%%%%%%%%%%%%%%%%%%%
When an external simple shear flow is superimposed on the Kirchhoff vortex, the configuration is known to have three kinds of impact on the elliptic vortex patch: (i) rotation: full rotation of ellipse with periodically changing aspect ratio and angular velocity, (ii) nutation: back and forth oscillatory angular motion of elliptic patch, and (iii) elongation: irreversible elongation of vortex patch due to strong external straining \citep[see][]{kida1981motion,neu1984dynamics,dritschel1990stability}. Though the area of the ellipse is preserved, the unsteady rotation of the ellipse with changing aspect ratio creates an unsteady flow-field around, even in the co-rotating frame \citep{kida1981motion}, which is referred to as the `Kida vortex' in this paper. The original motivation for studying sheared vortical patches was to understand vortex interactions better. However, when analyzed from the perspective of tracer transport, sheared elliptic vortices exhibited chaotic Lagrangian trajectories \citep{polvani1990chaotic,dahleh1992exterior}. Even a minor imposed shear induces periodic unsteadiness in the deformation of the Kida vortex, disrupting the Hamiltonian integrability of the Kirchhoff vortex. The hyperbolic fixed points and heteroclinic connections of the Kirchhoff vortex (see Section \ref{sec2}) experience perturbations, making them susceptible to transverse intersections and thus allow for the possibility of chaotic dynamics \citep{smale1967differentiable,bertozzi1987extension}. A comprehensive investigation was conducted into the impact of unsteady perturbations on tracer transport in an otherwise integrable system of a pair of oppositely signed point vortices by \cite{rom1990analytical}. In the absence of perturbation, the vortex pair translates with a constant velocity, resulting in a steady flow field in the co-moving frame with the vortices. However, the introduction of an external periodic strain field, even in the co-moving frame, makes the flow field unsteady, causing the tangling heteroclinic orbits in the flow field and leading to the chaotic transport of certain passive tracers. This phenomenon also results in fluid entrainment by the vortex system, enhancing mixing. Notably, this study represents one of the early applications of tools such as the Melnikov analysis from dynamical systems to analyse and quantify chaos and mixing in a fluid flow problem. In the current context of inertial particle transport in the Kida vortex, we apply some of the techniques derived from their work.

The dispersed phase embedded in the coherent structures in the various geophysical and astrophysical flows is rarely inertialess - dust, bubbles, planktons. This raises the question of how particle inertia alters particle dispersion in the neighbourhood of vortices. Here, we are interested in the dynamics of small, heavy inertial particles, thus ignoring the additional physics of added mass effects, the Basset history term, convective inertia, and Faxen corrections. Studies have shown that particle inertia can suppress the chaotic transport in vortical flows \citep{angilella2010dust,angilella2014inertial}. However, we have demonstrated recently \citep{nath2024irregular} that particle inertia can induce chaotic dynamics and lead to non-ergodic dynamics due to the `scattering' interaction of inertial particles with an ordered array of stagnation points. Particle inertia modifies the fluid tracer fixed points and their homoclinic/heteroclinic connections, which subsequently play a prominent role in long-time particle transport. In this paper, we study the transport of heavy inertial particles near a non-axisymmetric vortex patch - first in the Kirchhoff vortex and then in its strained variant, the Kida vortex. The configuration chosen is a simple scenario of an isolated vortex where we can study the modification of the heteroclinic tangles by particle inertia analytically and comment on the competing roles of background shear and particle inertia to promote or suppress chaotic transport. 

An earlier investigation on particle transport in a strained elliptical vortex is documented in the work by \cite{chavanis1999trapping}. This study specifically focuses on the trapping of dust by anticyclonic vortices in Keplerian proto-planetary disks, proposing it as a mechanism for planet formation. In this context, the vortices experience Keplerian shear, leading to the survival of only anticyclonic vortices that achieve a steady elliptic configuration. The strength of the vorticity is determined by Keplerian shear, employing a solution provided by \cite{moore1971structure}. The analysis of inertial dust particle transport is conducted in a frame co-rotating with the vortex. The findings indicate that particles with small inertia approximately follow an elliptic path, drifting inwards due to drag and the Coriolis force, ultimately being captured by the vortex centre. On the other hand, particles with larger inertia exhibit an epicyclic motion but eventually sink into the vortex. Particles with substantial inertia may even escape the vortex. Despite some apparent similarities with the second part of our work, which involves the transport of inertial particles by a strained elliptic vortex, there are notable differences. Our study considers an elliptic vortex model for a coherent vortex in turbulence, allowing for any general value of the strain rate it experiences. Consequently, the vortex is not in a steady state but  unsteady motion, leading to the intriguing particle dynamics discussed in this paper.

The remainder of this paper is organised as follows. Section \ref{sec2} provides an overview of the dynamics of heavy inertial particles in the Kirchhoff vortex, detailing their clustering in various fixed points and presenting a stability analysis of these fixed points. In Section \ref{sec3}, we primarily focus on the modification of these dynamics in a strained Kirchhoff vortex due to an imposed weak pure-strain flow, i.e., in a rotating Kida vortex. We analyse the perturbative changes from stable fixed points to stable limit cycles. Additionally, a Melnikov analysis is employed on saddle points to demonstrate the existence of chaotic dynamics for inertial particles with sufficiently small inertia. Large-time Lyapunov exponents and fractal dimension calculations are used to confirm the presence of chaotic dynamics. A small discussion on the inertial particle dynamics in nutating and elongating Kida vortices is added at the end of Section \ref{sec3}. The large-time dispersion characteristics of heavy inertial particles in Kirchhoff and Kida vortices are discussed in Section \ref{sec4}. Additionally, the trapping of particles around a Kida vortex is studied by evaluating their residence time in the neighbourhood of the vortex.

% usually they are suspended with particles. More often the suspended particles will be heavy, i.e. their density is much higher than the ambient fluid density, as well as inertial, i.e., their particle relaxation time is very high compared to flow time scale. 
% Transport of inertial particles in rotating/vortex flows is of importance
% Inertial aprticle transport in rotating flows
% maxey riley in rotating
% divergence 
% cyclonic anticyclonic
% Kirchhoff vortex
% Accretion  disc

%%%%%%%%%%%%%%%%%%%%%%%%%%%%%%%%%%%%%%
%%%%%%%%%%%%%%%%%%%%%%%%%%%%%%%%%%%%%%%%%%%%
% \begin{figure}
%         \begin{subfigure}[b]{0.5\textwidth}
%                 \includegraphics[width=\linewidth]{Axes.pdf}
%                % \caption{A gull}
%                 %\label{fig001a}
%         \end{subfigure}%
%         \begin{subfigure}[b]{0.5\textwidth}
%                 \includegraphics[width=\linewidth]{fixedpoints_r=0p5.eps}
%                 %\caption{A mouse}
%                 %\label{fig:mouse}
%         \end{subfigure}
%         \caption{(a) Schematic showing the elliptic vortex patch with stationary reference frame $X'-Y'$ and the co-rotating reference frame $X-Y$. (b) The (steady) streamlines of the Kirchhoff vortex flow-field in the co-rotating reference frame.}
%         \label{Fig001}
% \end{figure}
%%%%%%%%%%%%%%%%%%%%%%%%%%%%%%%%%%%%%%%%%%%%%%%%%%%%%%%%%%%%%%%%%%%%%%%%%%%%%%%%%%%%
%%%%%%%%%%%%%%%%%%%%%%%%%%%%%%%%%%%%%%%%%%%%%
\section{\label{sec2} Dynamics of heavy inertial particles in an elliptic vortex}
\subsection{An isolated elliptic patch of uniform vorticity - the Kirchhoff vortex}
\begin{figure}
    \centering
    \includegraphics[width=\linewidth]{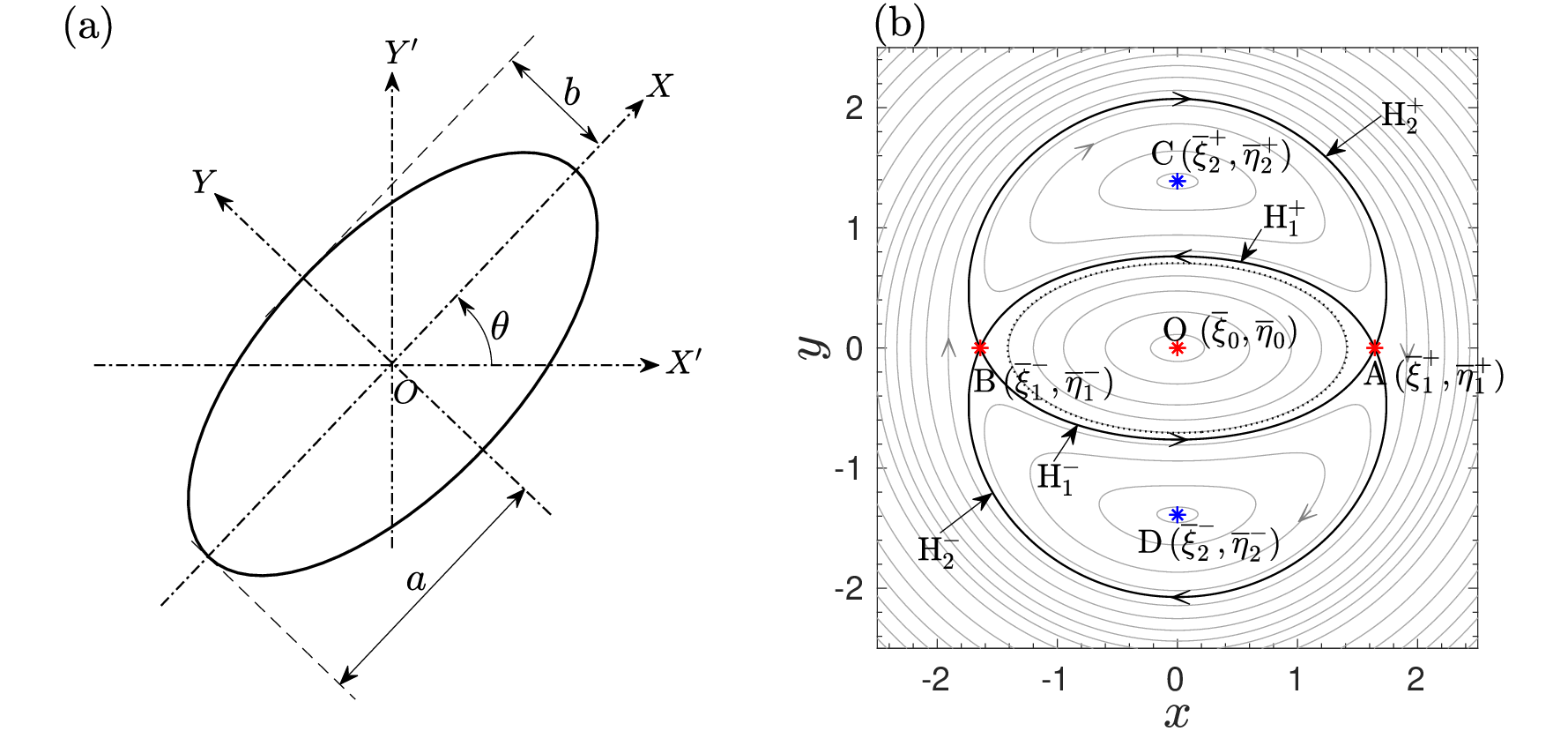}
    \caption{(a) Schematic showing the elliptic vortex patch with stationary reference frame $X'-Y'$ and the co-rotating reference frame $X-Y$. (b) The (steady) streamlines of the Kirchhoff vortex in the co-rotating reference frame.}
    \label{Fig001}
\end{figure}

Consider an elliptical patch of constant vorticity ($\omega_0$) in an irrotational background. As mentioned earlier, due to the non-axisymmetric vorticity distribution, the elliptic patch of uniform vorticity ($\omega_0$) will rotate with constant angular velocity $\Omega = \omega_0\, a\, b/(a+b)^2$, where $a$ and $b$ are semi-major and semi-minor axes of the elliptic patch (see figure \ref{Fig001}); the size and shape of the elliptic patch is preserved during the rotation. The configuration, known as the Kirchhoff vortex, is given by the stream function (as observed by a stationary observer)
\begin{equation}
\psi_v'= \left\{
\begin{array}{ll}
      -\frac{\omega_0}{2\, (a+b)}\, (b \, x^2 +a\, y^2), & \frac{x^2}{a^2}+\frac{y^2}{b^2} < 1 \vspace{0.35cm} \\
      
      -\frac{a\, b\, \omega_0}{4}\, \left(2\, \xi + e^{-2\, \xi}\, \cos(2\, \eta) \right), & \frac{x^2}{a^2}+\frac{y^2}{b^2} > 1~. \\
\end{array} 
\right. 
\label{eqn1p1}
\end{equation}
%%%%%%%%%%%%%%%%%%%%%%%%%%%%%%%%%%%%
In the co-rotating frame with the vortex, the stream function is $\psi = \psi'+\frac{\Omega}{2}\, (x^2+y^2)$, where $(x,y)$ and $(\xi,\eta)$ are respectively the Cartesian and elliptic coordinates measured in a co-rotating frame with the ellipse, which are inter-related as $x = \sqrt{a^2-b^2}\, \cosh \xi \, \cos \eta$ and $y = \sqrt{a^2-b^2}\, \sinh \xi \, \sin \eta$, with $\xi \geq 0$ and $\eta \in [0,2\pi)$. We use primed ($'$) variables to denote quantities in the stationary frame, whereas un-primed variables represent quantities in the co-rotating frame. We follow the same convention throughout this paper unless specified otherwise. The stationary reference frame ($X'-Y'$) and the co-rotating reference frame ($X-Y$) are shown schematically in figure \ref{Fig001}(a), making an instantaneous angle $\theta$, relates them as $x' = x\, \cos \theta - y \, \sin \theta$ and 
$y' = x \, \sin \theta + y \, \cos \theta$, where $\frac{d \theta}{d t} = \Omega$. An advantage of choosing a co-rotating frame is that the velocity field is steady in the co-rotating frame. The corresponding streamlines are shown in figure \ref{Fig001}(b). The tracer dynamics in the co-rotating frame is thus governed by a 2D, time-independent dynamical system, which guarantees non-chaotic fluid pathlines. Inside the ellipse, the flow field is a solid-body rotation, and far away in the outer region, it resembles a decaying point vortex. The flow-field is a tripole structure \citep[see][]{viudez2021robust,xu2023dynamics} with five fixed points (where the flow velocity is zero) A, B, C, D and O as marked in the figure. The origin O ($\overline{\xi}_0$, $\overline{\eta}_0$) is an elliptic fixed point; the pair A and B ($\overline{\xi}_1^{\pm}$, $\overline{\eta}_1^{\pm}$) located along the major axis line outside the ellipse are hyperbolic type fixed points; the pair C and D ($\overline{\xi}_2^{\pm}$, $\overline{\eta}_2^{\pm}$) located along the minor axis line outside the ellipse (inside the lobes) are elliptic type fixed points \citep[see][]{kawakami1999chaotic}. The hyperbolic fixed points are interconnected by two pairs of heteroclinic orbits, denoted as $H_1^{\pm}$ and $H_2^{\pm}$. As we proceed further into our analysis, we will learn the significance of these heteroclinic orbits and their possible perturbation, which is crucial for the onset of chaos, in Section \ref{sec3}. The fixed points are `fixed points' only for a co-rotating observer; however, for a stationary observer, they resemble `Lagrange points' of celestial mechanics \citep{fitzpatrick2012introduction}.
%%%%%%%%%%%%%%%%%%%%%%%%%%%%%%%%%%%%%%%%%%%%%%%%%%%%%%%%%%%%%%%%%%%%%%%%%%%%%%%%%%%%%%%%%%%%%%%%%%%%%
\subsection{Dynamics of inertial particles}
The dynamics of heavy inertial point particles in a background flow can be studied using the Maxey-Riley equation \citep[see][]{maxey1983equation}. In a rotating reference frame with the ellipse (of angular velocity $\Omega$), the modified form of the Maxey-Riley equation by accounting for the pseudo forces reads, in nondimensional form
\begin{equation}
        \dot{\textbf{v}} = \frac{\textbf{u}(\textbf{x})-\textbf{v}}{St}+\textbf{x}\, \Omega^2 - 2\, \Omega \, \hat{\textbf{e}}_z\times \textbf{v}~,
        \label{eqn2p1}
\end{equation}
where $\textbf{v}$ is the particle velocity, $\textbf{u}(\textbf{x})$ is the velocity of the Kirchhoff vortex (steady, in the co-rotating frame) evaluated at the particle location $\textbf{x}$, $\hat{\textbf{e}}_z$ is the unit vector along the $z$ axis perpendicular to the plane. We use the length scale $\sqrt{a\, b}$ and the rotation time scale $\omega_0^{-1}$ to nondimensionalize the system, where, as mentioned earlier, $a$ and $b$ are the semi-major and semi-minor axes of the ellipse, respectively, and $\omega_0$ is the uniform vorticity of the elliptical patch. Another relevant time scale in the problem is $\tau_p=2/9\rho_p a^2/\rho_g\nu$ - the relaxation time scale for a particle of characteristic size $a$ and density $\rho_p$ navigating in a carrier phase of density $\rho_g$ and kinematic viscosity $\nu$ respectively. The Stokes number ($St = \tau_p\, \omega_0$) quantifies the relative magnitude of the two time scales and thus provides a nondimensional measure of particle inertia. We denote the nondimensional time derivative using overdot ($\, \, \dot{}$ ). The nondimensional quantities are represented with the same notation as dimensional quantities, as we deal only with nondimensional quantities from here onwards (unless specified explicitly). The elliptic vortex is characterized by the nondimensional aspect ratio $r = b/a$,
%, and without loss of generality, we only focus here on the cases of 
where we consider $0 < r < 1$.

The first term on the right-hand side of equation (\ref{eqn2p1}) is the Stokes drag, the second is the centrifugal force, and the third is the Coriolis force. The Euler force ($-\dot{\Omega} \, \hat{\textbf{e}}_z \times \textbf{x}$)—a fictitious tangential force that arises in a rotating reference frame—does not appear here because the rotation rate is uniform (i.e., $\dot{\Omega} = 0$). However, we will see later, in the context of the Kida vortex, the Euler force needs to be accounted for due to its non-uniform rotation rate. It is assumed that the particles are point size and thus in a Stokes flow limit ($Re=\Omega a^2/\nu \ll 1$). Also, we are dealing with heavy particles (i.e., much denser than the background fluid ($\rho_p \gg \rho_g$), as is the case for dust particles or water droplets in the air). Thus, as was mentioned earlier, the physics of added mass force and the Basset history effect is negligible. However, suppose one were to study the role of fluid inertia with the motivation of the current study and explore the long-time dispersion dynamics of particles in the vicinity of coherent structures. In that case, the inclusion of convective inertia ($Re \neq 0$) is expected to play a more prominent role than the Basset history effect \citep[see][]{lovalenti1993hydrodynamic,dorgan2007efficient}. However, for the sake of simplicity, we restrict ourselves to the $Re\ll 1$ regime and ignore the physics of both unsteady and convective inertia. In addition, for the same reason of $\rho_p \gg \rho_g$, the indirect effects of the Coriolis and centrifugal forces acting on the fluid, which induces corrections into the particle equation, have been neglected in equation (\ref{eqn2p1}).

\begin{figure}
    \centering
    \includegraphics[width=1.0\linewidth]{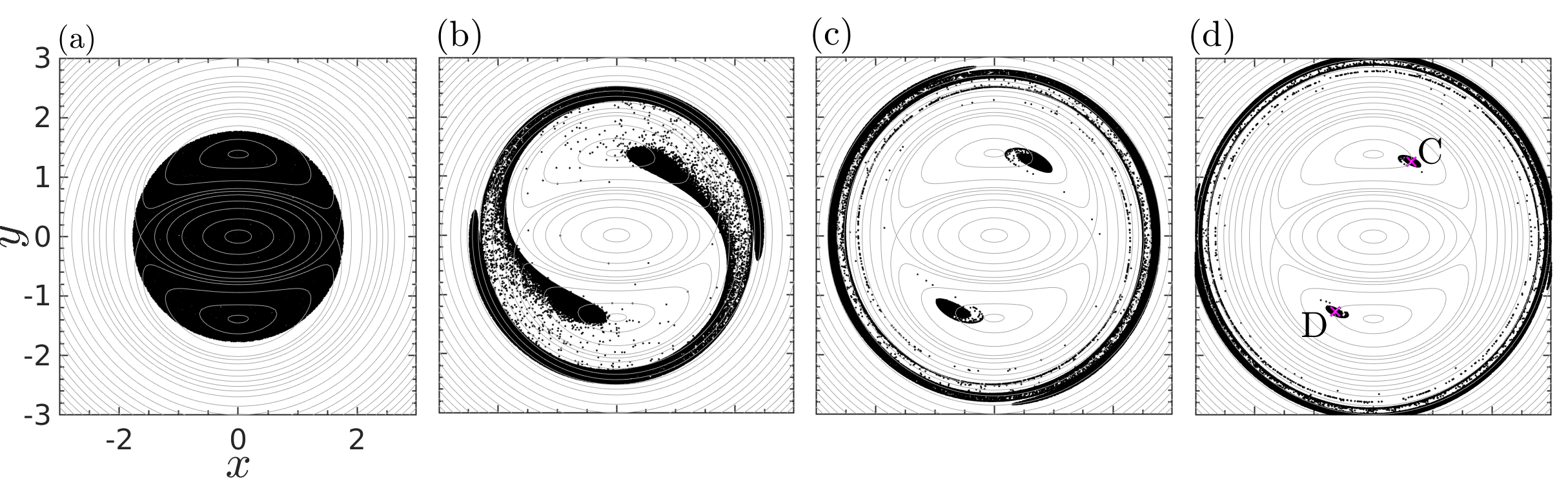}
    \caption{Snapshots showing the evolution of $10^5$ particles of $St = 0.5$ in a Kirchhoff vortex of $r = 0.5$ at (a) $t = 0$, (b) $t = 50$, (c) $t = 100$ and (d) $t = 150$ nondimensional time units, obtained from numerical simulation. The particles are initialized with zero velocity and randomly distributed inside a circle of nondimensional radius $1.76$, enclosing the ellipse.}
    \label{Fig002}
\end{figure}

In our study, we initialize a circular patch of inertial particles (randomly distributed) with zero initial velocity in the Kirchhoff vortex around the ellipse (see figure \ref{Fig002}(a)). We let the particles evolve and track them using the dynamic equation (\ref{eqn2p1}) along with the kinematic equation $\dot{\textbf{x}} = \textbf{v}$. We integrate the system using the ODE113 routine in Matlab with a relative error of $10^{-12}$, absolute error of $10^{-12}$ and a maximum time step of $1/10$ th of the Stokes number. The typical evolution of $St = 0.5$ particles in the Kirchhoff vortex of aspect ratio $r = 0.5$ observed from the co-rotating frame is shown as the snapshots in figure \ref{Fig002}.

As expected, the snapshots show that the inertial particles are getting centrifuged away from the central ellipse. However, unlike in the case of an axisymmetric vortex (like point vortex or Rankine vortex), here we see that some of the particles are getting attracted towards a pair of fixed points outside the ellipse, within each lobe denoted as C and D. Note that, the elliptic fixed points of fluid tracers in Kirchhoff vortex have already been denoted as C and D in figure \ref{Fig001}(b). The same choice of notation here for attracting fixed points will make sense in the analytical exploration of the system in the upcoming section. 

%%%%%%%%%%%%%%%%%%%%%%%%%%%%%%%%%%%%%%%
%%%%%%%%%%%%%%%%%%%%%%%%%%%%%%%%%%%%%%
%%%%%%%%%%%%%%%%%%%%%%%%%%%%%%%%%%%%%%%%%%
\subsection{\label{sec2p1} Analytical evaluation of fixed points and their stability}
The fixed points, as seen by an inertial particle in the Kirchhoff vortex, can be evaluated by setting its velocity and acceleration to be zero, i.e. $\textbf{v} = \dot{\textbf{v}} = \textbf{0}$. Substituting this in equation (\ref{eqn2p1}) gives that the locations of the fixed points are the solution of the equation $\textbf{u}(\textbf{x})+\textbf{x}\, St\, \Omega^2 = \textbf{0}$, i.e., the fixed points are formed by the balance between centrifugal force and Stokes drag. Note that the fixed point equation is modified from that of fluid tracers ($\textbf{u}(\textbf{x}) = 0$) due to the finite inertia effect as a $St$ dependent term. For analytical treatment to be made accessible, we may choose to rewrite the equation (\ref{eqn2p1}) in elliptic coordinates, in component form, as
\begin{subequations}
\begin{eqnarray}
    \ddot{\xi}&=&\frac{h\, k^{-1}\, u_\xi-\dot{\xi}}{St}+2\, \Omega\, \dot{\eta}+h^2\, \left\{\frac{\dot{\eta}^2-\dot{\xi}^2+\Omega^2}{2}\sinh 2\xi-\dot{\eta}\, \dot{\xi}\, \sin 2\eta \right\}~, \label{eqn2p2a}\\
    \ddot{\eta}&=&\frac{h\, k^{-1}\, u_\eta-\dot{\eta}}{St}-2\, \Omega\, \dot{\xi}-h^2\, \left\{\frac{\dot{\eta}^2-\dot{\xi}^2+\Omega^2}{2}\sin 2\eta+\dot{\eta}\, \dot{\xi}\,\sinh 2\xi\right\}~,  
    \label{eqn2p2b}
\end{eqnarray}
\label{eqn2p2}
\end{subequations}
where $x = k\, \cosh \xi\, \cos \eta$ and $y = k\, \sinh \xi\, \sin \eta$, and the fluid velocity components can be obtained from the corresponding stream function in co-rotating frame ($\psi$) as $u_\xi = \frac{h}{k}\, \frac{\partial \psi}{\partial \eta}$ and $u_\eta =  -\frac{h}{k}\, \frac{\partial \psi}{\partial \xi}$. For the Kirchhoff vortex (of stream function $\psi_v = \psi_v'+\frac{\Omega}{2}\, (x^2+y^2)$), these components can be evaluated as
\begin{subequations}
\begin{eqnarray}
    u_{\xi} &=& \left\{
\begin{array}{ll}
      \frac{h}{2\, k}\, \left(e^{-2\xi}-k^2\, \Omega \right)\, \sin 2\eta, & \quad \quad \quad \quad \quad \tanh \xi > r \vspace{0.25cm}\\
      \frac{h}{4}\, k^3\, \Omega\, (\Lambda-\cosh 2\xi)\, \sin 2\eta, & \quad \quad \quad \quad \quad \tanh \xi < r \\
\end{array} 
\right. \\
u_{\eta} &=& \left\{
\begin{array}{ll}
      \frac{h}{2\, k}\, \left(1-e^{-2\xi}\, \cos 2\eta-k^2\, \Omega \, \sinh 2\xi\right), & \quad \, \tanh \xi > r \vspace{0.25cm}\\
      \frac{h}{4}\, k^3\, \Omega\, (\Lambda-\cos 2\eta)\, \sinh 2\xi, & \quad \, \tanh \xi < r~, \\
\end{array} 
\right. 
\end{eqnarray}
\label{eqn2p3}
\end{subequations}
where the scale factor $h = (\cosh ^2\xi-\cos ^2\eta)^{-1/2}$, the parameters $k^2 = (1/r-r)$, $\Lambda = (1+r^2)/(1-r^2)$ and the angular velocity $\Omega = \dot{\theta} = r/(r+1)^2$. 
%The elliptic and the Cartesian co-ordinates are related as $x = k\, \cosh \xi\, \cos \eta$ and $y = k\, \sinh \xi \, \sin \eta$. 
Here, $\tanh \xi > r$ indicates the region outside the ellipse and $\tanh \xi < r$ indicates the region inside the ellipse since $\tanh \xi = r$ defines the boundary of the ellipse itself. The same equations in Cartesian coordinates are convenient in numerical simulations and can be found in Appendix \ref{appA}.

The system of equations (\ref{eqn2p2}), along with the appropriate velocity field, describes the trajectory of an inertial particle in an elliptic vortex. It is a nonlinear coupled dynamical system in a four-dimensional phase-space on the variables $\xi,\eta,\dot{\xi}$ and $\dot{\eta}$. The fixed points of the system can be obtained by solving the equations $\dot{\xi}=0$, $\dot{\eta}=0$, $\ddot{\xi} = 0$ and $\ddot{\eta}=0$ simultaneously. From equations (\ref{eqn2p2}), this gives the trivial criteria for all fixed points, i.e. $\dot{\overline{\xi}} = \dot{\overline{\eta}} = 0$; however, their locations $(\overline{\xi},\overline{\eta})$ should be obtained by solving the transcendental equations
\begin{subequations}
\label{eqn2p4}
\begin{eqnarray}
    2\, u_\xi + St\, h\, k\, \Omega^2\, \sinh 2\xi = 0~,\label{eqn2p4a}\\
    2\, u_\eta-St\, h\, k\, \Omega^2\, \sin 2\eta = 0~.  
    \label{eqn2p4b}
\end{eqnarray}
\end{subequations}
These equations must be solved separately inside and outside the ellipse to identify the fixed points since the velocity field is known in a piecewise manner. The identification of fixed points and their stability analysis are discussed in the following Sections \ref{sec2p1p1} and \ref{sec2p1p2}. Note that setting $St = 0$ should recover the dynamics of the passive fluid tracers. By doing so, one could see that the fixed point equations (\ref{eqn2p4}) reduce to $u_\xi = 0$ and $u_\eta = 0$ - which will retrieve the five classical fixed points of the Kirchhoff vortex mentioned in the introduction. 
%%%%%%%%%%%%%%%%%%%%%%%%%%%%%%%%%%%%%%%%%%%%%%%%%%%%%%%%%%%
% \begin{figure}
%         \begin{subfigure}[b]{0.5\textwidth}
%                 \includegraphics[width=\linewidth]{fixedpoints_r=0p5.eps}
%                % \caption{A gull}
%                 %\label{fig001a}
%         \end{subfigure}%
%         \begin{subfigure}[b]{0.5\textwidth}
%                 \includegraphics[width=\linewidth]{fixedpoints_r=0p5.eps}
%                 %\caption{A mouse}
%                 %\label{fig:mouse}
%         \end{subfigure}
%         \caption{(a) Fixed points for $r=0.5$, for log-spaced $St$ values. The merging of outer fixed points happens at $St_\textrm{cr}  \approx 0.772$. (b) The (steady) particle velocity filed (for $St = 0.2$) created by elliptic vortex patch of $r = 0.5$ in the co-rotating reference frame.}
%         \label{Fig003}
% \end{figure}
%%%%%%%%%%%%%%%%%%%%%%%%%%%%%%%%%%%%%%%%%%
%%%%%%%%%%%%%%%%%%%%%%%%%%%%%%%%%%%%%%%%%%%%%%%%%%%%%%%%%%%%%%%%%%%%%%%%%%%%%%%%%%%%%%%%%%%%%%%%%%%%%%%%%%%
\subsubsection{\label{sec2p1p1} Fixed points outside the ellipse}
\begin{figure}
    \centering
    \includegraphics[width=0.9\linewidth]{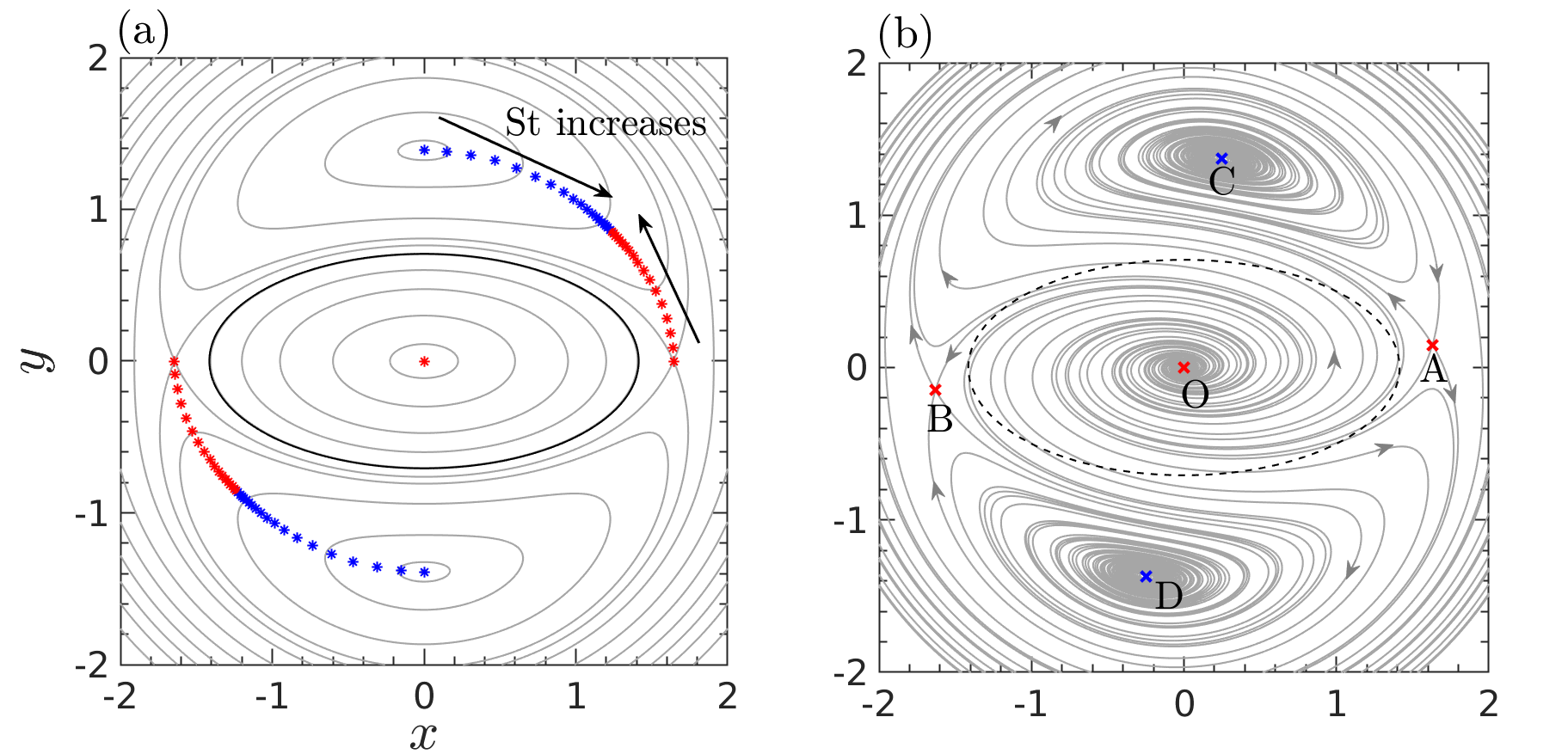}
    \caption{(a) Variation of the fixed points perceived by inertial particles in a Kirchhoff vortex of $r=0.5$ as $St$ changes (for log-spaced distribution of $St$). Elliptic and hyperbolic fixed points outside the ellipse merge at $St  \approx 0.772$. (b) The trajectories of $St = 0.2$ particles in a Kirchhoff of $r = 0.5$ in the co-rotating reference frame, obtained using slow manifold expression. Red indicates unstable fixed points, and blue indicates stable fixed points.}
    \label{Fig003}
\end{figure}

Outside the ellipse ($\tanh \xi > r$), the fixed point equations (\ref{eqn2p4}) can be written by substituting the appropriate velocity expressions from equations (\ref{eqn2p3}) as
\begin{subequations}
\begin{eqnarray}
    (St\, \Omega^2\, \sinh 2\xi+(k^{-2}\, e^{-2\xi}-\Omega)\, \sin 2\eta)\, h^2 = 0~,\\
    (St\, \Omega^2\, \sin 2\eta+\Omega\, \sinh 2\xi+k^{-2}\, (e^{-2\xi}\, \cos 2\eta-1))\, h^2 = 0~.
\end{eqnarray}
\label{eqn2p5}
\end{subequations}
These are modified equations for fixed points outside the elliptic vortex accounting for the effect of finite $St$. 
%Note that by taking the limit of $St \rightarrow 0$, these equations will reduce to the equations (14) in \citet{kawakami1999chaotic}, which gives the fixed points for fluid tracers. 
The solutions to equations (\ref{eqn2p5}) gives the fixed points ($\overline{\xi},\overline{\eta}$) with $p = \tanh \overline{\xi}$ and $q = \tan \overline{\eta}$ are given by
\begin{subequations}
\label{eqn2p6}
\begin{eqnarray}
    p^{\pm} &=& \frac{2-(\alpha \pm \beta)\, k^2\, \Omega^2}{2\,(1+k^2\, \Omega)}~,\\
    q^{\pm} &=& \frac{\alpha \mp \beta}{2\, St}~,
\end{eqnarray}
\end{subequations}
where $\alpha = k^2\, (1+St^2\, \Omega^2)$ and $\beta = \sqrt{\alpha^2-4 \, St^2}$. These solutions form a set of four fixed points outside the ellipse, located at ($\overline{\xi}_1^{+}$, $\overline{\eta}_1^{+}$) =$(\tanh^{-1}p^+,\tan^{-1}q^+)$, ($\overline{\xi}_1^{-}$, $\overline{\eta}_1^{-}$) =$(\tanh^{-1}p^+,-\pi+\tan^{-1}q^+)$, ($\overline{\xi}_2^{+}$, $\overline{\eta}_2^{+}$) =  $(\tanh^{-1}p^-,\tan^{-1}q^-)$ and ($\overline{\xi}_2^{-}$, $\overline{\eta}_2^{-}$) =$(\tanh^{-1}p^-,-\pi+\tan^{-1}q^-)$. In the limit of $St \rightarrow 0$, we may deduce that these fixed points coincide with the Kirchhoff vortex's four classical fixed points A, B, C and D. For simplicity, we choose to call these finite $St$ modified fixed points with the same name as that corresponding to the fluid tracers (A, B, C, and D). The finite $St$ symmetrically displaces these fixed points, as shown in figure \ref{Fig003}(a): A and B shift counter-clockwise, while C and D shift clockwise as $St$ increases. As $St$ increases, the fixed points A($\overline{\xi}_1^{+}$, $\overline{\eta}_1^{+}$) and C($\overline{\xi}_2^{+}$, $\overline{\eta}_2^{+}$) approaches each other. The same thing happens for the counterpart fixed points B($\overline{\xi}_1^{-}$, $\overline{\eta}_1^{-}$) and D($\overline{\xi}_2^{-}$, $\overline{\eta}_2^{-}$) as well. 

In the limit of small Stokes number ($St \ll 1$), the governing equation (\ref{eqn2p1}) can be reduced to the slow-manifold form as
\begin{equation}
    \textbf{v} = \textbf{u}-St\, \left\{\frac{\partial \textbf{u}}{\partial t}+\textbf{u} \cdot \boldsymbol{\nabla}\textbf{u} - \textbf{x}\, \Omega^2 + 2 \,\Omega\, \hat{\textbf{e}}_z \times \textbf{u}\right\}+\textit{O}(St^2)~.
    \label{eqn2p7}
\end{equation}
using which the particle trajectories are obtained and shown in the figure \ref{Fig003}(b). By comparing these particle trajectories with the streamlines for fluid tracers shown in figure \ref{Fig001}(b), it is evident that the inertial particles perceive different fixed points compared to fluid tracers. They match exactly with the solution of equation (\ref{eqn2p5}) we obtained analytically (as marked in blue and red). The trajectories also imply that the fixed points A and B remain hyperbolic fixed points (saddles); however, the fixed points C and D behave as stable spirals for inertial particles. The divergence of the system (\ref{eqn2p1}) in the four-dimensional phase space is given by $\boldsymbol{\nabla} \cdot \textbf{v}+\boldsymbol{\nabla}_{\textbf{v}} \cdot \dot{\textbf{v}} = -2/St$, a negative quantity, indicating the dissipative nature of the dynamical system for any finite inertial particles. Here, $\boldsymbol{\nabla} \cdot ()$ represents the divergence in physical space, while $\boldsymbol{\nabla}_{\textbf{v}} \cdot ()$ represents the divergence in velocity/momentum space. Even in the $St \ll 1$ limit, the divergence of the reduced system (\ref{eqn2p7}) in the two-dimensional phase space yields $\boldsymbol{\nabla} \cdot \textbf{v} =\textit{O}(St) \neq 0$, indicating compressible nature of inertial particle flow in phase space unlike the fluid tracers where $\boldsymbol{\nabla} \cdot \textbf{u} = 0$. The dissipative nature of the system for finite inertia particles is responsible for the elliptic fixed points becoming spiral attractors. However, the hyperbolic fixed points remain intact, though their position get displaced.

To analytically show this, we use the linear stability analysis and systematically study the effect of finite $St$ on the stability of the fixed points. The Jacobian matrix for the system of differential equations~(\ref{eqn2p2}), evaluated at the fixed point ($\overline{\xi},\overline{\eta}$) is
\begin{equation}
\mathsfbi{J} = 
\begin{pmatrix}
0 & 0 & 1 & 0\\
0 & 0 & 0 & 1\\
\Gamma & -\Delta & -\frac{1}{St} & 2\, \Omega\\
\Delta-\frac{2\, k^{-2}\,(1-p)\, (p+q^2) }{St\, (p^2+q^2)} & \Gamma+\frac{2\, k^{-2}\, q\, (1-p)^2}{St\, (p^2+q^2)} & -2\, \Omega & -\frac{1}{St}
\end{pmatrix}~,
\label{eqn2p8}
\end{equation}
%%%%%%%%%%%%%%%%%%%%%%%%%%%%%%%%%%%%%%%%
%fixed_points_and_particle_streamfunction
%%%%%%%%%%%%%%%%%%%%%%%%%%%%%%%%%%%%%%%%
%%%%%%%%%%%%%%%%%%%%%%%%%%%%%%%%%%%%%
%EigenValues_r=0p5_new
\begin{figure}
    \centering
    \includegraphics[width=\linewidth]{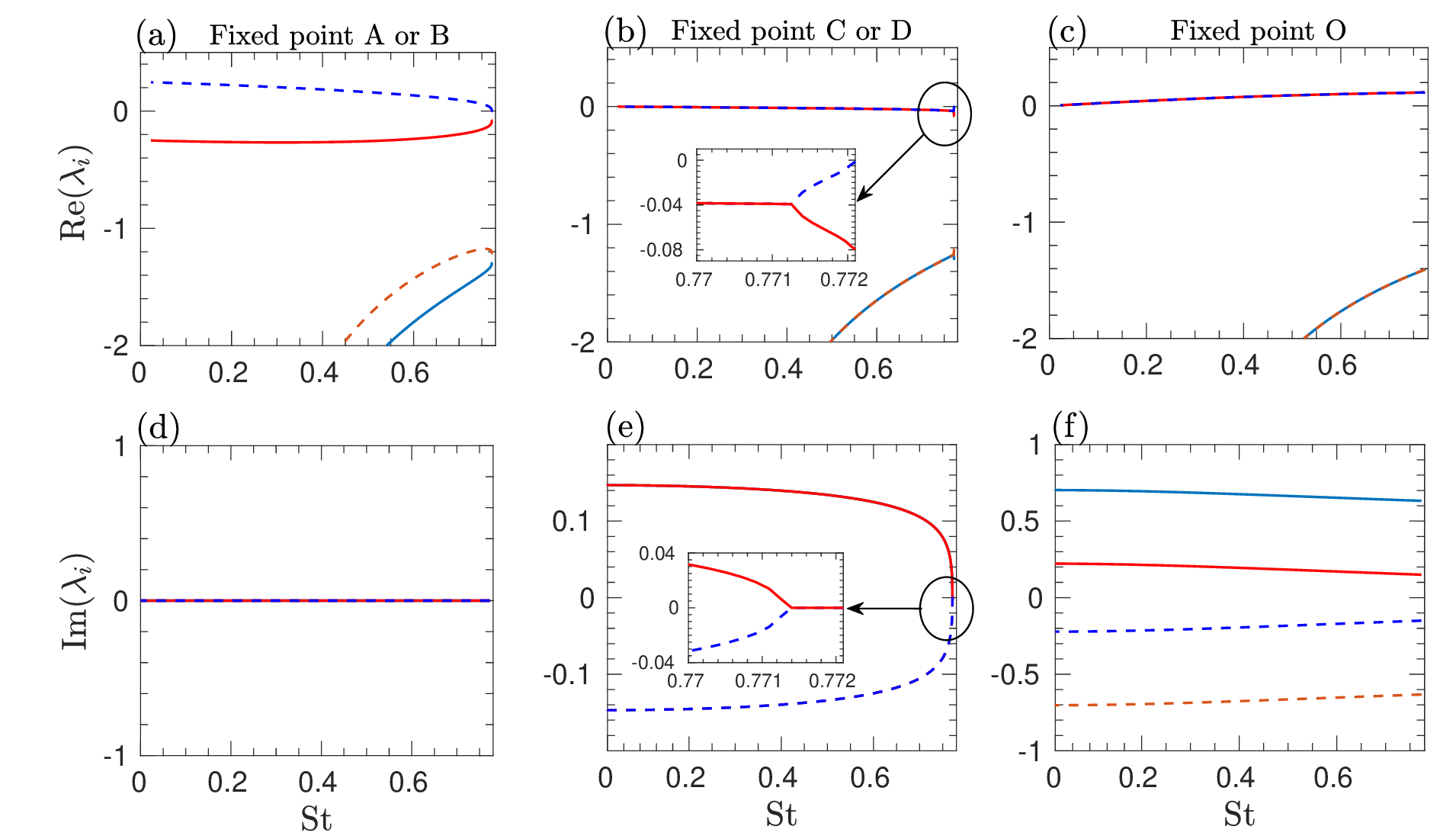}
    \caption{Variation of all eigenvalues with $St$ corresponding to the typical fixed points of inertial particles in a Kirchhoff vortex of $r=0.5$ : (a,d) Real and imaginary parts of the eigenvalues of fixed points A or B. (b,e) Real and imaginary parts of the eigenvalues of fixed point C or D. (c,f) Real and imaginary parts of the eigenvalues of fixed point O.
    %(Note that the variation of eigenvalues for fixed points B and D resemble that of the A and C respectively, and thus not explicitly shown here.)
    }
    \label{Fig004}
\end{figure}
%%%%%%%%%%%%%%%%%%%%%%%%%%%%%%%%%%%%%%%%%
where $\Gamma=(p^2-1)\{\Omega\, (1+q^2)\, (St\, \Omega\, (p^2-q^2)-2\,p\,q)+2\,k^{-2}\,(p+q^2)\,q\,(1-p)\}/\{St\, (p^2+q^2)^2\}$ and $\Delta = (1-p)\, (1+q^2)\, \{\Omega\, (1+p)\,(p^2-q^2+2\, p\, q\, St\, \Omega)-k^{-2}\,(1-p)\, (p^2-q^2) \}/\{St\, (p^2+q^2)^2\}$. The same matrix evaluated in Cartesian coordinates can be found in Appendix \ref{appA}. By evaluating the eigenvalues for the matrix $\mathsfbi{J}$, we can identify that the fixed point pairs A and B are saddles, and the pairs C and D are stable spirals for a finite inertial particle. For instance, the variation of all the eigenvalues with Stokes number for each fixed point type is shown in figure \ref{Fig004}. From figures \ref{Fig004}(a) and (d), it can be seen that any finite $St$ particle will perceive fixed points A \& B with purely real-valued eigenvalues with signature $(+,-,-,-)$ indicating saddles in the four-dimensional phase space. However, for the fixed point C/D, as shown in figures \ref{Fig004}(b) and (e), the eigenvalue will be a pair of complex conjugates with a negative real part, indicating a stable spiral in the four-dimensional phase space. Thus, we conclude that the suspended heavy inertial particles will spiral and cluster towards the fixed points C and D outside the elliptic vortex as time progresses, which we have observed in numerical simulations (see figure \ref{Fig002}). Note that, for the fixed point C/D, for $r=0.5$, when $St \gtrsim 0.771$ (shown in the insets of figures \ref{Fig004}(b) and (e)), we may see that there occurs a bifurcation and the eigenvalues are no more conjugate pairs, rather purely real-valued with signature $(-,-,-,-)$, indicating stable node/sink in the four-dimensional phase space. Consequently, the particles will radially move towards the fixed points instead of executing a spiral motion.
% For an elliptic vortex of aspect ratio $r = 0.5$, this critical value of Stokes number can be numerically identified as $St \approx 0.771$. However, by investigating the discriminant of the solutions of quartic eigenvalue polynomial for $\mathsfbi{J}$ (which can be obtained from equation (\ref{eqn2p8}) or equation (\ref{eqnA4})), the general expression for the Stokes number at which this behaviour change happens can be obtained.
For an elliptic vortex of arbitrary aspect ratio $r$, the critical Stokes number $\mathcal{S}_1$ at which the spiral to node transition happens can be identified from the discriminant of the solutions of quartic eigenvalue polynomial for $\mathsfbi{J}$ (which can be obtained from equation (\ref{eqn2p8}) or equation (\ref{eqnA4}), by setting $\lvert \mathsfbi{J}-\lambda\, \mathsfbi{I} \rvert = 0$). By examining the discriminant, one may find its behaviour changes when $St^2$ satisfies the cubic equation
\begin{equation}
\begin{split}
        St^6+\frac{3\,(1+r)^4\,(1+4\,r+r^2)}{2\, r^3}\, St^4+\frac{(1+r)^8\,(1+6\,r+r^2)\,(5+22\,r+5\, r^2)}{16\, r^6}\,St^2 \\-\frac{(1+r)^{12}\,(1-r)^2\,(3+r)\,(1+3\,r)}{16\, r^8}=0~.
    \label{eqn2p9}
    \end{split}
\end{equation}
The real-valued, non-negative solution of the equation (\ref{eqn2p9}) gives the critical Stokes number $St = \mathcal{S}_1$ for any $r$. For the case of $r = 0.5$, one could verify that this solution is $\mathcal{S}_1 \approx 0.7713$, matching with the prediction from figures \ref{Fig004}(b) and (e).

As $St$ increases, saddle type and stable spiral type fixed points approach closer, as mentioned earlier. If we keep increasing $St$, we will find that they merge (i.e., $p^+ = p^-$ and $q^+=q^-$) and vanish at a critical inertia value. From equation (\ref{eqn2p6}), one could deduce that this happens only if $\beta = 0$. Using the expression $\beta^2 = k^4\, (1+St^2\, \Omega^2)^2-4\, St^2$ and solving for Stokes number, one would find the critical value at which merging happens is $St = \mathcal{S}_2 = (1+r)^2\, (1-\sqrt{r})/(r\, (1+\sqrt{r}))$. i.e., the fixed points outside the elliptic vortex (A, B, C and D) exist only for particles with $0 \leq St < \mathcal{S}_2$. For an elliptic vortex with $r = 0.5$, the critical value can be evaluated as $\mathcal{S}_2 \approx 0.7721$, i.e. A merges with C and B merges with D at this critical value and disappears. Thus, we have only shown the eigenvalues in the figure \ref{Fig004} for Stokes number until $0.7721$.

% An analytical exercise on equations (\ref{eqn3p8}) yields that the fixed points can reappear for larger values of Stokes number if $St \in \left(\frac{(1+r)^2\, (1+\sqrt{r})}{r\, (1-\sqrt{r})},\frac{(1+r)^2\, \sqrt{(3+r)\, (1+3\, r)}}{r\, (1-r)}\right)$. However, this range corresponds to very large values of Stokes numbers, practically violating our assumptions. For example, in the case of $r = 0.5$, this range is $St \in (26.228,26.622)$. Thus, we ignore them in our analysis. 
%%%%%%%%%%%%%%%%%%%%%%%%%%%%%%%%%%%%%%%%%%%%%%%%%%%%%%%%%%%%%%%%%%%%%%%%%%%%%%%%%%%%%%%%%%%%%%%%%%%%%%%%%%%%%%%%%%%%%%%%%%%%%%%%%%%%%%%%%%%%%%%%%%%%%%%%%%%%%%%%%%%
\subsubsection{\label{sec2p1p2} Fixed points inside the ellipse}
The origin O located inside the ellipse ($\tanh \xi < r$) continues to exist for inertial particles (see figure \ref{Fig003}(b)). However, for inertial particles, O behaves as an unstable spiral. To show this, we follow the same procedure mentioned in the previous Section \ref{sec2p1p1}, using the flow field inside the ellipse. By substituting appropriate expressions for $u_{\xi}$ and $u_{\eta}$ from equations (\ref{eqn2p3}) in the fixed point equations (\ref{eqn2p4}), we obtain 
\begin{subequations}
    \begin{eqnarray}
    (2\, St\, \Omega\, \sinh 2\xi+k^{2}\,(\Lambda -\cosh 2\xi)\, \sin 2\eta)\, h^2 = 0~,\\
    (2\, St\, \Omega\, \sin 2\eta-k^2\, (\Lambda -\cos 2\eta)\, \sinh 2\xi)\, h^2 = 0~.
\end{eqnarray}
\label{eqn2p10}
\end{subequations}
 In the $St \rightarrow 0$ limit, these equations yield the elliptic fixed point for fluid tracers at the origin. Moreover, irrespective of the value of $St$, equations (\ref{eqn2p10}) has a single real-valued solution $(\overline{\xi}_0,\overline{\eta}_0) = (0,\pi/2)$, indicating that the origin O remains to be the fixed point for any inertial particle as well. The Jacobian matrix for linear stability for the fixed point at origin O is,
\begin{equation}
\mathsfbi{J} = 
\frac{1}{St}\,\begin{pmatrix}
0 & 0 & St & 0\\
0 & 0 & 0 & St\\
St\,\Omega^2 & -r\, \Omega & -1 & 2\, St\,\Omega\\
\Omega/r & St\, \Omega^2 & -2\, St\, \Omega & -1
\end{pmatrix}~,
\label{eqn2p11}
\end{equation}
which has two pairs of complex conjugate eigenvalues. One of these pairs has a non-negative real part (see figures \ref{Fig004}(c) and (f)), indicating unstable spiral behaviour for any nonzero $St$. Thus, the particles will be centrifuged away from the origin spirally. Some of them, starting within certain basins (see Section \ref{sec2p3}) accumulate in the stable fixed points C and D outside the ellipse, as they are stable fixed points, as shown in the previous Section \ref{sec2p1p1}. Note that as we increase the particle inertia, even after the mutual annihilation of fixed points A, B, C, and D, fixed point O exists and remains an unstable spiral. Thus, beyond the $\mathcal{S}_2$, one can observe that all particles merely centrifuge away without getting trapped in any Lagrange points.

% We randomly initialized $10^4$ particles inside a circular region of unit radius (non-dimensional) around the origin, with zero initial velocity. We tracked each of them by numerically integrating equations (\ref{eqn3p5}) (we used the Cartesian version of the equations for this purpose, which can be found in Appendix \ref{appA}) using a fourth-order Runge-Kutta scheme. For particles of $St = 0.5$ in an elliptic vortex of $r = 0.5$, their distribution after non-dimensional time $t = 150$ units is shown in figure \ref{Fig002}(a). A fraction of the particles are found to be spirally clustering towards the fixed points C and D, and the remaining fraction is spiralling away to the unbounded fluid. 

% figure \ref{Fig002}(b) shows the filled contourplot of the Okubo–Weiss parameter $\mathcal{Q}$ in the rotating reference frame (see equation (\ref{eqn2p4})) for the elliptic vortex of aspect ratio $r=0.5$. The figure shows that the parameter is positive inside the elliptic region, indicating a source behaviour, and negative outside the elliptic region, indicating a sink behaviour. The negative value of $\mathcal{Q}$ fulfils the requirements for attracting fixed points outside the elliptic vortex. % Near the edge of major axes, the value of $\mathcal{Q}$ is the smallest, indicating clustering

%%%%%%%%%%%%%%%%%%%%%%%%%%%%%%%%%%%%%%%%%%%%%%%%%%%%%%%%%%%%%%%%%%
%FixedPointTransitionsSchematic
\begin{figure}
    \centering
\includegraphics[width=0.8\linewidth]{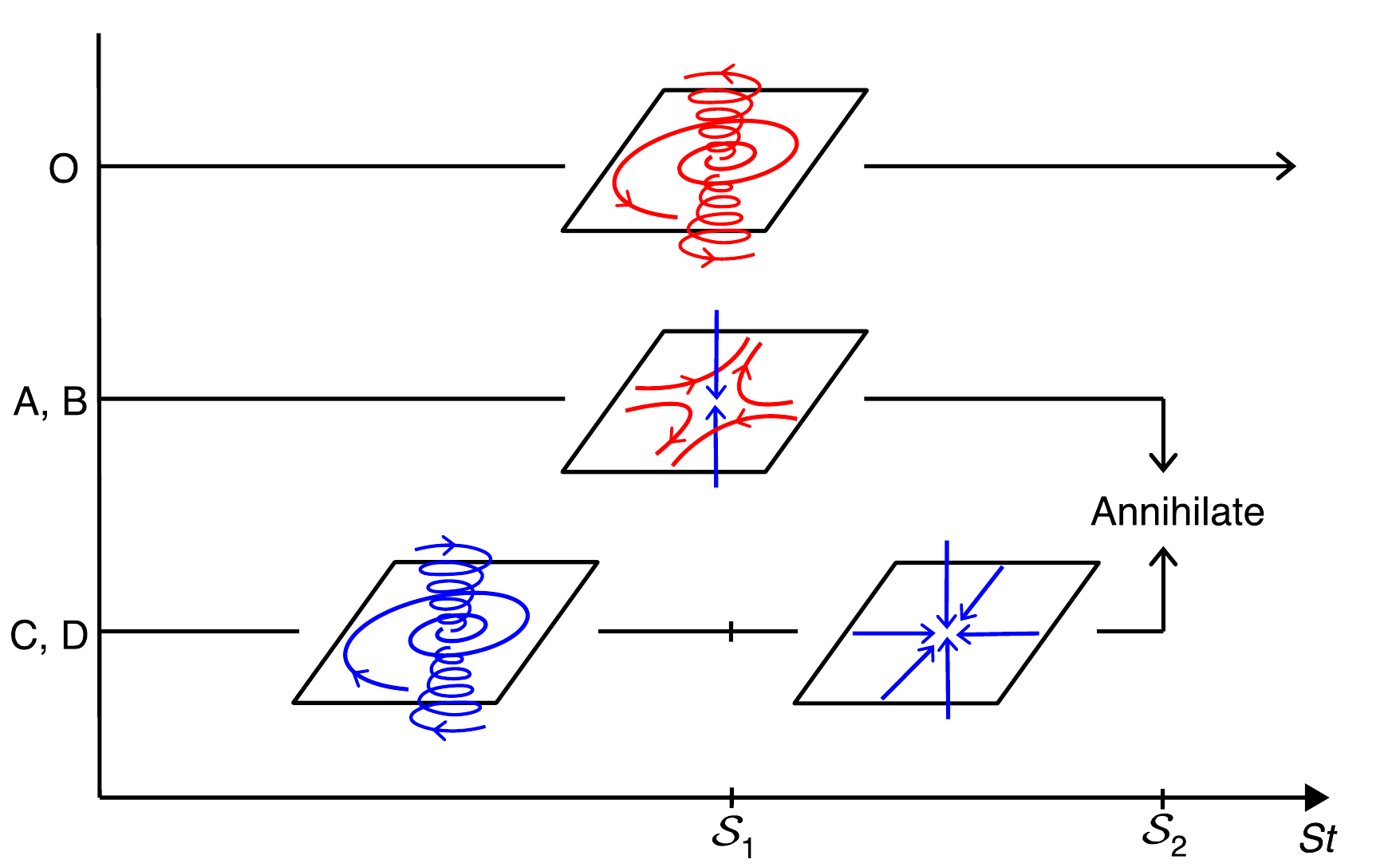}
    \caption{A schematic showing the projection of a four-dimensional phase space topology into three dimensions, illustrating various fixed points and their transitions at critical Stokes numbers.}
    \label{Fig005}
\end{figure}
%%%%%%%%%%%%%%%%%%%%%%%%%%%%%%%%%%%%%%
%%%%%%%%%%%%%%%%
The nature of fixed points and phase space topology is depicted schematically in figure \ref{Fig005} as the Stokes number varies. It is important to note that the phase space is four-dimensional; however, the schematic only presents a three-dimensional projection. The fixed points C/D exhibit a `2-spiral sink' behavior when $St < \mathcal{S}_1$ and transition to a `sink' when $\mathcal{S}_1 < St < \mathcal{S}_2$. On the other hand, fixed points A/B demonstrate a `3:1 saddle' behaviour for $St < \mathcal{S}_2$. Additionally, fixed points A (B) annihilate with C (D) and cease to exist when $St = \mathcal{S}_2$. Meanwhile, fixed point O persists and behaves as a `2-spiral saddle' for all Stokes numbers. For more on terminology, see \cite{hofmann2018visualization}.
%%%%%%%%%%%%%%%%%%%%%%%%%%%%%%%%%%%%%%%%%%%%%%%%%%%%%%%%%%%%%%%%%%
%reg_fps.eps
\begin{figure}
    \centering
\includegraphics[width=0.9\linewidth]{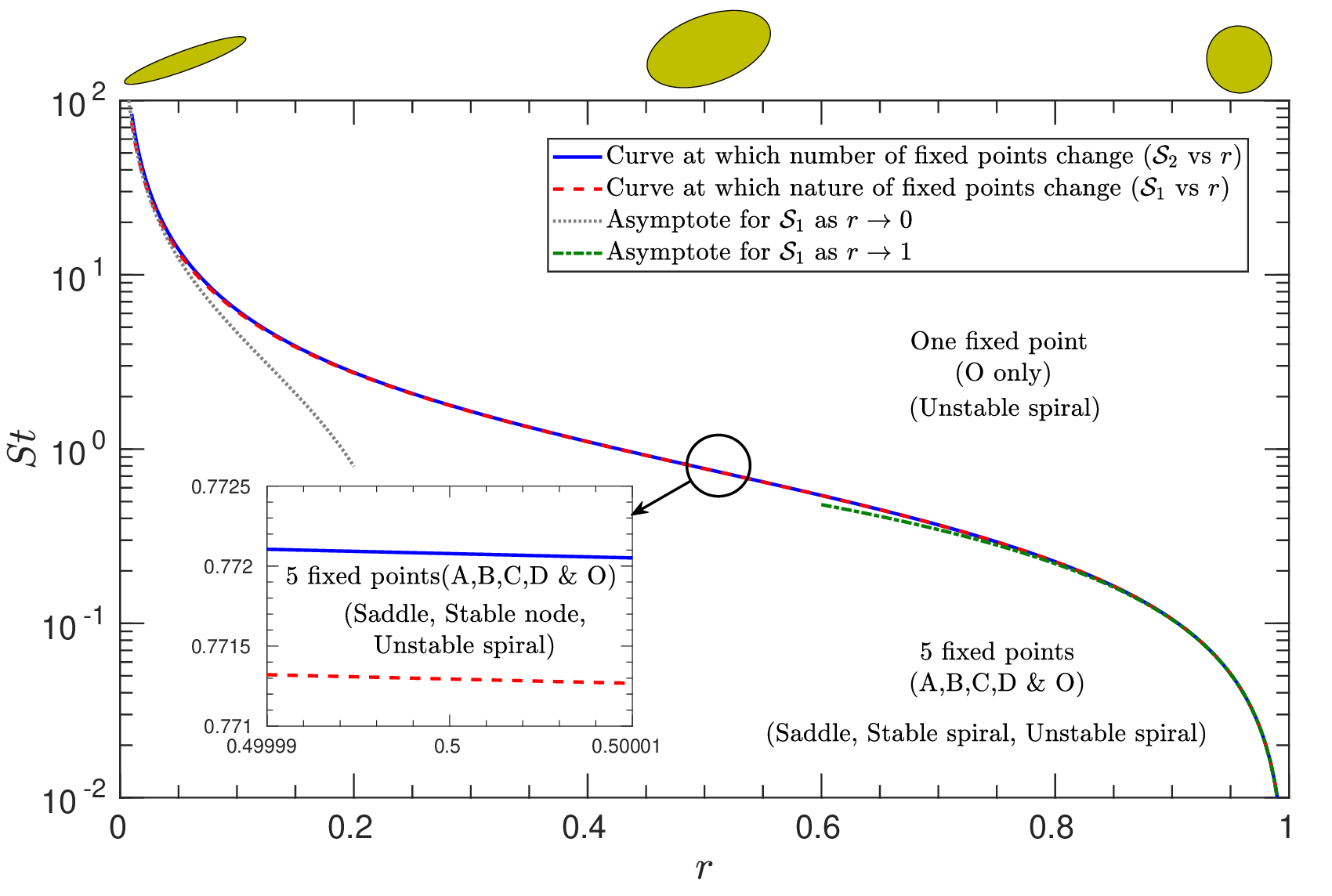}
    \caption{The curves in the $r-St$ plane demarcate regions where the change in the number of fixed points (blue colour) and the change in the nature of stable fixed points (red colour) happens. The representative elliptic patch of vorticity is also shown in greenish yellow color for three different aspect ratios.}
    \label{Fig006}
\end{figure}
%%%%%%%%%%%%%%%%%%%%%%%%%%%%%%%%%%%%%%
%%%%%%%%%%%%%%%%
%%%%%%%%%%%%%%%%%%%%%%%%%%%%%%%%%%%%%%%%%%%%%%%%%%%%%%%%%%%%%%%%%%%%%%%%%%%%%%%%%%%%%%%%%%%%%%%%%%%%%%%%%%%%%%%%%%%%%
\subsection{\label{sec2p2} Effect of the aspect ratio on the distribution of fixed points}
Until now, all the analyses have been performed explicitly for the elliptic vortex with aspect ratio $r = 0.5 $. However, if the vortex has a different aspect ratio, the distribution and stability nature of fixed points will change. Solving equations (\ref{eqn2p5}) and \ref{eqn2p10}), we obtain the fixed points inside and outside the ellipse, respectively, for various $St$ and $r$ values. There is no qualitative change in the distribution of fixed points, such that the fixed point inside the ellipse will always be at the origin, and there will always be four fixed points outside the ellipse below $\mathcal{S}_2$. In the $r - St$ plane (see figure \ref{Fig006}), it is shown that the critical curve (blue colour) which demarcates the region where all the five fixed points (A, B, C, D and O) coexists and the region where only the one at the origin only exists.

By analysing the stability of the fixed points, we have already seen that the fixed point at the origin remains an unstable spiral for any $St >0$ in a Kirchhoff vortex of $r=0.5$. One can verify that this will also be valid for any $r \in (0,1)$. The pair A and B remain saddles for all aspect ratios provided they exist (indicating the robustness of hyperbolic fixed points). However, the stable spiral fixed points can change their behaviour if the aspect ratio is above some critical value. The same fact has already been discussed towards the end of Section \ref{sec2p1p1} specifically for the aspect ratio $r = 0.5$. For any particular aspect ratio, some critical Stokes number $\mathcal{S}_1$ exists above which the stable spirals become stable nodes/sinks. As mentioned in Section \ref{sec2p1p1}, by solving for the non-negative real-valued solution of equation (\ref{eqn2p9}), one can find the critical pairs of $St$ and $r$ at which this behaviour change happens, and is shown in the $r - St$ plane (see figure \ref{Fig006}, red colour). From asymptotic analysis of equation (\ref{eqn2p9}), one may show that $\mathcal{S}_1 = r^{-1}\, \sqrt{3/5}-298/(25\, \sqrt{15})+\textit{O}(r)$ for $r \rightarrow 0$ and $\mathcal{S}_1 = (1-r) + \frac{1}{2}\, (1-r)^2+\textit{O}(1-r)^3$ for $r \rightarrow 1$, and the asymptotes are also shown in the figure. On the other hand the exact expression for $\mathcal{S}_2 = (1+r)^2\, (1-\sqrt{r})/(r\, (1+\sqrt{r}))$, mentioned at the end of Section \ref{sec2p1p1} has asymptotic forms, $\mathcal{S}_2 = r^{-1} -2/\sqrt{r}+4+\textit{O}(\sqrt{r})$ for $r \rightarrow 0$ and $\mathcal{S}_2 = (1-r) + \frac{1}{2}\, (1-r)^2+\textit{O}((1-r)^3)$ for $r \rightarrow 1$.
We find that $\mathcal{S}_1<\mathcal{S}_2$ for any $r \in (0,1)$. Note that the red curve is very close to the blue curve; however, they never intersect, indicating that for any fixed $St > 0$, there always exists a narrow parameter regime bounded by both red and blue curves ($\mathcal{S}_1 < St < \mathcal{S}_2$) inside which the fixed points C and D will behave as nodes/sinks.

For any finite $r$ value, one may note that both $\mathcal{S}_1$ and $\mathcal{S}_2$ values are finite. However, as $r \rightarrow 0$, both diverge as $\textit{O}(1/r)$, becoming a larger value.  This divergence suggests that particles must possess high inertia for critical dynamical behavioural changes to occur. However, note that when $r \rightarrow 0$, the vortex rotates slowly ($\Omega = r/(1+r)^2 \rightarrow 0$). Thus, the relevant timescale in the problem becomes the angular velocity $\Omega$ rather than the vorticity of the patch. Consequently, the Stokes number based on the angular velocity of the vortex, $St_{\Omega} = St \, r/(r+1)^2$, evaluated for critical behaviours $\mathcal{S}_{\Omega,1} \sim \sqrt{3/5}$ and $\mathcal{S}_{\Omega,2} \sim 1$, remains constant and bounded in the limit of $r \rightarrow 0$, indicating that the divergence was merely an artefact of the adopted scaling.

%%%%%%%%%%%%%%%%%%%%%%%%%%%%%%%%%%%%%%%%%%%%%%%%%%%%%%%%%%%%%%%%%%%%%%%%%%%%%%%%%%%%%%%%%%%%%%%%%%%%%%%%%%%%%%%%%%%%%%%%%
\subsection{\label{sec2p3} The basin of attraction of the fixed points}
\begin{figure}
    \centering
    \includegraphics[width=\linewidth]{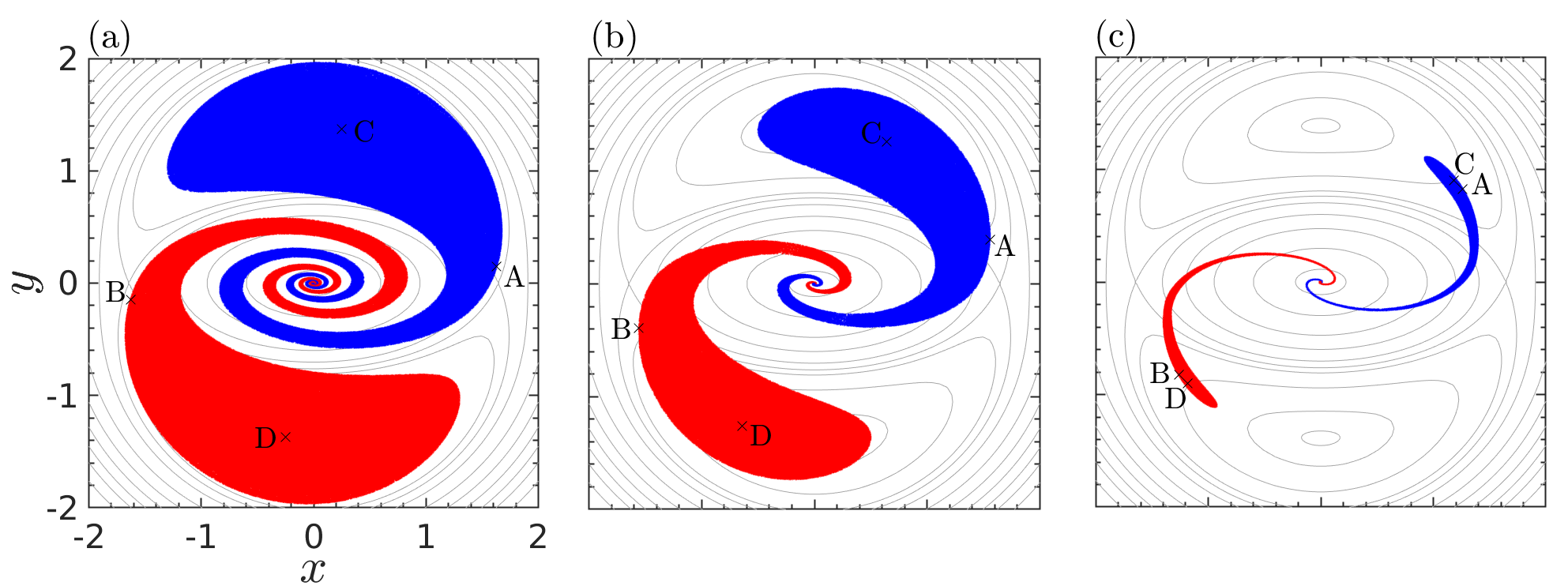}
    \caption{The basin of attraction for inertial particles of (a) $St = 0.2$, (b) $St = 0.5$ and (c) $St = 0.77$ in a Kirchhoff vortex of $r=0.5$. The corresponding fixed points A, B, C and D are marked in each figure using the `$\times$' symbol. The streamlines of the Kirchhoff vortex are shown in grey in the background.}
    \label{Fig007}
\end{figure}

From numerical simulations, we have seen that all the particles repelled away from the unstable fixed points (A, B and O) are not attracted to the stable fixed points (C and D); instead, some spiral away to infinity. The initial particle locations that would result in them getting attracted to any of the two stable fixed points are shown in figure \ref{Fig007}, representing the basin of attraction of those fixed points. The basins are coloured to distinguish each other. The region outside these coloured regions indicates the basin of attraction of infinity; i.e., particles initialized in these regions will eventually spiral away to infinity and never get attracted to any of the stable fixed points. We have used $10^5$ particles randomly initialized in the flow field to generate the figure. We tracked their evolution numerically and identified those that would approach any fixed points over a long time. We marked their initial locations using specific colours to distinguish between the basins of each fixed point. As visible in figure \ref{Fig007}, the basin of attraction of the stable fixed points shrinks as $St$ increases and vanishes. It can be verified that the basins will disappear beyond the critical value $\mathcal{S}_2$, indicating the absence of stable fixed points. The figure shows that the stable fixed points C and D are enclosed within the corresponding basin of attraction, which is obvious. However, the unstable fixed points A and B appear to fall at the edge of the basins. The regular nature of the basins shows non-chaotic dynamics of inertial particles in the elliptic vortex. Close to origin O, the basin of attraction of stable fixed points (coloured regions), and that of infinity (empty region) forms an inter-twisting pattern. As a result, a particle starting closer to the origin will have dynamics sensitive to the initial condition. A small change in the initial position could lead the particle to end up either in the fixed points C or D or spiral away to infinity.

%It is appropriate to mention the study of 
Here,  we would like to highlight prior studies on the dynamics of inertial particles in the neighbourhood of a pair of like-signed point vortices by \citet{angilella2010dust,ravichandran2014attracting,zhao2024trapping}. The system of like-signed point vortices bears similarities with that of the Kirchhoff vortex. 
%The system has strikingly identical behaviour to that of the Kirchhoff vortex. 
The inertial particles exhibit qualitatively identical behaviour near the point vortex pair. The types of fixed points, stability characteristics, and the basin of attractions show qualitative similarities to those observed in the case of the elliptic vortex. A recent study by \citet{kapoor2024trapping} extends previous studies near the like-signed point vortex pair by focusing on the dynamics of inertial particles of any density ratio. The study reveals interesting dynamical behaviours, with the existence of periodic and chaotic dynamics of particles depending on their inertia and density ratio. Additionally, in the limit of a large density ratio, the particle dynamics show similarity to that in a Kirchhoff vortex. Though it seems like an elliptic vortex can be simply replaced by a pair of like-signed vortices, it is worthwhile to focus on the subtle differences. The flow field in both cases matches well in the far field only. The Kirchhoff vortex and the pair vortices have different features in the near-field. Notably, the origin behaves as a saddle in the case of pair vortices, whereas in the elliptic vortex, it acts as a center/spiral. 

%As a result, the number of heteroclinic orbits is also higher. However, one may notice that the flow-fields will resemble each other better as the separation of point vortices reduces (during a merging process). Thus, we hope that the qualitative similarity in the dynamics of inertial particles observed here is adequately justified. 
%%%%%%%%%%%%%%%%%%%%%%%%
%%%%%%%%%%%%%%%%%%%%%%%%%%%%%%%%%%%%%%%
%basins_r=0p5_new
%%%%%%%%%%%%%%%%%%%%%%%%%%%%%%%%%%%%%%%%%%%%%%%%%%%%%%%%%%%%%%%%%%%%%%%%%%%%%%%%%%%%%%%%%
\section{\label{sec3} Dynamics of heavy inertial particles in a strained elliptic vortex}
\begin{figure}
    \centering \includegraphics[width=0.8\linewidth]{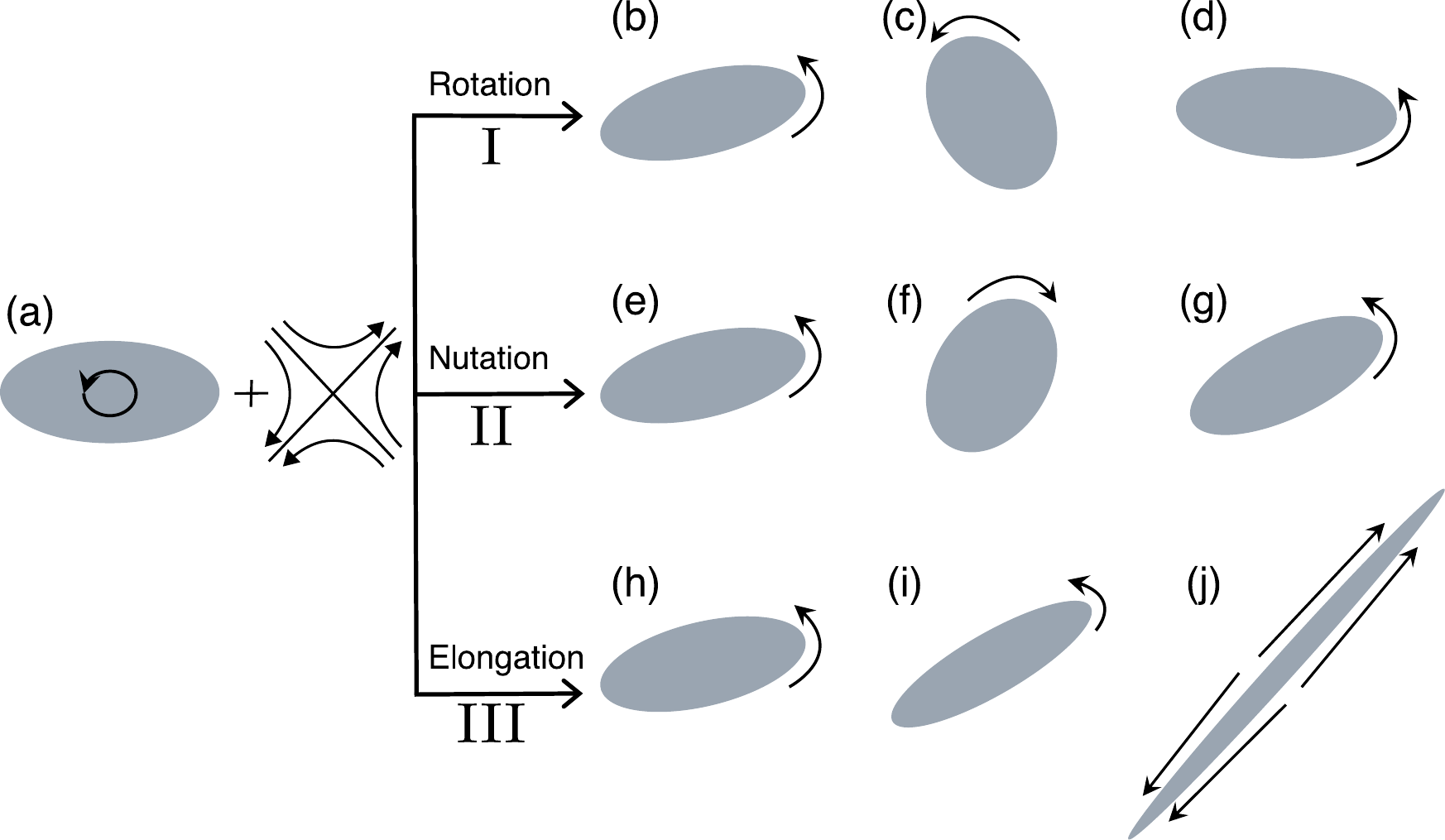}
    \caption{Schematic showing the three major types of dynamics exhibited by a strained elliptical vortex: (a) an initial elliptical vortex patch with anti-clockwise vorticity content superimposed with an external straining flow, (b-d) rotation: full rotation and periodic variation in the aspect ratio of the ellipse, (e-g) nutation: back-and-forth oscillation and periodic variation in the aspect ratio of the ellipse, and (h-j) elongation: the ellipse irreversibly stretches and aligns along the straining axes of the flow.}
    \label{Fig008}
\end{figure}
%Schematic showing the dynamics exhibited by a strained elliptical vortex: (a) initial vortex patch with anti-clockwise vorticity content strained by an external planar extensional flow (irrotational), (b-d) rotation: full rotation and periodic change in the aspect ratio of the ellipse due to weak external straining, (e-g) elongation: due to strong external straining, the vortex patch asymptotically aligns with and gets indefinitely stretched along the unstable axes of the extensional flow.
%%%%%%%%%%%%%%%%%%%%%%%%%%%%%%%%%%%%%%%%%%%%%%%%%%%%%%%%%%%%%%%%%%%%%%%%%%%%%%%%%%%%%%%%%
%$ - \frac{\gamma}{4}\, (x'^2 + y'^2)$, 
%and $\gamma$ is the vorticity,
    %+\frac{\gamma}{2}~, 
%and vorticity ($\gamma$)

When an external uniform shear flow is superimposed on the Kirchhoff elliptical vortex ($\psi'_v$), \cite{moore1971structure} was the first to obtain stationary elliptical vortices of uniform vorticity as exact solutions of the steady Euler equations. Later, \cite{kida1981motion} obtained the exact solutions of the unsteady Euler equations, resulting in unsteady elliptical vortices in a uniform shear flow, now known as the Kida vortex. In a special case of an external pure-strain flow of the form $\psi'_e = \frac{s}{4}\, (-x'^2 + y'^2)$, where $s$ is the strain rate, the dynamics of the elliptic vortex of uniform vorticity is governed by
\begin{subequations}
    \begin{eqnarray}
    \dot{r} &=& -s\, r\, \sin 2\theta~,\\
    \dot{\theta} &=& \Omega+\frac{s}{2}\,\Lambda\, \cos 2\theta~,
\end{eqnarray}
    \label{eqn3p1}
\end{subequations}
where, $r$ is the instantaneous aspect ratio and $\theta$ is the instantaneous orientation of the ellipse. The strain rate ($s$) in equations (\ref{eqn3p1}) is scaled with the uniform vorticity of the ellipse. The expressions for $\Omega = r/(r+1)^2$, $\Lambda = (1+r^2)/(1-r^2)$, and $k^2 = (1-r^2)/r$ remain consistent with those defined for the Kirchhoff vortex. However, for the Kida vortex, these parameters are defined using the instantaneous aspect ratio $r = r(t)$, which evolves over time. It is evident from equations (\ref{eqn3p1}) that the dynamics of the Kida vortex depend on the initial aspect ratio ($r(t=0) = r_0$), initial orientation ($\theta(t=0) = \theta_0$), and the strain rate ($s$) of the external flow. The dynamical system described in equations (\ref{eqn3p1}) possesses a conserved integral, as identified by \citet{neu1984dynamics,baylay1996three,meacham1997hamiltonian}, which is 
\begin{eqnarray}
    \mathcal{H}=\log\left(\frac{(1+r)^2}{4 \, r}\right)+s \frac{(1-r^2)}{2\, r}\, \cos 2\, \theta~.
    %+\gamma\,\frac{(1-r)^2}{2\, r}~. 
    \label{eqn3p2new}
\end{eqnarray}
The numerical value of $\mathcal{H}$ is determined by $r_0$, $\theta_0$ and $s$, and remains constant throughout the time evolution of the vortex.

Analysis of the system in equations (\ref{eqn3p1}) reveals that the unsteady Kida vortex can exhibit three distinct dynamical behaviours: rotation, nutation, and elongation \citep[see][]{kida1981motion,neu1984dynamics,dritschel1990stability}. Figure \ref{Fig008} schematically illustrates these dynamics of the Kida vortex. A general Kida vortex, as shown in \cite{kida1981motion}, which encounters a broader class of background linear flow (that has both straining and rotational components, for, e.g. simple shear flow), will also exhibit all these dynamical behaviours. However, in this study, we limit ourselves to the case of a Kida vortex under pure-strain (irrotational) background flow. An elliptical vortex patch conserves its area throughout its dynamical evolution, while its aspect ratio, angular velocity, and angular orientation vary over time. The ellipse undergoes rotation similar to the Kirchhoff vortex for low strain rates, albeit with periodically changing aspect ratio and angular velocity. However, the ellipse tends to align with the straining axis at high strain rates and elongate indefinitely. The ellipse can oscillate with a periodically varying aspect ratio under suitable conditions.

\begin{figure}
    \centering
\includegraphics[width=1.0\linewidth]{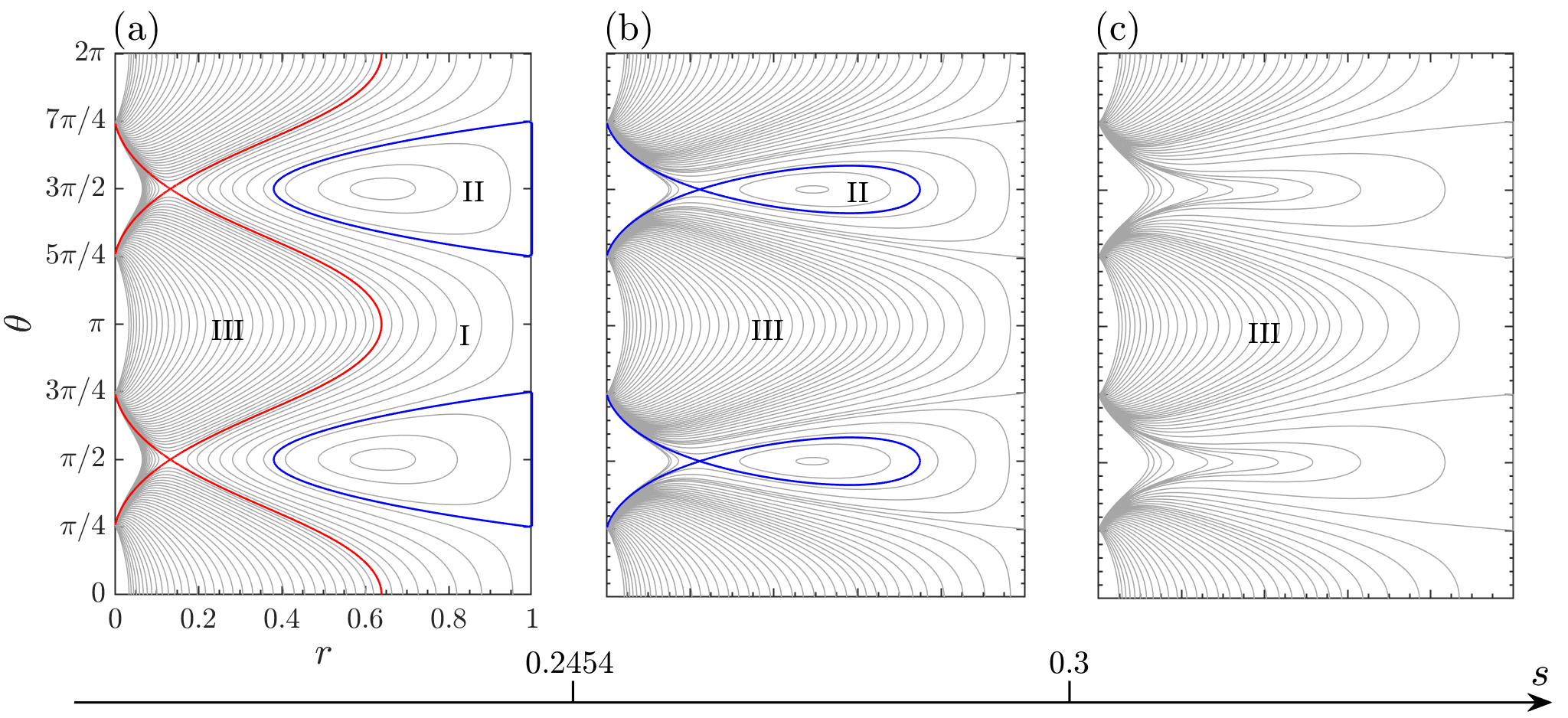}
    \caption{The phase portraits show contour lines of constant $\mathcal{H}$ in the $r$ versus $\theta$ plane for a Kida vortex for (a) $s = 0.2$, (b) $s = 0.27$, and (c) $s = 0.32$. The various dynamical regimes are marked using Roman numerals: \textrm{I} for rotation, \textrm{II} for nutation, and \textrm{III} for elongation. The blue and red curves demarcate these dynamical regimes. The critical values of strain rates at which the transition in dynamical nature occurs are marked on the $s$ axis. }
    \label{Fig009}
\end{figure}

To elaborate, consider the phase portrait in figure \ref{Fig009}, which depicts contours of constant $\mathcal{H}$ for three different $s$ values. Depending on the initial values $r_0$ and $\theta_0$, which determine the value of $\mathcal{H}$, the subsequent dynamics of the vortex will flow along the contour line of the same $\mathcal{H}$ in the phase portrait. Figure \ref{Fig009}(a) corresponds to $s = 0.2$, encompassing all three dynamical regimes. In region-\textrm{I}, a typical $\mathcal{H}$ contour extends from $\theta = 0$ to $\theta = 2\pi$, while $r$ varies periodically over a restricted range, indicating rotation dynamics of the ellipse. In region-\textrm{II}, both $r$ and $\theta$ vary periodically over a restricted range, indicating nutation dynamics where the ellipse oscillates with a changing aspect ratio instead of completing a rotation. In region-\textrm{III}, the contour lines show $r$ asymptotically approaching zero while $\theta$ approaches $\pi/4$ or $5\pi/4$ (i.e., the straining axes), indicating irreversible elongation of the ellipse. When $s = 0.27$, as shown in figure \ref{Fig009}(b), only nutation and elongation dynamics exist, with rotation dynamics disappearing. The critical value of $s$ at which this transition occurs is approximately $s \approx 0.2454$. Further increasing $s$ to 0.32, as depicted in figure \ref{Fig009}(c), results in only elongation dynamics, as nutation also disappears for $s \gtrsim 0.3$. Note that these critical values of $s$ for the disappearance of rotation and nutation dynamics have been previously discussed by \cite{kida1981motion} and \cite{neu1984dynamics}, though their values differ by a factor of two due to different definition conventions (i.e., 0.1227 and 0.15, respectively). Therefore, for $s \gtrsim 0.3$, all elliptical vortices in a pure-strain flow will elongate indefinitely, as noted by \cite{moore1971structure}. For larger strain rates, the phase portrait qualitatively resembles figure \ref{Fig009}(c) with only elongation dynamics. Conversely, when $s$ is very small, the phase portrait resembles figure \ref{Fig009}(a) but with a larger region-\textrm{I}, indicating a higher likelihood of rotation dynamics.

%%%%%%%%%%%%%%%%%%%%%%%%%%%%%%%%%%%%%%%%%%%%%%%%%%%%%%%%%%%%%%%%%%%%%%%%%%%%%%%%%%%
%phase_space_Kida_r_theta.eps
%%%%%%%%%%%%%%%%%%%%%%%%%%%%%%%%%%%%%%%%%%%%%%%%%%%%%%%%%%%%%%%%%%%%%%%%%%%%%%%%%%%%%%%%%
%%%%%%%%%%%%%%%%%%%%%%%%%%%%%%%%%%%%%%%%%%%%%%%%%%%%%%%%%%%%%%%%%%%%
%r_vs_th_Kida1
\begin{figure}
    \centering
\includegraphics[width=1.0\linewidth]{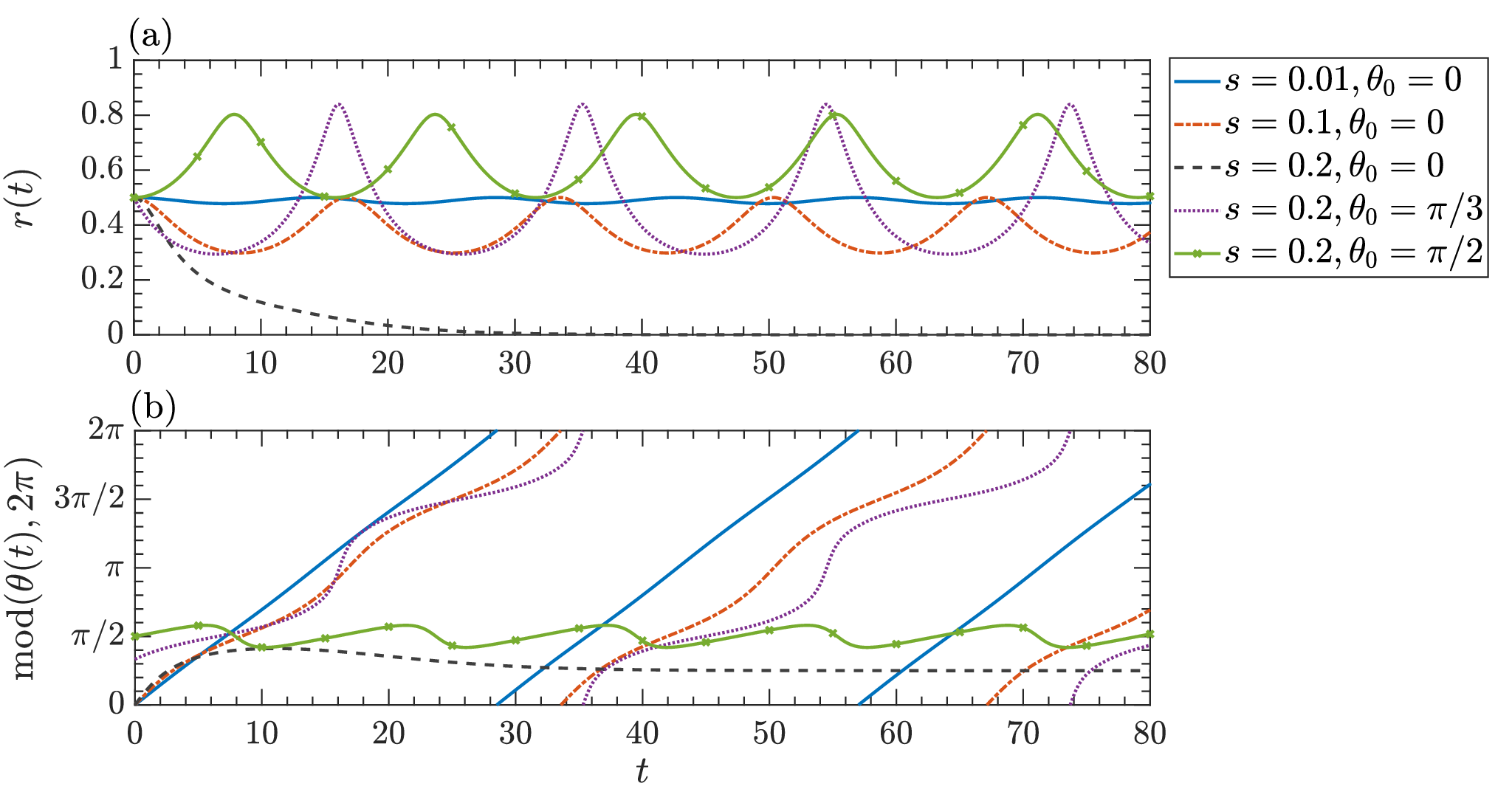}
    \caption{
The time variation of the (a) aspect ratio $r$ and (b) angular orientation $\theta$ of a Kida vortex with $r_0 = 0.5$ is shown for various non-dimensional strain rates $s$ and initial orientations $\theta_0$. The angular variation is represented modulo $2\pi$ to indicate the completion of a full ellipse rotation. Different colors indicate the various dynamics of the Kida vortex: blue, orange, and violet for rotation; black for elongation; and green for nutation. The period of rotation differs for each case, as inferred from the figure.}
    \label{Fig010}
\end{figure}
%%%%%%%%%%%%%%%%%%%%%%%%%%%%%%%%%%%%%%%%%%%%%%%%%%%%%%%%%%%%%%%%%%%%%%%%%%%%%%%%%%%%%%%%%

Figure \ref{Fig010} shows the typical time evolution of $r$ and $\theta$ of a Kida vortex with $r_0 = 0.5$, for various $s$ and $\theta_0$ values, obtained by numerically solving equations (\ref{eqn3p1}). For the smallest strain rate ($s = 0.01$) and $\theta_0 = 0$, the aspect ratio of the ellipse barely changes, but it undergoes rotation as $\theta$ varies from 0 to $2\pi$ periodically. For a slightly larger strain rate ($s = 0.1$) with the same $\theta_0$, the ellipse completes a rotation again but with a larger periodic variation in the aspect ratio. However, when $s$ is increased to $0.2$ with the same $\theta_0$, the ellipse asymptotically elongates ($r \rightarrow 0$) and aligns with the straining axes ($\theta \rightarrow \pi/4$), losing its periodic dynamics. This occurs because the critical $s$ value for the specific initial conditions ($r_0 = 0.5$ and $\theta_0 = 0$) is numerically determined to be $0.174$, beyond which only elongation dynamics exist. Note that this critical value is less than $0.3$ mentioned earlier, which is the critical value for this transition irrespective of the initial conditions. Now, fixing $s = 0.2$ but adjusting the initial $\theta$: when $\theta_0 = \pi/3$, the ellipse undergoes rotation; however, when $\theta_0 = \pi/2$, the ellipse nutates. The figure summarizes the typical cases of the Kida vortex considered in this study to analyse the dispersion of particles. Our primary focus here is on the dynamics of inertial particles in a slightly strained ($s \ll 1$) Kida vortex with $r_0 = 0.5$ and $\theta_0 = 0$, as discussed in the upcoming subsection. Later (see \S \ref{sec3p2}), we also consider the case of sufficiently strong external straining ($s = 0.2$) on the same ellipse but with $\theta_0 = 0$ and $\pi/2$, to discuss particle dispersion in nutating and elongating Kida vortices.

% \textcolor{blue}{   The critical strain rate depends on the initial aspect ratio and orientation of the ellipse. For $ r(t=0) = r_0 = 0.5 $ and $ \theta(t=0) = \theta_0 = 0 $, from equations (\ref{eqn3p1}), we can numerically identify that the critical strain rate (non-dimensional) is $ s \approx 0.174 $. The numerical solutions of equations (\ref{eqn3p1}) showing the variation of $ r $ and $ \theta $ of the ellipse with time for various $s$ values are shown in figure \ref{Fig010}. For smaller values of $ s $, we observe a periodic variation of the aspect ratio and angular orientation with time. However, for larger values of $ s $, at large times, the aspect ratio asymptotes to zero (i.e., the ellipse stretches indefinitely) and the orientation asymptotes to $ \pi/4 $ (i.e. aligning with the straining axes of the external planar extensional flow). Note that, as the external linear flow is purely straining and has no vorticity component in the opposite sense of the elliptical patch's vorticity content, nutation is impossible in our configuration.}

% \textcolor{blue}{In this study, we mainly focus on the small straining limit ($s \ll 1$), where the Kida vortex undergoes a full rotation as it is slightly affected by the external straining. Later, we also consider the case of strong external straining when discussing the large time dispersion of particles in Section \ref{sec4p1}.}
\subsection{Dynamics of inertial particles in a weakly strained, rotating Kida vortex }
\label{sec3p1}
 This section focuses on the dynamics of inertial particles in a Kida vortex, which is weakly strained ($s \ll 1$) by a pure-strain flow. We consider the case where the ellipse undergoes a rotary dynamics. The inertial particles in the co-rotating frame with the ellipse are tracked using the modified form of the Maxey-Riley equation 
\begin{equation}
        \dot{\textbf{v}} = \frac{\textbf{u}-\textbf{v}}{St}+\textbf{x}\, \dot{\theta}^2 - 2\, \dot{\theta} \, \hat{\textbf{e}}_z\times \textbf{v} - \ddot{\theta} \, \hat{\textbf{e}}_z\times \textbf{x} ~.
        \label{eqn3p2}
\end{equation}

%%%%%%%%%%%%%%%%%%%%%%%%%%%%%%%%%%%%%%%%%%%%%%%%%%%%%%%%%%%%%%%%%%%%%%%%%%%%%%%%%%%%%%%%%%%%%%%%%
%Evolution_particles_in_Kida
\begin{figure}
    \centering
    \includegraphics[width=1.0\linewidth]{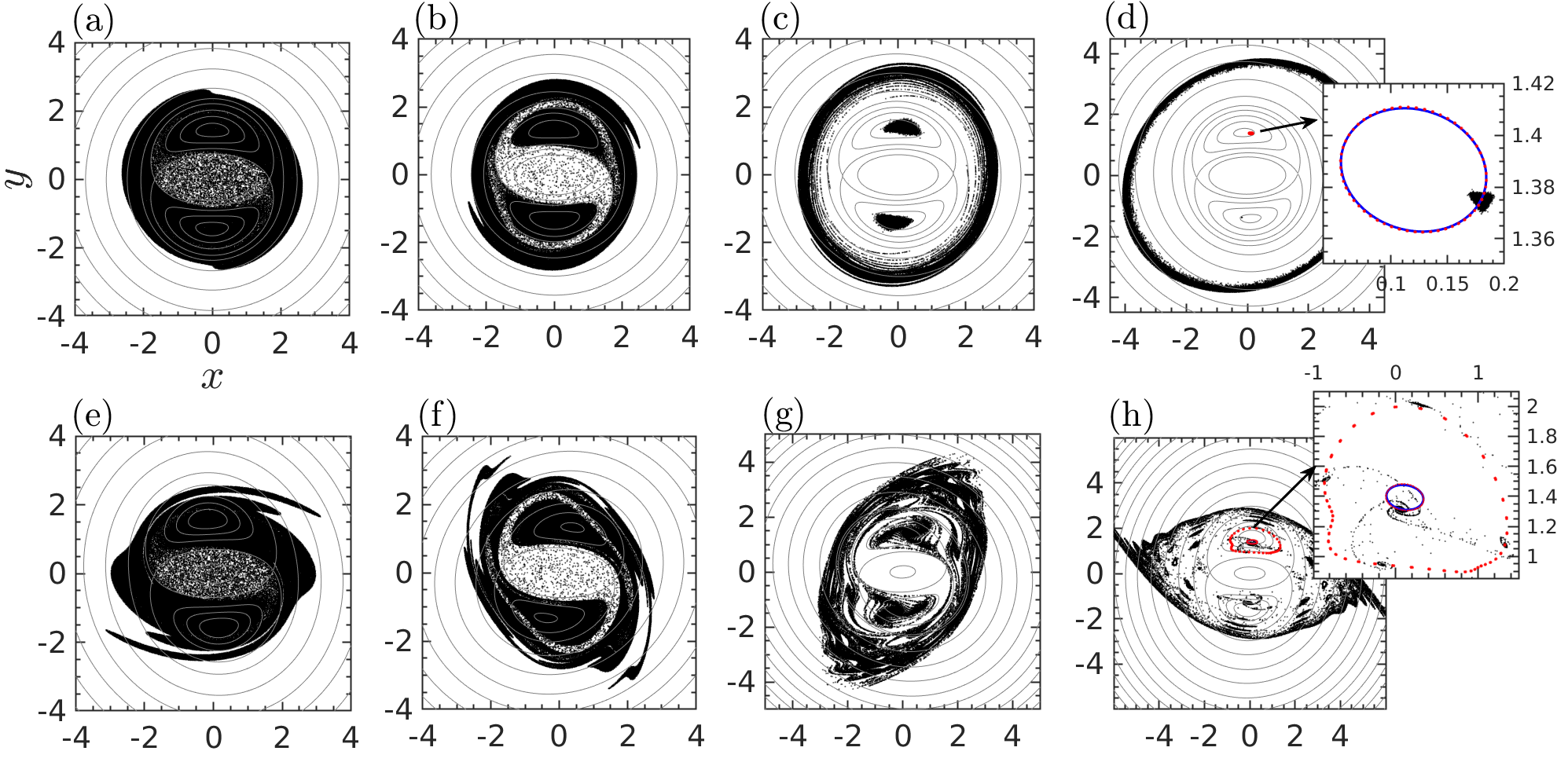}
    \caption{Snapshots showing the evolution of $2 \times 10^5$ inertial particles of $St = 0.1$ in a Kida vortex of $r_0 = 0.5$, $\theta_ 0 = 0$ for (a-d) $s = 0.01$ and (e-h) $s = 0.035$, observed in a co-rotating frame, obtained from numerical simulation. The time stamps are: (a,e) $t = 50$, (b,f) $t = 100$, (c,g) $t = 500$, (d,h) $t = 2000$.The particles are initialised with zero velocity inside a circular region of radius $2.5$, enclosing the ellipse. The instantaneous streamlines of the corresponding Kida vortex are shown in grey in the background. The insets of (d) and (h) are the enlarged views at $t = 2000$ showing the accumulation of particles in limit cycle trajectories, obtained numerically, shown in red. The blue colour indicates the limit cycles evaluated using a perturbative approach as in Section \ref{sec3p1p1}.}
    \label{Fig011}
\end{figure}
%%%%%%%%%%%%%%%%%%%%%%%%%%%%%%%%%%%%%%
%%%%%%%%%%%%%%%%%%%%%%%%%%%%%%%%%%%%%%%%
Note that the unsteady nature of the Kida vortex, even in the co-rotating frame, is accounted for in the Stokes drag term, centrifugal term and Coriolis term on the right-hand side as  $\textbf{u} = \textbf{u}(\textbf{x},t)$ and $\dot{\theta}=\dot{\theta}(t)$. In addition, the non-uniform rate of rotation of the frame, which is responsible for the Euler force, is represented by the last term on the right-hand side. We simulate the dynamics of $St = 0.1$ point particles in a Kida vortex of initial aspect ratio $r_0 = 0.5$ and orientation $\theta_0 = 0$. The simulation is performed for two different strain rate values, and the particles' evolution is shown in figure \ref{Fig011} as snapshots, as observed in a co-rotating frame. For a smaller value of strain rate $s = 0.01$, the overall dynamics resemble that in the case of the Kirchhoff vortex, as can be seen from figure \ref{Fig011}(a-d); most of the particles are getting centrifuged away, and a fraction of them attracted towards a pair of stable attractors. Unlike the Kirchhoff vortex, here the attractors are not fixed points but limit cycles in extended phase space (including the time axis), which can be seen as the red curve in the inset of figure \ref{Fig011}(d), obtained by tracking the particles for a long time. 

For a relatively larger value of the strain rate $ s = 0.035 $ (which is still less than the critical value of 0.174, allowing for the rotation of the ellipse), the particle dynamics become intricate, as evident in the corresponding figures \ref{Fig011}(e-h). We notice that some particles are drawn towards an additional, larger-sized limit cycle, depicted in red in the inset of figure \ref{Fig011}(h). Contrary to a simple spiralling dispersion, the remaining particles become trapped in stationary attractors (for a co-rotating observer, which appears to be counter-rotating as in the snapshots). Further details regarding this attractor and particle dispersion due to it can be found in Sections \ref{sec3p1p2} and \ref{sec4}, respectively. Additionally, the particle dynamics exhibit chaotic behaviour, consistent with the effect of external strain on fluid tracers \citep{polvani1990chaotic,ngan1996elliptical,kawakami1999chaotic}. In this context, a systematic analytical study is presented in the upcoming Sections \ref{sec3p1p1} and \ref{sec3p1p2} to illustrate that the interplay of inertia ($St$) and straining ($s$) leads to complex dynamics, as depicted in figure \ref{Fig011}.
%%%%%%%%%%%%%%%%%%%%%%%%%%%%%%%%%%%%%%%%%%%%%%%%%%%%%%%%%%%%%%%%%
\subsubsection{\label{sec3p1p1} Modification to the fixed points of the Kirchhoff vortex due to external straining}
For the analytical treatment to be accessible, the governing equation (\ref{eqn3p2}) is written in component form using elliptic coordinates as
\begin{subequations}
\begin{eqnarray}
            \ddot{\xi} &=& \frac{h\, k^{-1}\,u_\xi-\dot{\xi}}{St}-\frac{2\, \dot{k}}{k}\, \dot{\xi}+2\, \dot{\theta}\, \dot{\eta} -\frac{h^2\, \dot{k}}{2\, k\,St}\, \sinh 2\xi  \nonumber\\      
    &+&h^2\, \left\{ -\left(\dot{\eta}\, \dot{\xi}+\frac{\dot{k}\, \dot{\theta}}{k}+\frac{\ddot{\theta}}{2}\right)\, \sin 2\eta+\left(\frac{\dot{\eta}^2-\dot{\xi}^2+\dot{\theta}^2}{2}-\frac{\ddot{k}}{2\, k}\right)\, \sinh 2\xi\right\}~,\\
        \ddot{\eta} &=& \frac{h\, k^{-1}\, u_\eta-\dot{\eta}}{St}-\frac{2\, \dot{k}}{k}\, \dot{\eta}-2\, \dot{\theta}\, \dot{\xi} +\frac{h^2\, \dot{k}}{2\,k\, St}\,  \sin 2\eta\nonumber\\      
    &-&h^2\, \left\{ \left(\dot{\eta}\, \dot{\xi}+\frac{\dot{k}\, \dot{\theta}}{k}+\frac{\ddot{\theta}}{2}\right)\, \sinh 2\xi+\left(\frac{\dot{\eta}^2-\dot{\xi}^2+\dot{\theta}^2}{2}-\frac{\ddot{k}}{2\, k}\right)\, \sin 2\eta\right\}~.
    \end{eqnarray}
    \label{eqn3p3}
\end{subequations}
Note that when $\dot{k} = \ddot{\theta} = 0$, these equations will reduce to the case of Kirchhoff vortex as in equations (\ref{eqn2p2}); however, which is not the case for any finite strain $s \neq 0$. The flow velocity fields can be obtained from the total stream function ($\psi = \psi_v'+\psi_e'+\frac{\dot{\theta}}{2}\, (x^2+y^2)$) as $u_\xi = \frac{h}{k}\, \frac{\partial \psi}{\partial \eta}$ and $u_\eta =  -\frac{h}{k}\, \frac{\partial \psi}{\partial \xi}$. One must remember that the ellipse's evolution equations (\ref{eqn3p1}) must also be integrated along to obtain the particle trajectory. Since we know that the attractors of the system are not fixed points but limit cycles from the simulations, we may search for them using the approach by \citet{ijzermans2006accumulation} as a periodic modification to the fixed points of the unperturbed system. Let ($\overline{\xi},\overline{\eta}$) denote the fixed points for any inertial particle in Kirchhoff vortex, a time-dependent perturbation of the form $\xi = \overline{\xi}+s\, \xi'(t) + \textit{O}(s^2)$ and $\eta = \overline{\eta}+s\, \eta'(t) + \textit{O}(s^2)$ may denote the limit cycles for any small straining $s \ll 1$.

 The system behaves as a perturbed Kirchhoff vortex in an external straining flow of $s \ll 1$. A time-periodic, strain-dependent term will perturb the Kirchhoff vortex's Hamiltonian (stream function).  
For low strain rates, a regular perturbation approach yields the solution of equations (\ref{eqn3p1}) as $r = r_0+s\, r_1+\textit{O}(s^2)$ and $\theta = \theta_0+\Omega_0\, t+s\, \theta_1+\textit{O}(s^2)$ where
    \begin{subequations}
    \begin{eqnarray}
        r_1 &=&  -  \frac{r_0}{\Omega_0}\, \sin(\Omega_0\, t)\, \sin(\Omega_0\, t+2\, \theta_0) ~,\\
        \theta_1 &=& \frac{(1+r_0^3)}{ r_0\, (1-r_0)}\,  \sin(\Omega_0\, t)\, \cos(\Omega_0\, t+2\, \theta_0)-\frac{t}{2}\, \frac{(1-r_0)}{(1+r_0)}\, \cos 2\theta_0 ~,
        \end{eqnarray}
            \label{eqn3p4}
    \end{subequations}
with $r_0 = r(t = 0)$, $\theta_0 = \theta(t = 0)$ and $\Omega_0 = r_0/(r_0+1)^2$. Note that the leading order solutions (i.e., solutions when $s = 0$) match the Kirchhoff vortex, indicating a constant aspect ratio and uniform rotation rate. Since the parameter $k$ is dependent on aspect ratio $r$, $k$ can also be expanded as a perturbation in $s$ as $k = k_0 - s\, k_0\, \Lambda_0\, r_1/(2\, r_0)+\textit{O}(s^2)$, where $k_0^2 = (1/r_0-r_0)$ and $\Lambda_0 = (1+r_0^2)/(1-r_0^2)$. Similarly, the velocity field $(u_\xi,u_\eta)$ can also represent a strain perturbation to that of the Kirchhoff vortex.

By using these perturbation expansions along with the perturbation ansatz for the limit cycle co-ordinates ($\xi = \overline{\xi}+s\, \xi'(t) + \textit{O}(s^2)$ and $\eta = \overline{\eta}+s\, \eta'(t) + \textit{O}(s^2)$), substituted in the governing equations (\ref{eqn3p3}) will reduce to the matrix equation
% \begin{subequations}
% \begin{eqnarray}
% St \, \ddot{\xi}'+\dot{\xi}'-2\, St \, \Omega_0\, \dot{\eta}'+K_{11}\, \xi'+K_{12}\, \eta' \nonumber \\
% +M_{11}\, \cos (2\, \Omega_0\, t+2\, \theta_0)+M_{12}\, \sin (2\, \Omega_0\, t+2\, \theta_0) + N_1 = 0\\
% St \, \ddot{\eta}'+\dot{\eta}'+2\, St \, \Omega_0\, \dot{\xi}'+K_{21}\, \xi'+K_{22}\, \eta' \nonumber \\
% +M_{21}\, \cos (2\, \Omega_0\, t+2\, \theta_0)+M_{22}\, \sin (2\, \Omega_0\, t+2\, \theta_0) + N_2 = 0
%     \end{eqnarray}
%     \label{}
% \end{subequations}
\begin{equation}
    St\, \ddot{\chi}+\mathsfbi{L}\,\dot{\chi}+\mathsfbi{K}\,\chi = -\mathsfbi{M}\, \lambda -N
    \label{eqn3p5}
\end{equation}
where $\chi = [\xi'(t),\eta'(t)]^\textrm{T}$, $\lambda = [\cos (2\, \Omega_0\, t+2\, \theta_0), \sin (2\, \Omega_0\, t+2\, \theta_0)]^\textrm{T}$, $N = [N_1, N_2]^\textrm{T}$ are vectors, and
$\mathsfbi{K}$, $\mathsfbi{L}$ and $\mathsfbi{M}$ are $2 \times 2$ matrices. The entries $N$, $\mathsfbi{K}$, $\mathsfbi{L}$ and $\mathsfbi{M}$ are determined by the Kirchhoff vortex and its fixed points ($\overline{\xi},\overline{\eta}$) and are listed in the Appendix \ref{appB}. The equation (\ref{eqn3p5}) represents a pair of coupled, forced, second-order, linear ordinary differential equations. We are interested in the large-time/steady-state solutions as they would tell us about the limit cycles. For this, we assume the periodic solution of the form  $\chi = F_0 +\mathsfbi{F}_1\, \lambda$ and a substitution yields two independent matrix equations (which are related to the particular solution of equation (\ref{eqn3p5}))
    \begin{subequations}
    \begin{eqnarray}
     \mathsfbi{K}\, F_0 &=& -N~,\\
 \left(4\, \Omega_0^2\, St\, \mathsfbi{I}-2\, \Omega_0\, \mathsfbi{L}\, \boldsymbol{\Sigma}-\mathsfbi{K} \right)\, \mathsfbi{F}_1 &=& \mathsfbi{M}~,
 \end{eqnarray}
 \label{eqn3p7}
 \end{subequations}
where $\mathsfbi{I} = \big(\begin{smallmatrix}
  1 & 0\\
  0 & 1
\end{smallmatrix}\big)$ is $2 \times 2$ identity matrix and $\boldsymbol{\Sigma} = \big(\begin{smallmatrix}
  0 & -1\\
  1 & 0
\end{smallmatrix}\big)$. 
%%%%%%%%%%%%%%%%%%%%%%%%%%%%%%%%%%%%
%Limitcycles_r0=0p5_th0=0_s=3p5e-2_analyticApprox
\begin{figure}
    \centering
    \includegraphics[width=0.5\linewidth]{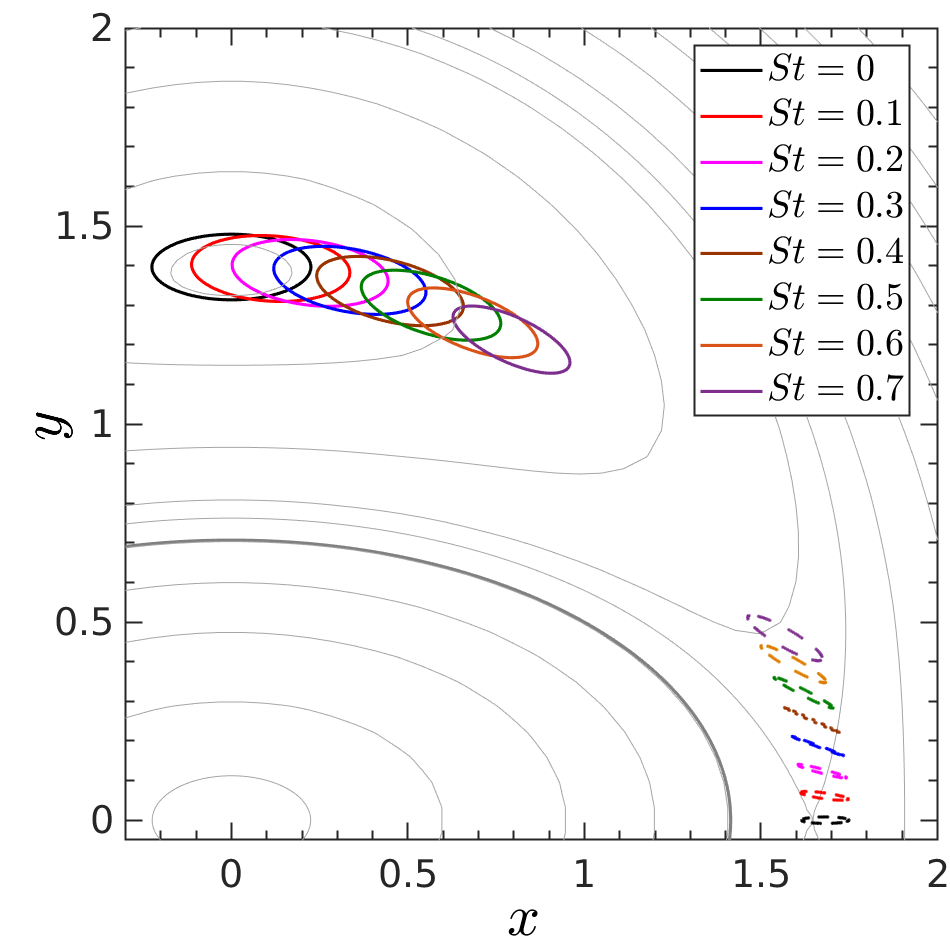}
    \caption{The variation of limit cycles corresponds to the fixed points of inertial particles of various $St$ in a Kida vortex of $r_0 = 0.5$, $\theta_0 = 0$ and $s = 0.035$, obtained using the perturbative approach (c.f. figure \ref{Fig003}(a)). The stable limit cycles are represented using continuous lines, and the unstable limit cycles are represented using dashed lines. Only the first quadrant is shown here. A similar pair of limit cycles also exists in the third quadrant (not shown here). The perturbative approach does not capture the larger limit cycles and thus is not shown here. The grey curves in the background indicate the streamlines of the Kida vortex, and the black curve indicates the ellipse at $t = 0$.}
    \label{Fig012}
\end{figure}
%%%%%%%%%%%%%%%%%%%%%%%%%%%%%%%%%%%%%%%%%%%%%%%%%%%%%%%%%%%%%%%%%%%%
%%%%%%%%%%%%%%%%%%%%
These equations are linear and a simple inversion yields the unknowns $F_0$ and $\mathsfbi{F}_1$, thus the limit cycles. However, the stability of the limit cycle is determined by the transient dynamics of the solutions of equation (\ref{eqn3p5}). It is determined by the eigenvalues of the matrix $\mathsfbi{P}$ which satisfy the quadratic matrix equation $St\, \mathsfbi{P}^2+\mathsfbi{L}\cdot \mathsfbi{P}+\mathsfbi{K} = \mathsfbi{0}$ (related to the homogeneous solution of equation (\ref{eqn3p5})). Any eigenvalue with a positive real part indicates an unstable limit cycle; if all the eigenvalues have a negative real part, it is stable. The limit cycles as a perturbation to the fixed points of Kirchhoff vortex ($[\xi,\eta] = [\overline{\xi},\overline{\eta}]^T + s\, \chi+\textit{O}(s^2)$) are then evaluated for the case of $r_0 = 0.5$ and $\theta_0 = 0$ and is shown in figure \ref{Fig012} as $St$ varies. Note that the originally stable fixed points (C and D) transform into stable limit cycles, as confirmed through numerical verification by tracking particles over an extended period, as depicted in the insets of figure \ref{Fig011}(d) and (h). While theoretically predicting that the hyperbolic fixed points (A and B) transition into unstable limit cycles under shear perturbation, it may be more accurate to state that the hyperbolic fixed points retain their hyperbolic nature. However, the stable and unstable manifolds connecting them are now susceptible to the oscillatory effects induced by this unstable limit cycle, potentially leading to their intersections and the formation of tangles. A detailed analysis of the perturbations to the hyperbolic fixed points is presented in the upcoming Section \ref{sec3p1p2}.

The perturbation to the fixed point at the origin O yields all the entries of $\mathsfbi{M}$ and $N$ to be zero, indicating that no limit cycle perturbation exists (either stable or unstable) to the origin. The origin remains an unstable spiral for inertial particles in the Kida vortex, which centrifuges them away from the centre. 

In the inset of figure \ref{Fig011}(d) and (h), the perturbatively obtained limit cycle solutions are shown in blue and agree with the respective actual limit cycles (red colour, obtained numerically). However, for the case of $s=0.035$, an additional pair of larger limit cycles exists, as can be seen in the inset of figure \ref{Fig011}(h) that this method could not capture.
%%%%%%%%%%%%%%%%%%%%%%%%%%%%%%%%%%%%%%%%%%%%%%%%%%%%%%%%%%%%%%%%%%%%%%%%%%%%%%%%%%%%%%%%%%%%%%%%%%%%%%%%%
\subsubsection{\label{sec3p1p2} Non-integrable perturbation to saddles and criteria for chaos}
The external straining ($s \ll 1$) introduces a time-periodic perturbation (non-integrable) to the Kirchhoff vortex (integrable Hamiltonian system). This perturbation leads to Lagrangian chaos of fluid tracers ($St = 0$) in the Kida vortex \citep{polvani1990chaotic}. The perturbations affect the hyperbolic fixed points of the Kirchhoff vortex, leading to the heteroclinic orbits to split into stable and unstable manifolds, which transversely intersect and form tangles, as shown by the zero crossings of the corresponding Melnikov functions. The result is a topological similarity to a horseshoe map and a dense set near the intersections \citep[see][]{smale1967differentiable,ott2002chaos}, which guarantees the existence of a strange attractor and chaotic dynamics. Though the original theorem is formulated for homoclinic orbits, a later extension for heteroclinic orbits conveys a similar concept \citep[see][]{bertozzi1987extension}. Here, we study the modification to the dynamics of suspended heavy inertial particles in the Kida vortex due to their finite inertia using the same approach. A competing effect may arise for inertial particles due to their finite inertia and suppress chaos as seen from similar studies \citep{angilella2010dust,angilella2014inertial}. To analytically study this, we consider modifying the particle dynamics in the Kida vortex (of $s \ll 1$) due to weak inertial particles ($St \ll 1$).

Using the slow manifold expansion of equations (\ref{eqn3p3}) for $St \ll 1$, we obtain the modified kinematic equations for $\dot{\xi}$ and $\dot{\eta}$ as perturbations in $St$. By combining the $s \ll 1$  expansion of instantaneous aspect ratio $r$ and angle of rotation $\theta$ with this slow manifold equation, we can express the modified kinematic equations describing the trajectory of an inertial particle in Kida vortex as
\begin{subequations}
    \begin{eqnarray}
    \dot{\xi} = \hat{f}_1(\xi,\eta)+s\, \hat{g}_1(\xi,\eta;t)+St\, \hat{\phi}_1(\xi,\eta)+\textit{O}(s^2, St^2, s\, St)~,\\
    \dot{\eta} = \hat{f}_2(\xi,\eta)+s\, \hat{g}_2(\xi,\eta;t)+St\, \hat{\phi}_2(\xi,\eta)+\textit{O}(s^2, St^2, s\, St)~,
\end{eqnarray}
 \label{eqn3p8}
\end{subequations}
where the relevant functions $\hat{f}_1$ to $\hat{\phi}_2$ outside and inside the ellipse are listed in Appendix \ref{appC}. %A similar set of equations has already been derived by \citet{kawakami1999chaotic} for the transport of fluid tracers ($St = 0$) in Kida vortex with $\theta_0 = \pi/4$ (however with some sign mistake, see Appendix \ref{appC}).
In equations (\ref{eqn3p8}), the functions $\hat{f}_1$,  and $\hat{f}_2$ describe the dynamics resulting from the dominant Kirchhoff vortex (integrable); $\hat{g}_1$ and $\hat{g}_2$ are time-periodic perturbations due to external straining flow; $\hat{\phi}_1$ and $\hat{\phi}_2$ are the time-independent perturbations due to particle inertia which is novel in our work. Thus, the dynamics of weak inertial particles in a Kida vortex will be perturbations to that of fluid tracers in a Kirchhoff vortex, where the perturbations are together contributed from external straining and particle inertia. 

The dynamics of fluid tracers in a Kirchhoff vortex (i.e., equations (\ref{eqn3p8}) with $s=0$ and $St = 0$) is integrable and thus represents a Hamiltonian system. The introduction of an external straining will not affect the integrability of the system inside the elliptic vortex as the flow-field inside is pure solid body rotation and devoid of any hyperbolic fixed points. However, the flow-field outside the ellipse has two hyperbolic fixed points, A and B, connected by the two pairs of heteroclinic orbits $\textrm{H}_1^{\pm}$ and $\textrm{H}_2^{\pm}$ as shown in figure \ref{Fig001}(b). The strain perturbation can lead to the formation of heteroclinic tangles. To identify that, we use Melnikov's method. %following \citet{kawakami1999chaotic}. 
According to the method, one must evaluate the Melnikov function for all hyperbolic fixed points, which measures the distance between stable and unstable manifolds in the Poincaré section. Odd zeros of the Melnikov function indicate a transverse crossing of the corresponding stable and unstable manifolds \citep[see][]{bertozzi1987extension,wiggins1990nonlinear}. For fluid tracers in the Kida vortex,  \citet{ngan1996elliptical,kawakami1999chaotic} have evaluated the Melnikov function, which is shown to satisfy the requirements for tangle formation for any nonzero straining. They also showed the space-filling nature of the Poincaré section and thus concluded that the transport of fluid tracers in the Kida vortex can be chaotic.

Here, we use the same analytical tools to identify the dynamical behaviour of heavy inertial particles in the Kida vortex. For very small values of Stokes number, the particles are expected to behave like fluid tracers and thus can be chaotic. On the contrary, when the strain rate is small, the inertial particles can simply be attracted to the fixed points of the dominant Kirchhoff vortex and thus won't be chaotic. It is when the Stokes number and strain rate are of the same order that the competition between these effects takes place, and complicated things can happen. Thus, for the case of $St \ll 1$ and $s \ll 1$, we consider $St = q\, s$, where $q$ is an $\textit{O}(1)$ quantity. By substituting the same in equations (\ref{eqn3p8}), the system can be re-written up to $\textit{O}(s^2)$ as
\begin{subequations}
    \begin{eqnarray}
    \dot{\xi} = \hat{f}_1(\xi,\eta)+s\, \left\{\hat{g}_1(\xi,\eta;t)+q\, \hat{\phi}_1(\xi,\eta)\right\}~,\\
    \dot{\eta} = \hat{f}_2(\xi,\eta)+s\, \left\{\hat{g}_2(\xi,\eta;t)+q\, \hat{\phi}_2(\xi,\eta) \right\}~.
\end{eqnarray}
 \label{eqn3p9}
\end{subequations}
The system is perturbed from the integrable Kirchhoff vortex in $s$ and thus may not be integrable in general. Following \citet{bertozzi1988heteroclinic},
%,li2001chaos,kawakami1999chaotic}
 the Melnikov functions associated with the hyperbolic fixed points of the system can be evaluated as
\begin{equation}
\begin{split}
        M_j^{\pm}(\phi) &= \int_{-\infty}^{\infty}\bigg[\hat{f}_1(\xi_j^{\pm}(t),\eta_j^{\pm}(t))\, \left\{ \hat{g}_2(\xi_j^{\pm}(t),\eta_j^{\pm}(t); t+\phi)+q\, \hat{\phi}_2(\xi_j^{\pm}(t),\eta_j^{\pm}(t))\right\} \\
        &-\hat{f}_2(\xi_j^{\pm}(t),\eta_j^{\pm}(t))\, \left\{ \hat{g}_1(\xi_j^{\pm}(t),\eta_j^{\pm}(t); t+\phi)+q\, \hat{\phi}_1(\xi_j^{\pm}(t),\eta_j^{\pm}(t))\right\}\bigg]\, e^{-\int_0^t \textrm{tr}(Df)\,dt'}\, dt~,
        \end{split}
        \label{eqn3p10}
\end{equation}
where the subscript $j (= 1, 2)$ indicates the heteroclinic orbits $(\xi_j^{\pm}(t),\eta_j^{\pm}(t))$ of the unperturbed system (Kirchhoff vortex) connected to the hyperbolic fixed points for which the Melnikov function is evaluated. Here the term $\textrm{tr}(Df) = \frac{\partial \hat{f}_1}{\partial \xi}+\frac{\partial \hat{f}_2}{\partial \eta}$ is the trace of the Jacobian matrix of the unperturbed system evaluated along the same heteroclinic orbit, which is not zero in general. One may note that the non-zero trace results from writing the dynamical system in non-standard variables $(\xi, \eta)$. However, the divergence of the corresponding velocity vector field in the elliptical coordinate system, evaluated as $\boldsymbol{\nabla}_{\xi,\eta} \cdot [u_{\xi},u_{\eta}] = h^2\, \left( \frac{\partial (\hat{f}_1/h^2)}{\partial \xi}+\frac{\partial (\hat{f}_2/h^2)}{\partial \eta} \right)$, is different from the trace mentioned above; $\boldsymbol{\nabla}_{\xi,\eta} \cdot [u_{\xi},u_{\eta}]=0$ in accordance with the Hamiltonian nature of tracer dynamics in Kirchhoff vortex.

%. This trace is nonzero when evaluated in elliptic coordinates (even though the Kirchhof vortex is a Hamiltonian system and thus the trace will be zero in Cartesian coordinates). This nonzero trace is not accounted for by \citet{kawakami1999chaotic} for tracers ($q = 0$). 
%Nevertheless, their findings are not affected much since they ultimately identified the zeros of the Melnikov function qualitatively and thus showed the existence of chaos. 

%Apart from this missing trace term, setting $q=0$ will reduce our Melnikov function expression (\ref{eqn3p10}) to that given in equation (36) of \citet{kawakami1999chaotic}. 
After splitting the integrals, it can be seen that the time-periodic perturbations (due to straining) $\hat{g}_1$ and $\hat{g}_2$ will result in a periodic term in $M$ of period $\pi/\Omega_0$, however the time-independent perturbations $\hat{\phi}_1$ and $\hat{\phi}_2$ will contribute as a constant term, i.e. $M(\phi) = m_0(\phi)+q\, m_1$ where $m_0(\phi+\pi/\Omega_0) = m_0(\phi)$ is the periodic function and $m_1$ is a constant, which are dependent on the aspect ratio of the ellipse. When the particles are inertialess (i.e., $q=0$), the resulting Melnikov function is periodic $m_0(\phi)$, known to have zeros for any finite $s \ll 1$.
%according to \citet{kawakami1999chaotic}. 
However, for inertial particles ($q \neq 0$), the term $m_1$ effectively shifts the periodic function $m_0(\phi)$ up or down when plotted against $\phi$, depending on the sign of $m_1$. Thus, there can exist a critical value of $q = \lvert\max(m_0(\phi))/m_1\rvert$ such that beyond which the Melnikov function $M(\phi)$ ceases to have transverse zeros. This indicates that there can be a critical value of $St$ for a given strain rate $s$ above which the heteroclinic tangles will disappear and possibly the chaos too. Since, in our system, there is more than one heteroclinic orbit, this critical value will be the greatest of all the critical values corresponding to each orbit so that the complete disappearance of heteroclinic tangles will be ensured.

Similar dynamical behaviour can also be observed in systems where two competing physics leads to a critical parameter value at which the appearance/disappearance of chaos occurs, for example, (i) the settling of inertial particles in a flow-field created by a pair of like-signed vortices \citep[see][]{angilella2010dust}, and (ii) the transport of passive fluid tracers in the same flow-field near a wall \citep[see][]{angilella2014inertial}. In either case, the zeros of the sinusoidal Melnikov function are affected by the vertical shift to the Melnikov function due to the competing parameter in the problem. 
%%%%%%%%%%%%%%%%%%%%%%%%%%%%%%%%%%%%%%%%%%%%%%%%%%%%%%%%%%%%%%%%%%%%%%%%%%5
%MelnikovFunction12_Kida_r0=0p5
\begin{figure}
    \centering
    \includegraphics[width=\linewidth]{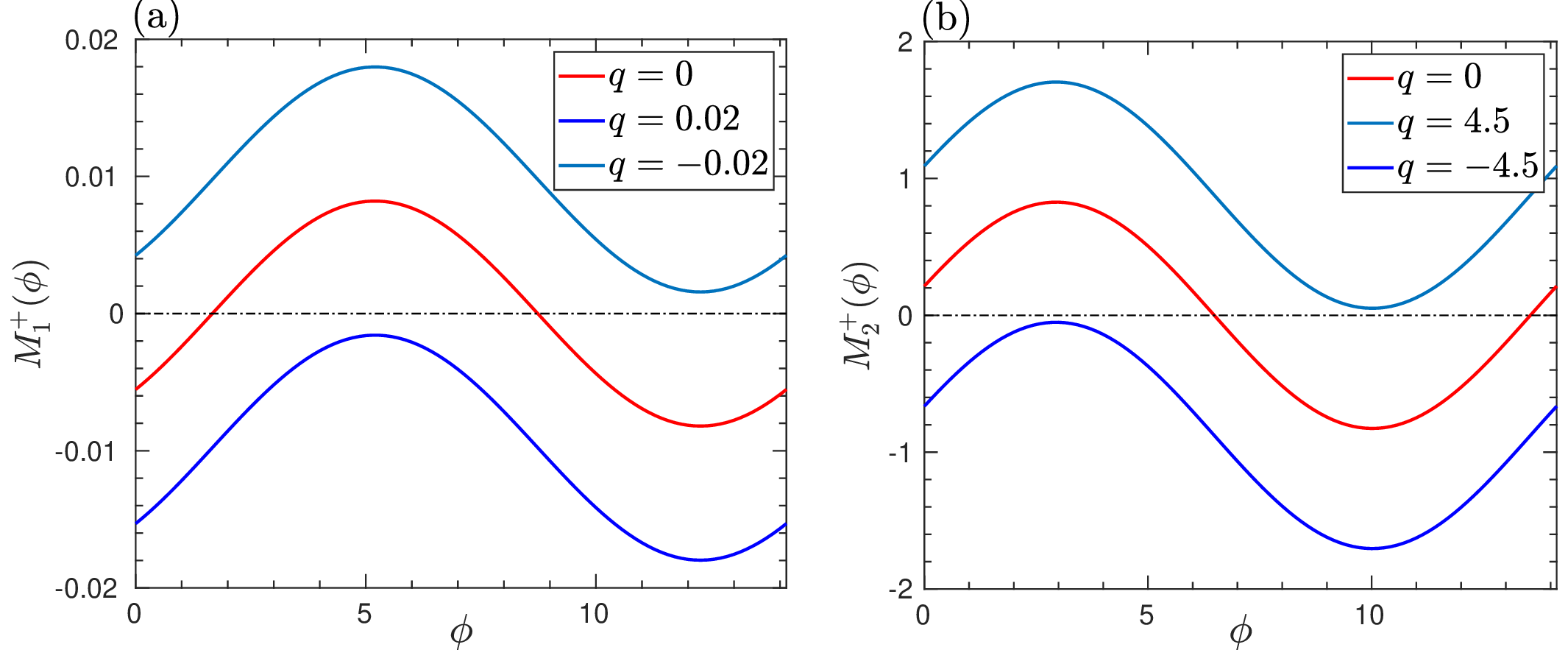}
    \caption{The Melnikov functions (a) $M_1^{+}(\phi)$ and (b) $M_2^{+}(\phi)$ for a Kida vortex of $r_0 = 0.5$ and $\theta_0 = 0$ are plotted against $\phi$ for various $q$ values. The abscissa is shown as a dashed-dotted line to identify the zeros of the Melnikov functions. }
    \label{Fig013}
\end{figure}
%%%%%%%%%%%%%%%%%%%%%%%%%%%%%%%%%%%%%%%%%%%%%%%%%
%%%%%%%%%%%%%%
%%%%%%%%%%%%%%%%%%%%%%%%%%%%%%%%%%%%%%%%%%%%%%%%%%%%%%%%%%%%%%%%%%%%
%Tangles_schematic
\begin{figure}
    \centering
    \includegraphics[width=1.0\linewidth]{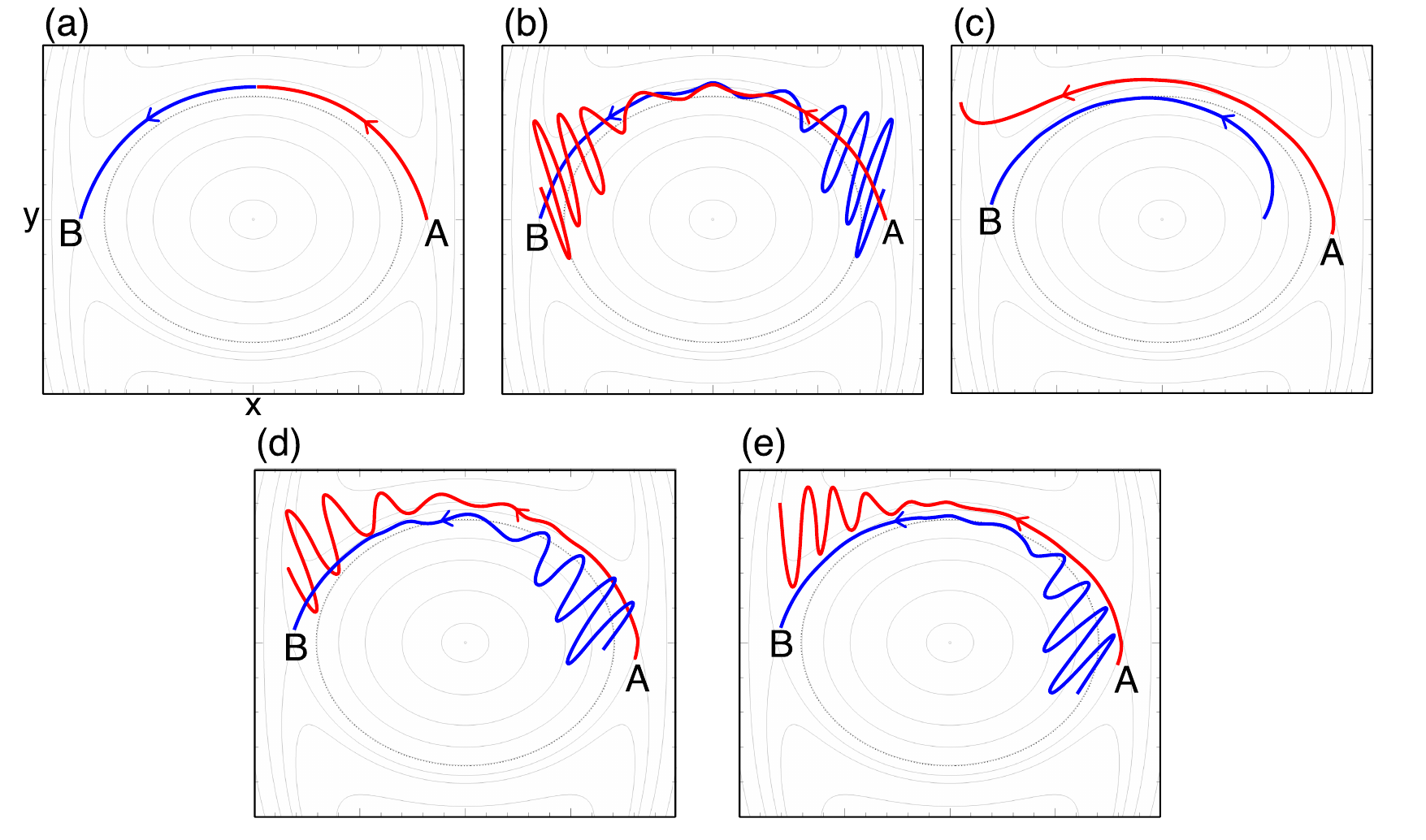}
    \caption{Schematic showing the unstable and stable manifolds of saddles A and B, respectively in the colour red and blue for various scenarios: (a) for tracers in Kirchhoff vortex ($s =0, St=0$), (b) for tracers in Kida vortex ($s \ll 1, St=0$), (c) for inertial particles in Kirchhoff vortex ($s =0, St \ll 1$), (d-e) inertial particles in Kida vortex, where (d) $s \ll 1, St < q_{\textrm{cr},1}\, s$ and (e) $s \ll 1, St > q_{\textrm{cr},1}\, s$.}
    \label{Fig014}
\end{figure}
%%%%%%%%%%%%%%%%%%%%%%%%%%%%%%%%%%%%%%%%%%%%%%%%%%%%%%%%%%%%%%%%%%%%%%%%%%%%%%%%%%%%%%%%%

For the Kida vortex, we explain the scenario using an example. Consider a Kirchhoff vortex with $r_0 = 0.5$ and $\theta_0 = 0$, perturbed with a pure-strain flow to form the Kida vortex. The heteroclinic orbits $\textrm{H}_1^{\pm}$ and $\textrm{H}_2^{\pm}$ are numerically obtained for the case. By substituting them in equation (\ref{eqn3p10}), we evaluate the Melnikov function for each orbit. For $M_1^{\pm}(\phi)$, the amplitude of the periodic function $m_0(\phi)$ is evaluated to be approximately $0.0082$ and $m_1 \approx -0.4889$. Similarly, for $M_2^{\pm}(\phi)$, the amplitude of corresponding $m_0(\phi)$ is evaluated to be approximately $0.8267$ and $m_1 \approx 0.1950$. The variation of the Melnikov functions with $\phi$ over a period $\pi/\Omega_0$ is shown in figure \ref{Fig013}.
%For a better understanding, we have fitted a sinusoidal curve for each of them and they respectively took the form: $M_1^+(\phi) = 0.008203\, \sin(0.4444\,\phi -1.368) - 0.48889\, q$,  $M_2^+(\phi) = 0.8267\, \sin(0.4444\,\phi -1.4214) +0.19503\, q$, $M_1^-(\phi) = 0.008203\, \sin(0.4444\,\phi -0.8288) - 0.48889\, q$ and $M_2^-(\phi) = 0.8267\, \sin(0.4444\,\phi +2.2594) +0.19503\, q$, where we may note that $2\, \Omega_0 = 0.4444..$ .  (Why there is phase difference between $M^+$ and $M^-$?). 
We can evaluate the critical value of $q$ at which the zeros of the Melnikov function disappear as $q_{\textrm{cr},1} \approx 0.0168$ and $q_{\textrm{cr},2} \approx 4.2395$, which is independent of whether it is $`+'$ or $`-'$ type heteroclinic orbit (due to their symmetry). Thus, we may conclude that the heteroclinic tangles of $H_1^{\pm}$ disappear if $St > q_{\textrm{cr},1}\, s$ and that of $H_2^{\pm}$ disappear if $St > q_{\textrm{cr},2}\, s$. When $St > q_{\textrm{cr},2}\, s$, the complete disappearance of heteroclinic tangles occurs.
% The result is in coherence with the observations from figure \ref{Fig012}, as explained earlier. 
In other words, there will not be any chaos when the straining is less than the critical value $s_{\textrm{cr},2} = St/q_{\textrm{cr},2}$. 

Figure \ref{Fig014} illustrates the schematic representation of tangles associated with the saddles A and B and the heteroclinic connection $H_1^{\pm}$ in various contexts. In each scenario, the red curve denotes the unstable manifold of saddle A, while the blue curve represents the stable manifold of saddle B. For tracers in the Kirchhoff vortex, as depicted in figure \ref{Fig014}(a), the stable and unstable manifolds coincide without any transverse intersection. With a small strain perturbation, as shown in figure \ref{Fig014}(b), the stable and unstable manifolds form folds and transversely intersect, leading to heteroclinic tangles. This tangle is responsible for the onset of chaotic dynamics of tracer particles in a Kida vortex. We now concentrate on the effect of particle inertia on the dynamics in the absence of external straining - inertial particles in the Kirchhoff vortex. Figure \ref{Fig014}(c) shows the effect of particle inertia, where inertial particles in the Kirchhoff vortex have wide-open, stable and unstable manifolds due to centrifugal effects. Folds are absent as there is no strain perturbation, preventing the formation of tangles. Figures \ref{Fig014}(d) and (e) examine the combined effect of strain perturbation and particle inertia. In figure \ref{Fig014}(d), particles with weak inertia ($St < q_{\textrm{cr},1}\, s$) allow for weak centrifuging, enabling the transverse intersection of the folds to form tangles. Conversely, in figure \ref{Fig014}(e), dominant particle inertia ($St > q_{\textrm{cr},1}\, s$) causes manifolds to spread apart and prevent the intersection of the folds.
%%%%%%%%%%%%%%%%%%%%%%%%%%%%%%%%%%%%%%%%%%%%%%%%%%%%%%%%%%%%5
%Basins_4d_St=0p1_s=2e-2_and_s=3e-2
\begin{figure}
    \centering
    \includegraphics[width=\linewidth]{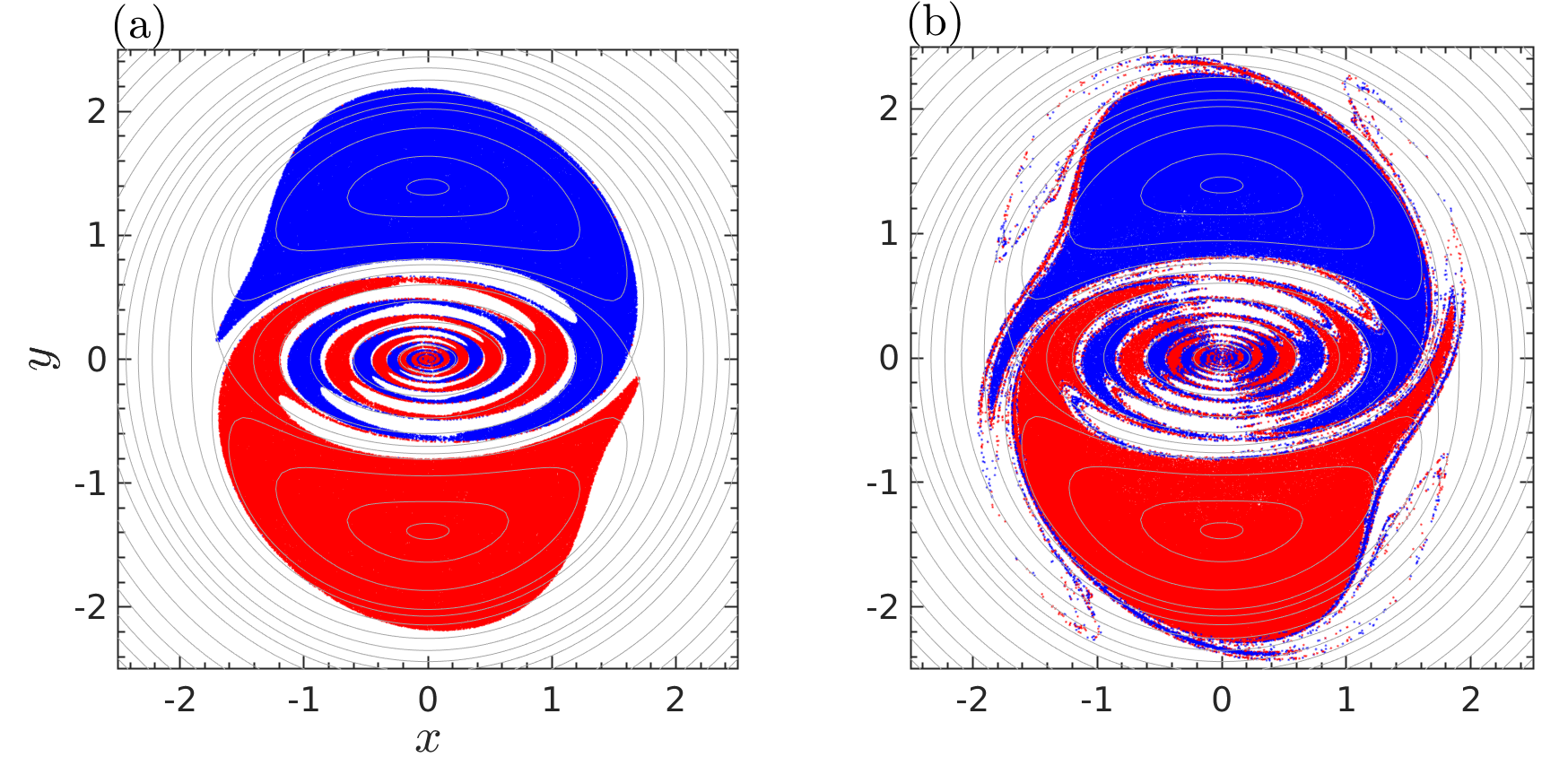}
    \caption{The basin of attraction of $St = 0.1$ particles in a Kida vortex (of $r_0 = 0.5$ and $\theta_0 = 0$) with strain rates (a) $s = 0.02$ indicating smooth basin boundaries and (b) $s = 0.03$ indicating fractal basin boundaries. Figures are generated using the full system of equations (\ref{eqn3p3}) along with equations (\ref{eqn3p1}). }
    \label{Fig015}
\end{figure}
%%%%%%%%%%%%%%%%%%%%%%%%%%%%%%%%%%%%%%%%%%%%%%%%%%%%%%%%%%%%%%%%%%%%%%%%%%%%%%%%%%%%%%%%%%%%%%%%%

\begin{figure}
    \centering
    \includegraphics[width=\linewidth]{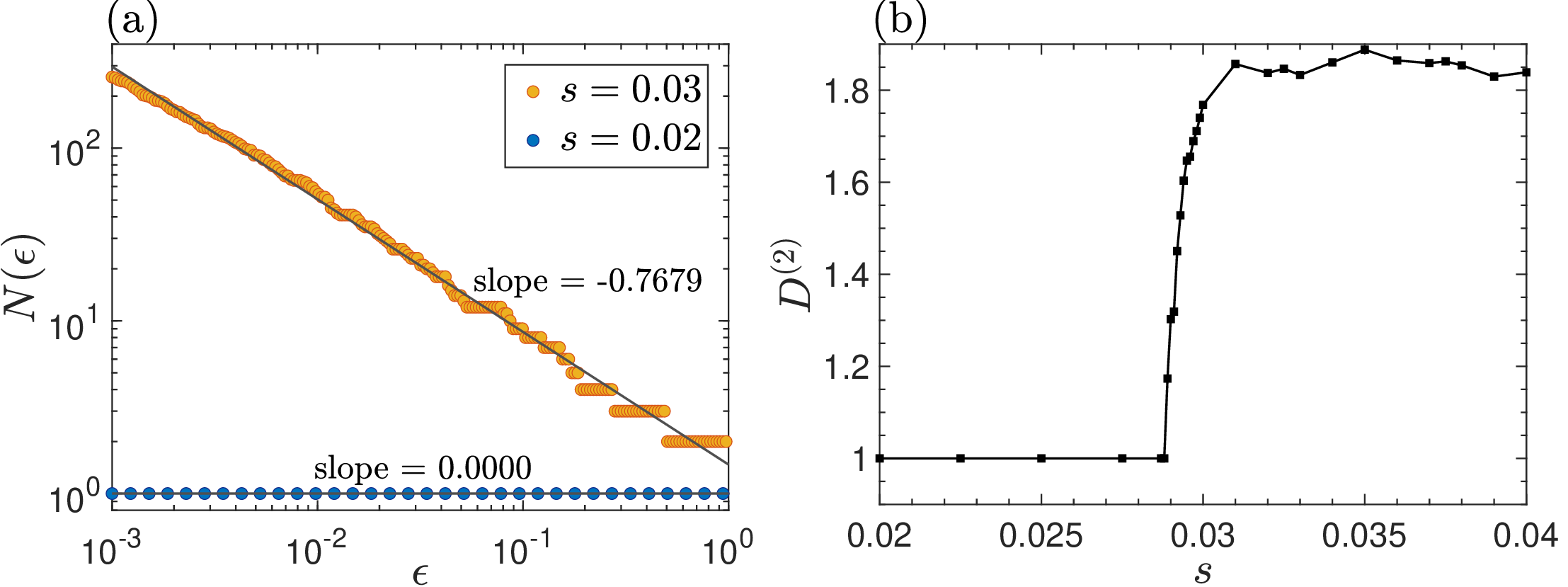}
    \caption{(a) Plot showing the variation of $\log(N(\epsilon))$ with $\log(\epsilon)$, where $N(\epsilon)$ is the number of boxes of side length $\epsilon$ required to cover the set of points in the basin boundary that intersects with a line segment at $x = 1.68$ and $-0.4 \leq y \leq 0.4$. The cases shown correspond to the two typical strain rates in figure \ref{Fig015}. The black straight lines represent the fitted curves with the specific slopes mentioned. (b) The variation of the dimension $D^{(2)}$ with the strain rate $s$. All cases here correspond to $St = 0.1$ particles in a Kida vortex with $r_0 = 0.5$ and $\theta_0 = 0$.}
    \label{Fig016}
\end{figure}

For inertial particles with $St = 0.1$,  the critical strain rates $s_{\textrm{cr}} = St/q_{\textrm{cr}}$ can be obtained as $s_{\textrm{cr},1}  \approx 5.9524 $ and $s_{\textrm{cr},2}  \approx 0.0236 $, such that when $s \lessapprox 0.0236$, there will be no chaotic dynamics. In figures \ref{Fig015}(a) and (b), we have shown the basin of attraction (evaluated using the full nonlinear system of equations (\ref{eqn3p3})) of $St = 0.1$ particles in Kida vortex with $r_0 = 0.5$, $\theta_0=0$ of $s = 0.02$ and $s = 0.03$ respectively. The basin of attraction has clear boundaries for the case of $s=0.02$, indicating regular dynamics; however, for $s = 0.03$, the boundaries are mixed/fractal, indicating an underlying chaotic set near the separatrices. The dimension of the basins is evaluated using a method similar to that used by \citet{angilella2014inertial}. We identify a set of points on a line intersecting the basin boundary. We chose the line segment $x = 1.68$ and $-0.4 \leq y \leq 0.4$, close to the hyperbolic fixed point A. Pairs of random points on the line segment are checked to see if they belong to different basins. If they do not, they are discarded. If they do, the bisection method is used to refine these points closer until their inter-distance reaches a threshold of $10^{-8}$, ensuring they approach the basin boundary. Once a few thousand such points are identified, lying on both the line and the basin boundary, we use a box-counting algorithm to identify the scaling law $N(\epsilon) \sim \epsilon^{-D^{(1)}}$ to obtain the dimension $D^{(1)}$ in the limit of $\epsilon \rightarrow 0$. Here, $N(\epsilon)$ is the number of boxes of side length $\epsilon$ required to cover the sampled points in the basin boundary. The dimension obtained ($D^{(1)}$) is the dimension of the intersection set between the basin boundary and the initial line segment. To obtain the basin boundary dimension in two-dimensional space, one needs to evaluate $D^{(2)} = 1 + D^{(1)}$. Figure \ref{Fig016}(a) shows the variation of $\epsilon$ versus $N(\epsilon)$ for the cases of $s=0.02$ and $s=0.03$, corresponding to the fractal basins in figure \ref{Fig015}. A curve fitting reveals that the respective slopes are zero and $-0.7679$, as mentioned in the figure, which indicates the respective fractal dimensions are $D^{(2)} = 1$ for the basin in figure \ref{Fig015}(a) and $D^{(2)} \approx 1.7679$ for the basin in figure \ref{Fig015}(b).
%%%%%%%%%%%%%%%%%%%%%%%%%%%%%%%%%%%%%%
%%%%%%%%%%%%%%%%%%%%%%%%%%%%%%%%%%%%%%%%%%%%
%frac_dim_box_counting1.eps
%%%%%%%%%%%%%%%%%%%%%%%%%%%%%%%%%%%%%%
%%%%%%%%%%%%%%%%%%%
Figure \ref{Fig016}(b) shows the variation of $D^{(2)}$ with $s$ for the basin boundaries corresponding to a fixed $St = 0.1$. The critical value of $s$ for the onset of a fractal basin boundary is $s_{\textrm{crit}}\approx 0.0288$. The prediction of $s_{\textrm{crit}}$ from the Melnikov analysis of the overdamped system (\ref{eqn3p9}) is $s_{\textrm{cr,2}} \approx 0.0236$. We would expect the predictions of the two methods to agree better for smaller values of $St$. The overdamped system is strictly valid in the asymptotic limit of $St\rightarrow 0$; $St=0.1$ is significantly large from the perspective of its validity (see \cite{nath2022transport} for discussion on flows with stagnation points). However, the overdamped system and the subsequent Melnikov analysis offer significant computational advantages over the full system, which tends to get very stiff for small $St$. It can be used to efficiently diagnose the chaotic dynamics when we are interested in first-order effects of particle inertia.

\begin{figure}
    \centering
    \includegraphics[width=1.1\linewidth]{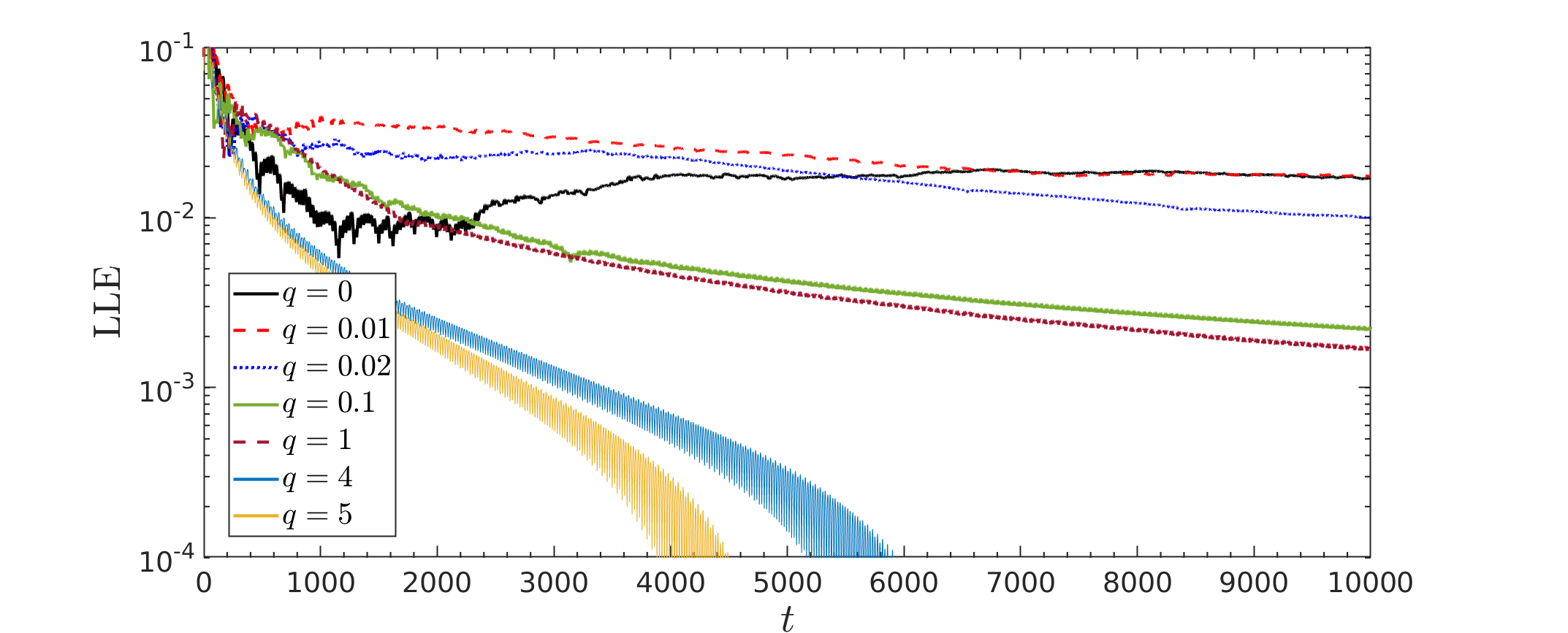}
    \caption{The variation of the largest Lyapunov exponent (LLE) with time for a Kida vortex with $r_0 = 0.5$, $\theta_0 = 0$ and $s = 10^{-2}$, corresponding to inertial particle trajectories of different $q$ (or $St$). Note that the case $q = 0$ indicates the LLE associated with tracers in the Kida vortex. The initial location of the particle is chosen to be the location of hyperbolic fixed point A at $t = 0$. }
    \label{Fig017}
\end{figure}

Note that the existence of zeros of the Melnikov function is a necessary criterion for chaos but not a sufficient one. Apart from a dense set, the heteroclinic tangles may also result in (i) a countable infinity of periodic orbits with arbitrarily high period and (ii) an uncountable infinity of periodic orbits \citep{smale1967differentiable}. Thus, for our system, whenever it is suspected, the existence of chaos needs to be shown using other methods, like Lyapunov exponents, frequency spectra, and Poincaré sections. However, from the investigations of chaotic dynamics of tracers in Kida vortex \citep{polvani1990chaotic,dahleh1992exterior,ngan1996elliptical,kawakami1999chaotic}, one can assume the same continues to exist for inertial particles until they have the critical inertia value. For further validation, we have evaluated the largest Lyapunov exponent (LLE) of the reduced dynamical system (\ref{eqn3p9}) to show the existence of chaos for weakly inertial particles and its disappearance for heavy inertial particles. We used the classical algorithm by \citet{benettin1980lyapunov} for this purpose. For a fixed strain rate of $s = 10^{-2}$, we evaluated the time variation of the LLE for various $St$ values for a point close to the hyperbolic fixed point A of the Kida vortex with $r_0 = 0.5$ and $\theta_0 = 0$. The results are shown in figure \ref{Fig017}. Note that fixed point A periodically changes location as the system changes periodically with time. Thus, we chose the location of point A  at $t = 0$ to evaluate the Lyapunov exponent. The figure shows that the system has a non-zero positive asymptotic value for LLE for small values of $q$ (or equivalently $St$, as $St = q \, s$). As $q$ increases beyond the first critical value $q_{\textrm{cr},1} \approx 0.0168$, the variation of LLE with time does not seem to saturate for large times, which indicates a weakening of chaotic behaviour. Similarly, as $q$ further increases to the next critical limit $q_{\textrm{cr},2} \approx 4.2395$, the LLE asymptotically approaches zero at large times, indicating a complete disappearance of chaos.

%%%%%%%%%%%%%%%%%%%%%%%%%
%%%%%%%%%%%%%%%%%%%%%%%%%%%%%%%%%%%%%%
%%%%%%%%%%%%%%%%%%%%%%%%%%%%%%%%%%%%%%5
%Lyaps_kida2d_elliptic_s=1e-2_r0=0p5_th0=0
%%%%%%%%%%%%%%%%%%%%%%%%%%%%%%%%%%%%%%
%%%%%%%%%%%%%%%%%%%%%%%%%%%%%%%%%%%%%%%%%%%%
%%%%%%%%%%%%%%%%%%%%%%%%%%%%%%%%%%%%%%
%%%%%%%%%%%%%%%%%%%%%%%%%%%%%%%%%%%%%%5
%St_cr_for_all_Hets
\begin{figure}
    \centering
    \includegraphics[width=1.0\linewidth]{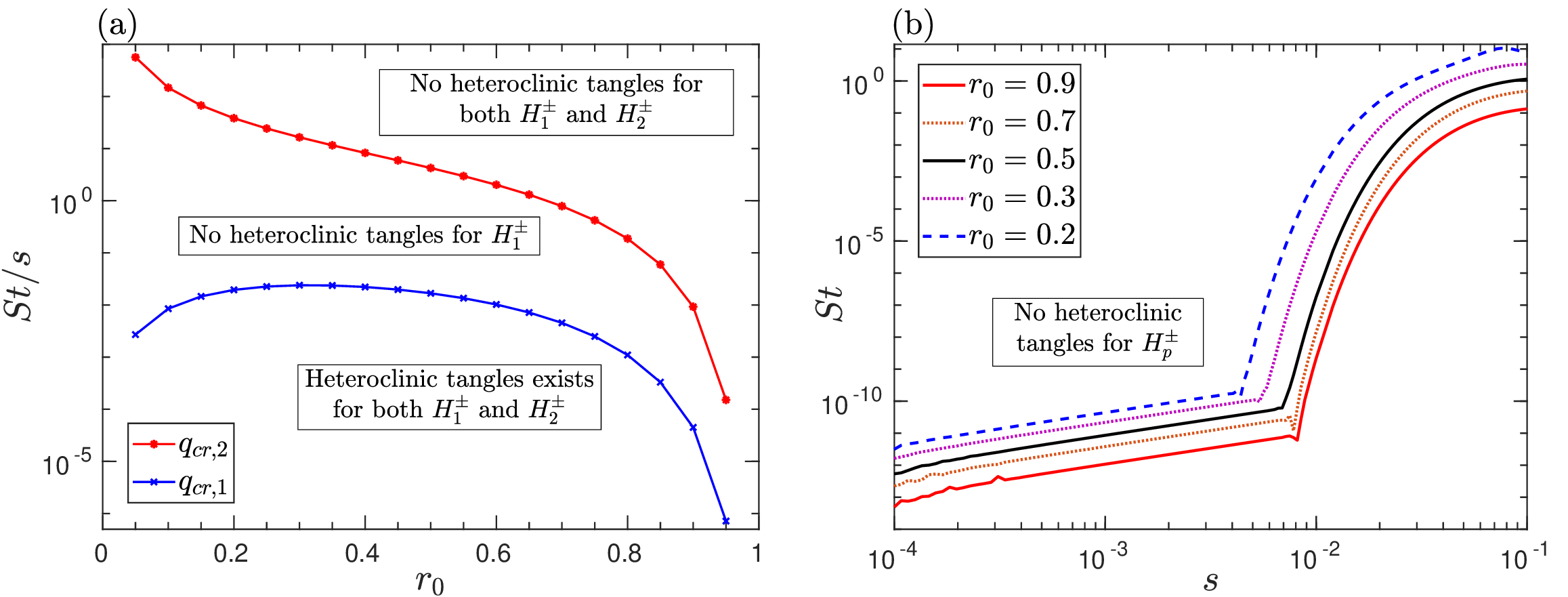}
    \caption{(a) The critical curves which demarcate the regions where the heteroclinic tangles can and cannot exist for $H_1^{\pm}$ and $H_2^{\pm}$ are shown in the parameter plot of $St/s$ versus $r_0$, for a Kida vortex, obtained from the Melnikov analysis. The initial orientation of Kida vortex $\theta_0$ will not affect these curves. (b) Curves in the $s - St$ plane demarcate the regions where  heteroclinic tangles can and cannot exist for $H_p^{\pm}$, for various $r_0$ values.}
    \label{Fig018}
\end{figure}
%%%%%%%%%%%%%%%%%%%%%%%%%%%%%%%%%%%%%%
%%%%%%%%%%%%%%%%%%%%%%%%%%%%%%%%%%%%%%%%%%%%
The critical ratio between $St$ and $s$ for the occurrence of heteroclinic tangles of both $H_1^{\pm}$ and $H_2^{\pm}$ have been evaluated numerically for various initial aspect ratios ($r_0$) of the elliptic vortex and plotted in figure \ref{Fig018}. The initial orientation ($\theta_0$) of the elliptic vortex will only phase shift the Melnikov functions; thus, these critical values won't be affected. The regions in the parametric plot where the heteroclinic tangles occur are identified and marked. The region above the red curve is where no heteroclinic tangles form, indicating complete suppression of chaos by inertia. However, below the red curve, chaotic transport may occur. The degree of chaos may vary across the blue curve. 

It is important to remember that these critical curves are derived from a perturbative method, assuming both $St \ll 1$ and $s \ll 1$. However, note that the red curve reaches a value of $St/s = \textit{O}(10^2)$ when $r_0 = 0.1$. For $s = \textit{O}(10^{-2})$, this results in the critical $St_{\textrm{cr},2} \sim 1$. The caveat here is that the Kirchhoff vortex's saddles disappear for $St>\mathcal{S}_2$; on the other hand, as mentioned earlier, the Kida vortex itself is elongated and destroyed beyond critical strain rate $0.3$. Thus, the validity of our analysis is restricted to sufficiently smaller $St$ and $s$, although their ratio can vary within the large range as shown in figure \ref{Fig018}(a). %\textcolor{magenta}{In addition, instead of a small $St$ perturbation analysis, if one uses a finite $St$ analysis for the system, one may find additional critical value of Stokes number corresponds to the saddle points \citep[see][]{levin1961studies,taylor1963scientific}, for example $St = 1/4$ as per the scaling in \cite{nath2022transport}, beyond which the phase space behaviour of saddles

%figures \ref{Fig015}(c) \& (d) shows the basin of attraction for the same cases generated using the full nonlinear system (equations (\ref{eqn4p4})), without the $s \ll 1$ \& $St \ll 1$ assumptions. There is a good qualitative agreement with the corresponding basins from both categories, especially for $s = 0.02$. We claim the agreement will be much better for smaller values of $s$ and $St$.
%%%%%%%%%%%%%%%%%%%%%%%%%%%%%%%%%%%%%%%%%%%%
\subsubsection{\label{sec3p1p3} Chaotic dynamics far away from the ellipse}
\begin{figure}
    \centering
    \includegraphics[width=1.2\linewidth]{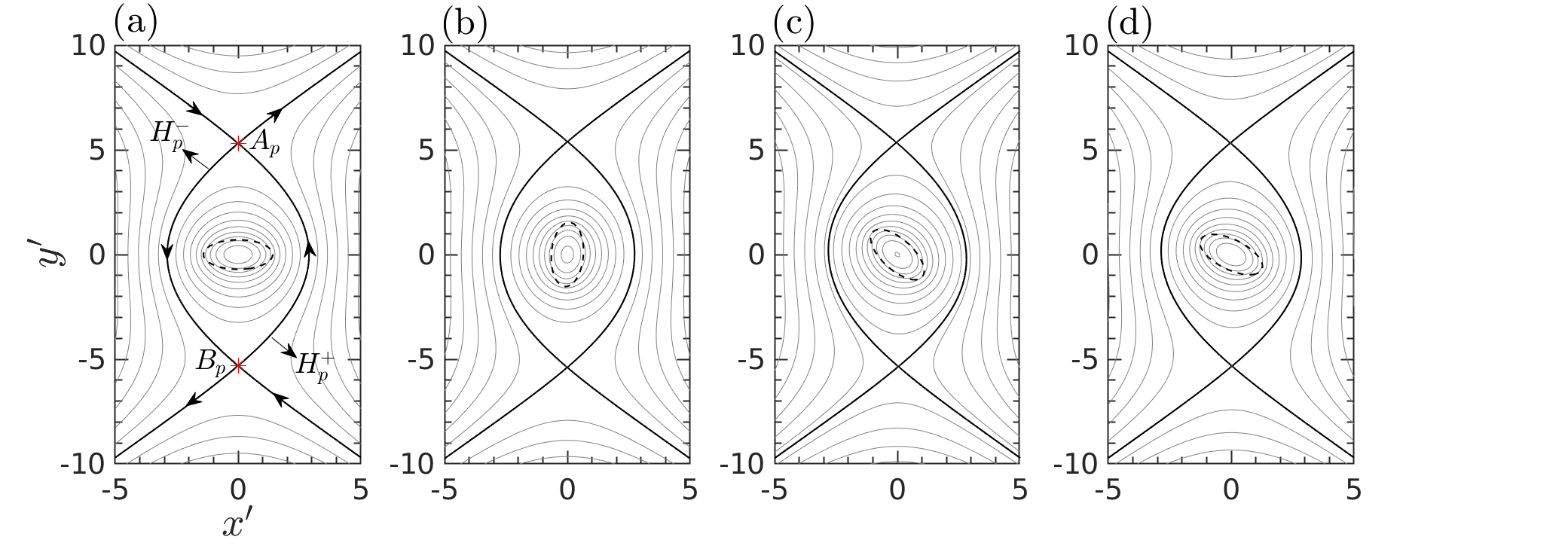}
    \caption{The typical streamlines of the Kida vortex, as observed in a stationary reference frame, show the hyperbolic fixed points far away for (a) $t = 0$, (b) $t = 7$, (c) $t = 11$ and (d) $t = 27$, shown in grey. The strain rate here is $s = 0.035$, and the period of revolution of the ellipse is $\approx 29.034$. The rotating ellipse at corresponding instants is shown as the black dashed curve.}
    \label{Fig019}
\end{figure}

Unlike the Kirchhoff vortex, far away from the Kida vortex (distance of  $\textit{O}(s^{-1/2})$), the decaying vortex field and external straining flow can balance each other and form a pair of hyperbolic fixed points/saddles. The saddles remain stationary when observed from the lab reference frame and exhibit counter-rotation when observed from a co-rotating perspective. The snapshots in figure \ref{Fig019} show the evolution of streamlines in the far-field of the Kida vortex over a period of revolution of the ellipse for $s=0.035$. The saddle points $\textrm{A}_p$ and $\textrm{B}_p$ are connected by a pair of heteroclinic orbits $H_p^{\pm}$ and are perturbed due to the unsteady oscillations of the central ellipse, which is shown to bring chaotic dynamics for fluid tracers, far away from the ellipse \citep{ngan1996elliptical,kawakami1999chaotic}. As done in the previous Section \ref{sec3p1p2}, we modify the tracer analysis by incorporating the effect of particle inertia. The detailed analysis of evaluating the Melnikov function for $H_p^{\pm}$ can be found in Appendix \ref{appD}. However, our findings are similar to the case for $H_1^{\pm}$ and $H_2^{\pm}$; i.e., finite inertia competes with strain perturbation and suppresses the chaotic transport.

Using the Melnikov analysis, the parametric regime in the $s - St$ plane, where $H_p^{\pm}$ can or cannot form a tangle, is identified and illustrated in figure \ref{Fig018}(b). For $r_0=0.5$ and $St = 0.1$, the critical value of the strain rate is found to be $s_{\textrm{cr},p} \approx 0.0341$, below which particles cannot exhibit chaotic dynamics near $H_p^{\pm}$. However, it is important to note a caveat: the Melnikov analysis for $H_p^{\pm}$ assumes that $St \sim 1$ and $s \ll 1$ (see Appendix \ref{appD}); nonetheless, the critical curve in figure \ref{Fig018}(b) covers even smaller values of $St$.
%%%%%%%%%%%%%%%%%%%%%%%%%%%%%%%%%%%%%%%%%%%%%%%%%%%%%%%%%%%%%%%%%%%%%%%%%%%%%%%%%%%%%%%%%%%%%%%%%%%%%%%%%%%%%%%%%%%%%%%
%%%%%%%%%%%%%%%%%%%%%%%%%%%%%%%%%%%%%%
%%%%%%%%%%%%%%%%%%%%%%%%%%%%%%%%%%%%%%5
%Streamlines_kida_snapshots_s=0p035_r0=0p5_th0=0_t=0_7_11_27_TP=29p0365
\subsection{Dynamics of inertial particles in a nutating and elongating Kida vortex }
\label{sec3p2}
\begin{figure}
    \centering
\includegraphics[width=1.05\linewidth]{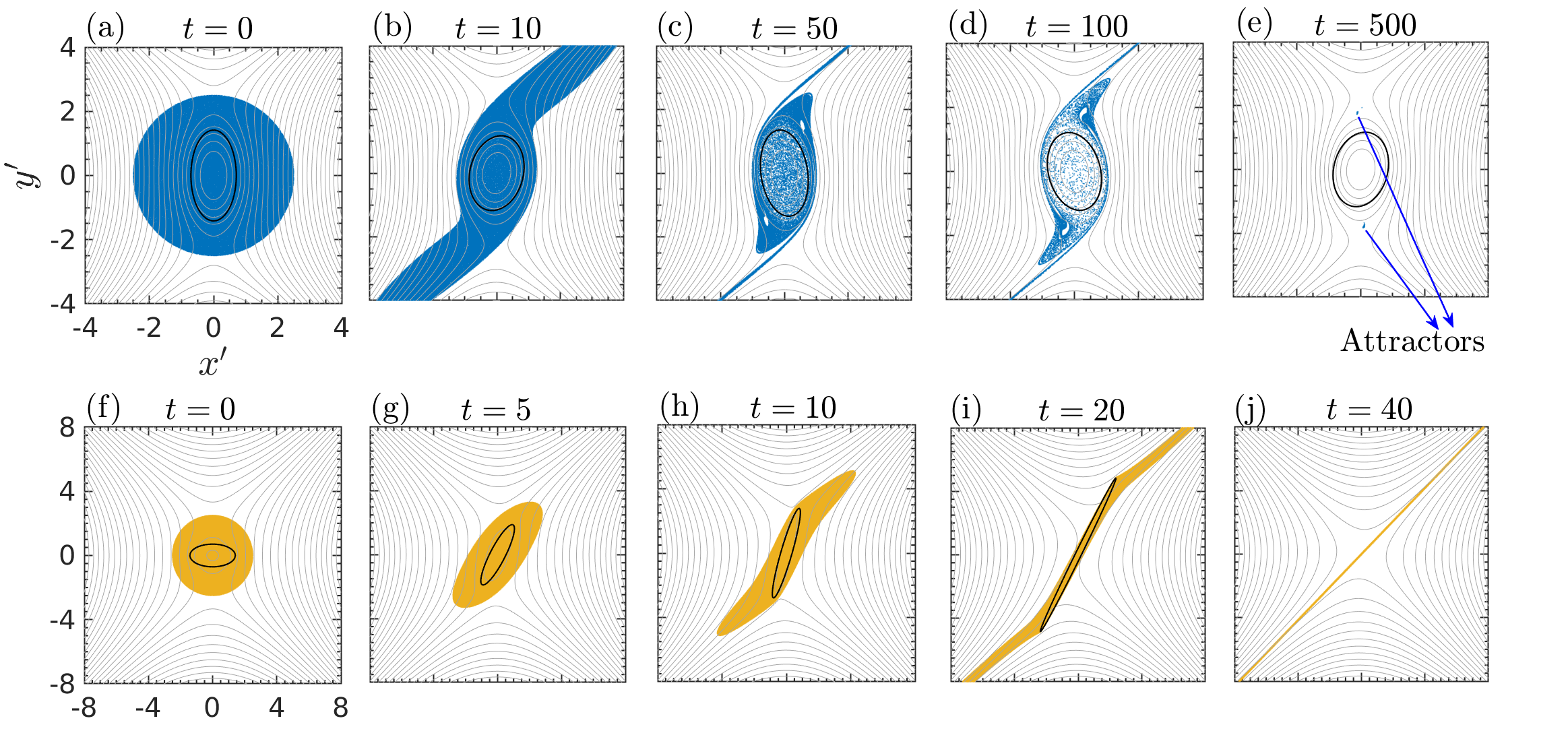}
    \caption{The time evolution of the dispersion of $2 \times 10^{5}$ inertial particles with $St = 0.1$ in a Kida vortex with $r_0 = 0.5$ and $s = 0.2$ is shown when it exhibits (a-e) nutation ($\theta_0 = \pi/2$) and (f-j) elongation ($\theta_0 = 0$). The plots are presented in the lab/stationary reference frame (i.e., using $x' - y'$ coordinates). The grey curves in the background depict the instantaneous streamlines in the lab reference frame, while the black curves represent the instantaneous state of the elliptical vortex. The axes limit for the nutation cases are fixed to [-4, 4], while for the elongation case, they are set to [-8, 8]. For better visualisation, the time stamps in both cases are taken incoherently.}
    \label{Fig020}
\end{figure}

So far, we have considered cases of a rotating Kida vortex under a weak pure-strain flow. Here, we discuss the role of a Kida vortex's nutating and elongating configurations on the dispersion of inertial particles. To be consistent with previous results, we consider an elliptical vortex with the same aspect ratio as before, i.e., $r_0 = 0.5$. With this aspect ratio, it must experience a larger strain rate to exhibit both nutation and elongation. We chose a strain rate of $s = 0.2$, with which the vortex can experience nutation if $\theta_0 = \pi/2$ and elongation if $\theta_0 = 0$ (see figure \ref{Fig010}). The governing equations (\ref{eqn3p3}) are solved along with equations (\ref{eqn3p1}) numerically for $St = 0.1$ particles in both the nutating and elongating cases of the vortex. The particles are initialised randomly in a circular patch with a radius of 2.5 nondimensional units. The time evolution of the particles is shown in figure \ref{Fig020}. It is important to bear in mind that as the study is restricted to a specific strain rate and initial aspect ratio, the results in this section only qualitatively represent the dynamics of inertial particles in a general class of nutating and elongating Kida vortices.

Figure \ref{Fig020}(a-e) shows the evolution of inertial particles when the vortex is nutating. It is observed that the straining axes of the background flow carry away the majority of particles far from the central vortex. Additionally, most particles initialised within the vortex are initially centrifuged out and then carried by the straining flow. However, over time, a smaller fraction of particles accumulate in a pair of attractors revolving around the vortex, similar to the case of a rotating Kida vortex (as shown in figure \ref{Fig011}). These particles can be seen in figure \ref{Fig020}(e) as two small blue dots. Note that these attractors revolve around the central ellipse, even though the ellipse only undergoes periodic oscillations.

Figure \ref{Fig020}(f-j) shows the results for the case of an elongating Kida vortex. The ellipse quickly aligns with the extensional axis of the straining flow, and so do the particles. All the particles are observed to move along the $\pm\pi/4$ axes. Note that a kink formation occurs in figure \ref{Fig020}(i) near the edge of the ellipse. This is because the particles near the ellipse move along with its instantaneous orientation (which is not $\pm\pi/4$), while the farther particles move along the straining axes ($\pm\pi/4$) of the background flow. At large times, these kinks disappear (see figure \ref{Fig020}(j)) as the ellipse aligns with the straining axes, and all the particles disperse in the same direction, $\theta = \pm \pi/4$.

%%%%%%%%%%%%%%%%%%%%%%%%%%%%%%%%%%%%%%%%%%%%%%%%%%%%%%%%%%%%
%Kida_large_s_particle-Dynamics
% \begin{figure}
%     \centering
%     \includegraphics[width=1.05\linewidth]{Fig020.eps}
%     \caption{\textcolor{blue}{The time evolution of the dispersion of $2 \times 10^{5}$ inertial particles (shown in red) with $St = 0.5$ in a Kida vortex with $r_0 = 0.5$, $\theta_0 = 0$, and $s = 0.2$ at nondimensional times (a) $t = 0$, (b) $t = 10$, (c) $t = 20$, and (d) $t = 40$ is illustrated. The plots are presented in the lab/stationary reference frame. The grey curves in the background depict the instantaneous streamlines in the lab reference frame, while the black curve represents the instantaneous state of the elliptical vortex. Panel (e) shows the mean square displacement of the particles evaluated in the lab frame over time for three different inertia values: $St = 0$, $0.1$, and $0.5$. The inset displays a zoomed version of the plot at large times. The asymptotes at small and large times are depicted using grey lines.}}
%     \label{Fig020}
% \end{figure}
%%%%%%%%%%%%%%%%%%%%%%%%%%%%%%%%%%%%%%%%%%%%%%%%%%%%%%%%%%%%%%%
%small_St_particles_in_nutating_and_elongating_Kida.eps
%%%%%%%%%%%%%%%%%%%%%%%%%%%%%%%%%%%%%%%%%%%%%%%%%%%%%%%%%%%%%%%%%%%%%%%
From the previous subsection, we inferred that for a fixed $St$ case, as $s$ increases, the particle dynamics can become chaotic beyond some critical strain rate, as demonstrated by Melnikov analysis—a perturbative analysis. However, for larger strain rates, as in the case here ($s = 0.2$), the applicability of Melnikov analysis is questionable. Thus, analysing the system in a large straining regime is analytically challenging and requires a fully numerical approach.
%%%%%%%%%%%%%%%%%%%%%%%%%%%%%%%%%%%%%%%%%%%%%%%%%%%%%%%%%%%%%%%%%%%%%%%%%%%%%%%%%%%%%%%%%%%%%%%%%%%%%%%%%%%%%%%%%%%%%%%%%%%%%%%%%%%%%%%%%%%%%%%%%%%%%%%%%%%%
\section{\label{sec4} Large time dispersion characteristics of particles}
\begin{figure}
    \centering
    \includegraphics[width=1.0\linewidth]{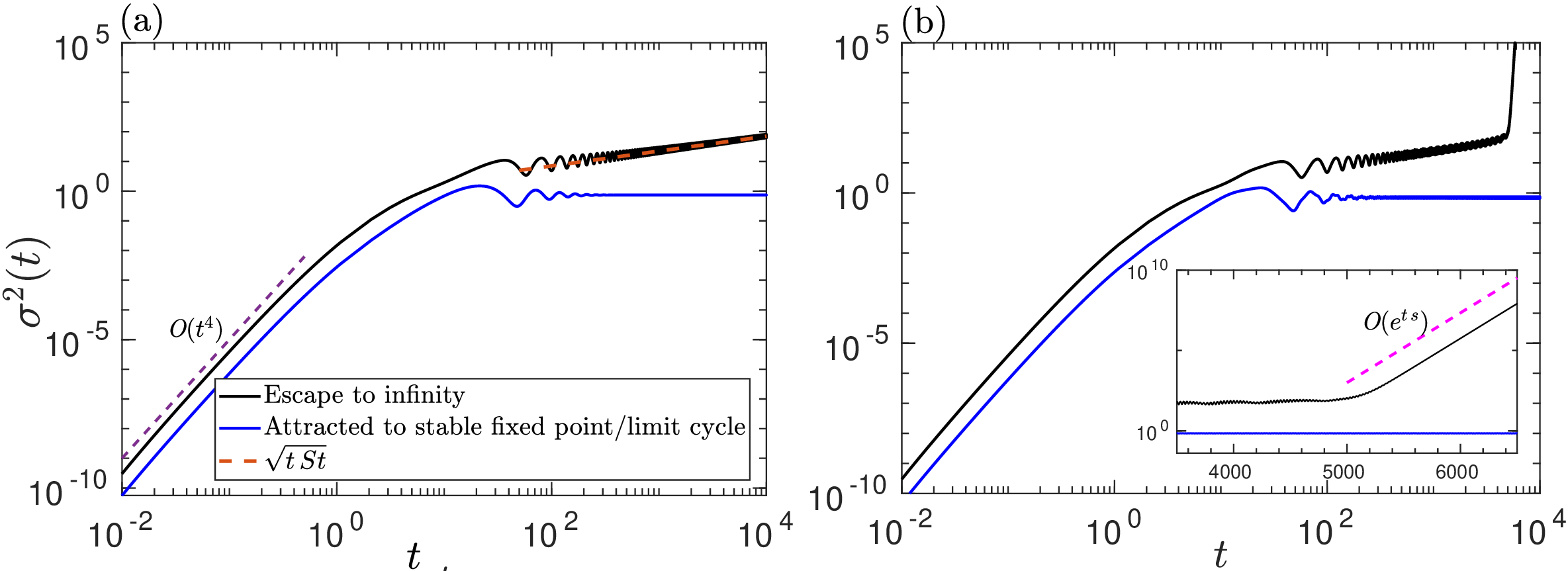}
    \caption{The variation of MSD (evaluated in a co-rotating frame) with time for typical inertial particles of $St = 0.5$ starting within and outside the basin of attraction of stable fixed points/limit cycles in (a) Kirchhoff vortex of $r = 0.5$ and (b) Kida vortex of $r_0 = 0.5$ and $s = 0.01$. The scaling behaviour of MSD is shown using dashed lines. }
    \label{Fig021}
\end{figure}

One of the motivations for studying the dynamics of particles embedded in a sea of coherent structures is to learn about their long-time dispersion and the rate of capture. As we have seen in the case of the Kirchhoff vortex, heavy inertial particles can have two fates at large times: those starting within the basin of attraction of a fixed point will get attracted towards the corresponding fixed point, and those starting outside will spiral off to infinity. We can quantify the dispersion of particles using a mean squared distance (MSD), $\sigma^2(t) = \langle \lVert \textbf{x}_i(t)-\textbf{x}_i(0) \rVert^2 \rangle$, which is the average squared distance the particles travelled from their initial location. Here, $\langle \cdot \rangle$ represents the average taken over all $i$, where $i$ indicates the index of (a few hundred of) trajectories of particles which started from the same basin. The variation of MSD with time can give information about the nature of dispersion the particle follows. MSD increasing quadratically with time indicates ballistic dispersion of the particle, while a linear in-time behaviour of MSD indicates diffusive dispersion of the particle; if MSD saturates with time, it suggests the particle is attracted to some fixed point and will have limited dispersion. The variation of MSD with time for typical particles in a Kirchhoff vortex is numerically obtained and shown in figure \ref{Fig021}(a). The MSD saturates at a large time for particles starting within a basin of attraction (blue curve), indicating the attraction towards the respective fixed point. However, MSD increases with time, indicating dispersion to infinity for a particle starting outside the basins (black curve). The oscillations in the MSD curves indicate the spiralling nature of trajectories. Despite that, the average growth of the black curve can be shown to scale as $\sigma^2(t) \sim \sqrt{t}$, indicating that the particle dispersion is slower than a diffusive process. The same scaling has already been obtained by \citep{ravichandran2014attracting} for the dispersion of inertial particles in a pair of like-signed vortices. In both cases, the particles perceive a point vortex field far away. Appendix \ref{appE} provides a detailed derivation for the scaling.
%%%%%%%%%%%%%%%%%%%%%%%%%%%%%%%%%%%%%%%
%SD_new_St=0p5_s=1e-2
%%%%%%%%%%%%%%%%%%%%%%%%%%%%%%%%%%%%%%%%%%%%%%%%%%%%%%%%%%%%

In the case of a Kida vortex, as the fixed points become limit cycles, corresponding MSD saturates but with sustained periodic oscillations, as shown in figure \ref{Fig021}(b) in blue. On the other hand, a particle centrifuged away will be affected by the heteroclinic orbits $H_p^{\pm}$. These orbits limit the centrifuging, trap the particles along its stable and unstable manifolds, and limit their dispersion. One may remember the far-field attractor mentioned at the beginning of Section \ref{sec3}. Since the fixed points $\textrm{A}_p$ and $\textrm{B}_p$ and the associated heteroclinic connections $H_p^{\pm}$ are stationary in the lab reference frame, for a co-rotating observer, this attractor seems to be counter-rotating, as seen in figure \ref{Fig011}(e-h). Below the associated critical Stokes number $St_{\textrm{cr},p}$, the particles may get transported chaotically and above it can be transported regularly near these orbits. However, at large time, the inertial particles can leak near through the saddles $\textrm{A}_p$ and $\textrm{B}_p$.
%if $2\, St\, s > 1$
This behaviour can be either due to the oscillations induced to the particle trajectory by the time-periodic perturbation by the central rotating ellipse or due to the under-damped oscillations of inertial particle trajectories near the stagnation points \citep{nath2022transport}. The leaked inertial particles are tempted to travel along the extensional axis of the straining flow exponentially with time and thus will have an enhanced dispersion later. The situation can be seen in the snapshot in figure \ref{Fig011}(h), where the centrifuged inertial particles are aligned along the orbits $H_p^{\pm}$; the particle trajectory oscillations near saddles $\textrm{A}_p$ and $\textrm{B}_p$, and their leakage along the extensional axis can be observed in a co-rotating frame. The black curve in figure \ref{Fig021}(b) shows the typical MSD for such particles. Due to the heteroclinic orbit barrier, the particle has a limited dispersion at intermediate times with large amplitude oscillations. At large times, the dispersion becomes exponentially fast (but anisotropic) as they get caught and confined along the unstable manifold of the heteroclinic orbit. This behaviour can be quantified by the scaling relation $\sigma^2(t) \sim \exp(t \, s)$ for small $St$.

One may note that at initial times, the dispersion of all particles starting from rest gets centrifuged from the centre, scaling identically as $\sigma^2(t) \sim t^4$ (super-ballistic), as illustrated in the figure. This behaviour is identical for both the Kirchhoff and Kida vortex. It occurs due to the constant initial acceleration experienced by the particles initially, akin to the scenario elucidated in \citet{nath2024irregular}. Since the particles start with zero initial velocity ($\textbf{v}(t=0) = \mathbf{0}$), equation (\ref{eqn3p2}) provides the initial acceleration $\textbf{a}_0 = \ddot{\textbf{x}}(t=0) = \textbf{u}(\textbf{x}_0,t=0)/St+\textbf{x}_0\, \Omega^2 - \dot{\Omega} \, \hat{\textbf{e}}_z\times \textbf{x}_0$, where $\textbf{x}_0 = \textbf{x}(t=0)$ denotes the initial particle location. Depending on the particle location, it may undergo an approximately constant initial acceleration contributed by the flow and the centrifugal force in a Kirchhoff vortex and, in addition, the Coriolis force in a Kida vortex. For $t \ll 1$, the approximate solution can be obtained as $\textbf{x}(t)\approx \textbf{x}_0 + \textbf{a}_0\, t^2/2$. This solution yields $\sigma^2(t) \approx \lvert \textbf{a}_0\rvert^2\,t^4/4$, i.e., $\sigma \sim t^4$ when $t \ll 1$.
%%%%%%%%%%%%%%%%%%%%%%%%%%%%%%%%%%%%%%%%%%%%%%%%%%%%%%%%%%%%%%%%%%
%%%%%%%%%%%%%%%%%%%%%%%%%%%%%%%%%%%%%%%%%%%%%%%%%%%%%%%%%%%%%%%%%%%%%%%%%%
\subsection{Residence time}
\label{sec4p1}
As shown earlier, in a Kirchhoff vortex, some of the heavy inertial particles can get trapped indefinitely at the fixed points C and D if the Stokes number is below some critical value (see Section \ref{sec2}). The remaining particles would be spirally centrifuged away at a rate $\sigma^2(t) \sim \sqrt{t\, St}$. However, no other particles, aside from those trapped at fixed points C and D, would get trapped by a Kirchhoff vortex. In the case of a Kida vortex, however, there exists a barrier, or separatrices ($H_p^{\pm}$), at the far field ($\textit{O}(1/\sqrt{s})$), formed by the balance between the external straining flow and the central vortex (see figure \ref{Fig019}). The separatrices act as a natural barrier and prevent any fluid tracer from crossing it. As a result, fluid parcels initialised inside the region bounded by the separatrices remain within it and become effectively trapped. However, in the case of finite inertia, particles can cross these separatrices, destroying the permanent trapping effect \citep[see][]{nath2022transport}. Nonetheless, it can be expected that once particles reach closer to the separatrices, they can be temporarily trapped or slowed down by these barriers.
%%%%%%%%%%%%%%%%%%%%%%%%%%%%%%%%%%%%%%%%%%%%%%%%%%%%%%%%%%%%
%Residence_time_small_St_Kida
\begin{figure}
    \centering
\includegraphics[width=1.05\linewidth]{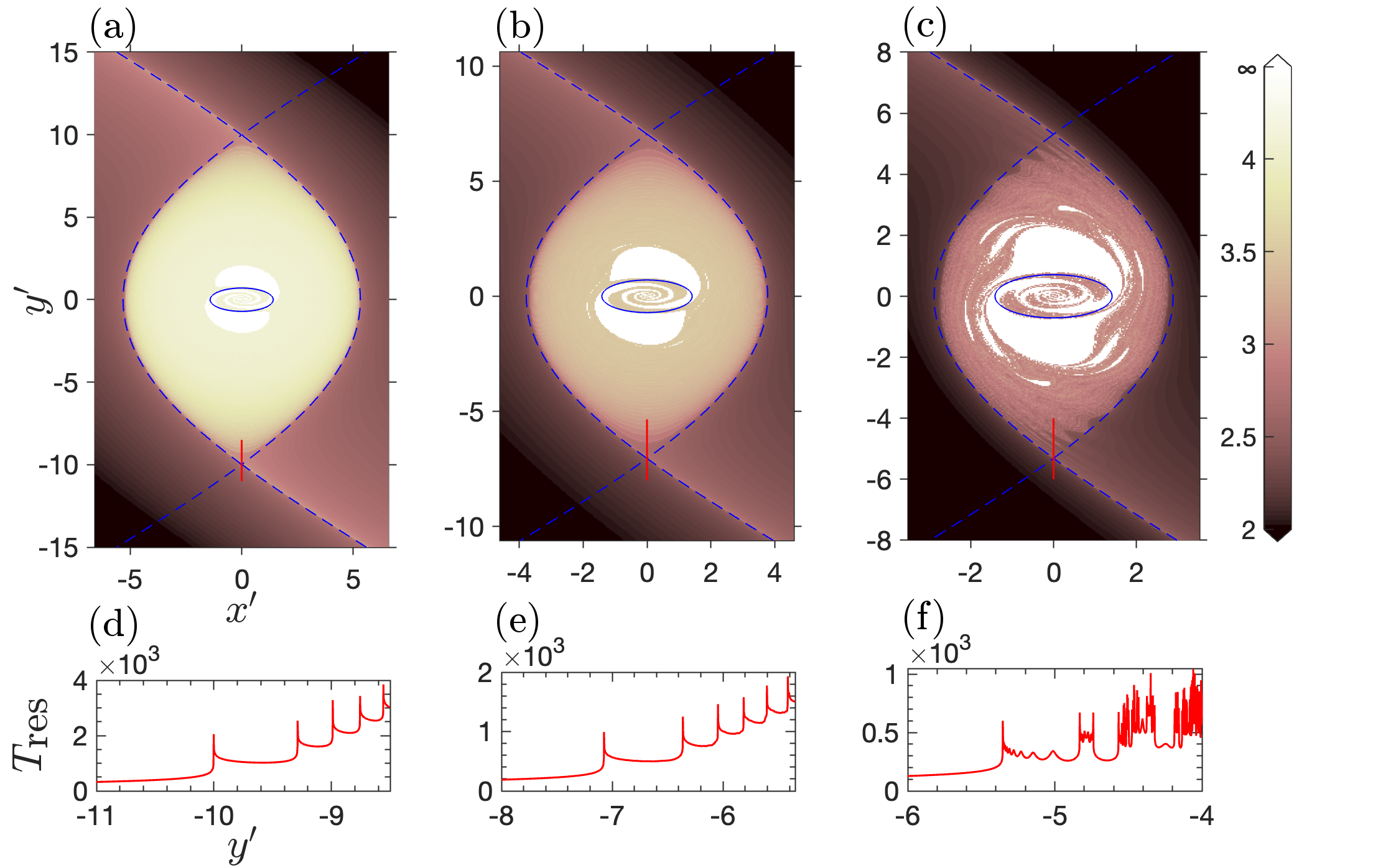}
    \caption{The field of residence time, in logarithmic scale $\log_{10}(T_{\textrm{res}})$, of inertial particles with $St = 0.2$ in the neighbourhood of a Kida vortex with $r_0 = 0.5$ and $\theta_0 = 0$ is shown as a coloured contour plot in the lab reference frame for (a) $s = 0.01$, (b) $s = 0.02$, and (c) $s = 0.035$. The boundary of the initial elliptical vortex patch is depicted as a solid blue curve, while the far-field separatrix at the initial time, separating the interior vortex flow and outer straining flow, is shown as a dashed blue curve. The variation of $T_{\textrm{res}}$ over a vertical section (shown as a red line in (a), (b), and (c)) across the saddle point $\textrm{B}_p$ is shown in plots (d), (e), and (f), respectively, where $x' = 0$ for the sections.}
    \label{Fig022}
\end{figure}
%%%%%%%%%%%%%%%%%%%%%%%%%%%%%%%%%%%%%%%%%%%%

To confirm this, we use a quantitative analysis by calculating the residence time ($T_\textrm{res}$) of particles. For any inertial particle in a Kida vortex, we track its dynamics by solving the equations (\ref{eqn3p9}) and integrating up to $1000$ time periods of the vortex revolution. The time taken for the particle to leave a square box of size $20 \times 20$ is considered the residence time, as particles once escape this box are expected to travel exponentially fast in a strain-dominant regime. We use $11 \times 10^{4}$ regularly arranged particles to obtain the field of $T_\textrm{res}$, shown in figure \ref{Fig022}(a-c) for particles with $St = 0.2$ at three different strain rates. The white regions indicate areas of infinite residence time, while the darkest regions indicate the shortest residence time or the quickest escape time. It can be observed that the white regions in the figure resemble the shape of the basins of attractors (fixed points or limit cycles) near C and D. This is expected, as they indicate the initial locations of particles that will never leave the $20 \times 20$ box. For the smallest value of $s = 0.01$, the interior of the separatrices shows a bright yellow region, indicating finite time trapping; but the inertial particles eventually escape as time progresses. As the strain rate increases, the brightness of this interior region reduces, indicating a shorter residence time. It can be seen that there are no white regions other than the basins, indicating that indefinite trapping can only occur near the attractors at C and D. The stable manifolds of the separatrices, extending to infinity, also show a slightly bright yellow colour, indicating that particles near these manifolds can be temporarily trapped. However, no particles, especially those far away, are permanently trapped by any other attractors.

In addition, for these same cases, we have also plotted the variation of residence time with a vertical section (i.e., a range of $y'$ values at a fixed $x' = 0$). The sections are shown as red vertical lines passing through the saddle point $\textrm{B}_p$ in figures \ref{Fig022}(a-c). The corresponding $T_\textrm{res}$ versus $y'$ is shown in figures \ref{Fig022}(d-f). The section is subdivided into $11 \times 10^4$ uniformly distributed points, and the corresponding residence times of these points are evaluated. It can be observed that,  for smaller strain rates,  there is a regular increase in $T_\textrm{res}$ as $y'$ approaches the vortex centre (see figures \ref{Fig022}(d \& e)). For the larger strain rate $s = 0.035$, however, the $T_\textrm{res}$ shows irregular dependence on $y'$, especially within the region bounded by separatrices - another indication of the underlying fractal nature (see figure \ref{Fig022}(f)).
%%%%%%%%%%%%%%%%%%%%%%%%%%%%%%%%%%%%%%%%%%%%%%%%%%%%%%%%%%%
\section{\label{sec5} Conclusion}
% two fluid model; dusty flows in rotating background
% Kida
% ? poly-disperse system: different fixed points

% ? Not all particles attracted to centres, some may go to infinity

% ? Figs showing evolution

% ? Annihilation of fixed points
We have studied the dispersion of heavy inertial particles in (i) an elliptic Kirchhoff vortex and (ii) a Kida vortex. This study has offered insights into particle transport in two-dimensional flows relevant to geophysical and astrophysical scenarios, which can be approximated as the resultant flow from isolated vortex patches and their interactions. The Kirchhoff vortex is an elliptic patch that self-rotates and creates an unsteady flow around it. The stability characteristics of the fixed points govern the dynamics of the inertial particles in this flow field. When observed from a co-rotating frame with the vortex, some particles are attracted to stable fixed points in the flow field, while the remaining ones are centrifuged to infinity. The location of the fixed points depends on the inertia of the particles. Thus, in a polydisperse system, each particle will cluster into its own fixed points, which can result in the segregation of particles according to their inertia. A critical Stokes number exists above which all the particles only get centrifuged to infinity. 

The introduction of a weak extensional flow to the system (owing to the interaction from other vortices), known as the Kida vortex, disrupts the particle trajectories in the Kirchhoff vortex. Inertial particles in the Kida vortex can have limit cycle trajectories as well as chaotic trajectories in an extended phase space over time. However, a Melnikov analysis revealed that particles with sufficiently high inertia are less prone to chaotic transport.

The present study offers insights into the clustering and dispersion of particles in an ambient vortical flow field. As mentioned earlier, this has been of interest to understanding the trapping of dust, which has the potential for forming planetesimals. However, we have exclusively considered one-way coupling between the particle and fluid phases in the current study. The feedback force from the dispersed phase to the carrier phase can be significant, particularly during the clustering phase. Allowing for two-way coupling introduces additional instabilities, such as streaming instabilities in the context of proto-planetary disks \citep{youdin2005streaming,youdin2007protoplanetary}. In the axisymmetric vortex system context, two-way coupling results in the expulsion of particle clumps in a spiralling arm manner, even in the limit of vanishing particle inertia and leads to a dramatic destabilization of an otherwise stable isolated vortex due to a particle-induced baroclinic torque \citep{shuai2022instability}. As mentioned earlier, a non-axisymmetric vortex system formed by like-signed point vortices exhibits topological and dynamical similarities to the system under consideration. While it is known that like-signed vortex pairs are prone to merging below a critical separation \citep{griffiths1987coalescing,dritschel2002vortex,cerretelli2003physical}, the inclusion of two-way coupling introduces novel vorticity dynamics before their merging \citep{shuai2024merger}. At the leading order, momentum coupling from fluid to particle induces clustering dynamics, as demonstrated in this paper. However, once clustering is achieved, the particle concentration at those locations becomes sufficiently high that the feedback from the particle phase to the fluid phase cannot be neglected. A recent study by the authors \citep[see][]{nath2024instability} reiterates the same argument: a dusty simple shear flow is unstable when the dust particles are distributed non-uniformly. In two-dimensional dusty turbulence, this feedback has produced a non-universal energy spectrum that depends on particle inertia and mass loading \citep{pandey2019clustering}. Therefore, it is crucial to approach the results presented in this paper with caution, acknowledging that their applicability may be limited as long as the particle feedback to the fluid is negligible, particularly in a dilute mixture, in the earlier stages of dynamics before particle clustering. After the onset of clustering, the feedback force would alter the morphology of the underlying coherent structures; subsequently, delineating a causal effect between clustering and two-way coupling may become difficult. 

While the present study focuses on the transport of very heavy particles, there may be potential applications for studying particle transport in oceanic flows where the particle-to-fluid density ratio is finite. Understanding particles' dispersion and duration of residence in a region is of immense interest when studied in the oceanic context. Studies on the transport of marine particulate matter often overlook the effects of finite inertia, treating the particles as tracers. Recently, however, the Maxey-Riley equation has been used to study transport in scenarios relevant to oceanic flows, be it in canonical flows like the unsteady double gyres \citep{sudharsan2016lagrangian} or the flow-field relevant to the feeding of jellyfish \citep{peng2009transport}. In most of these scenarios, coherent structures dominate the flow field and thus play an overarching role in the transport. Accounting for the inertial nature of particles is crucial for understanding the motion of drifting buoys, macroscopic algae, or debris, as shown in \citet{miron2020clustering}. Satellite observations show that the transport of floating matter on the ocean surface near coherent eddies is dissipative, contrary to the expected conservative fluid trajectories. \citet{beron2015dissipative} have shown that particle inertia effects, rather than the turbulent nature of the background flow, can better explain the divergence of nearby buoy trajectories observed in satellite data. Thus, future studies can explore the trajectories of particles in the vicinity of long-lived coherent structures in the ocean when particle inertia is accounted for in the dynamics, with appropriate modification to the Maxey-Riley equation \citep{beron2019building}.

% \textcolor{blue}{This study may also find applications in search and rescue operations in the ocean, where particle transport in eddies within a rotating background is prevalent. Recent studies have approached this problem using the Maxey-Riley equations \citep[see][]{beron2019building}. Our study can predict the paths and dispersive traits of particles that are not fully submerged and in the vicinity of ocean eddies. To achieve accurate estimates of particle tracking on the large scales of ocean transport, the Earth's rotation may also need to be accommodated in the Coriolis term in addition to the inherent rotation of eddies.}

%%%%%%%%%%%%%%%%%%%%%%%%%%%%%%%%%%%%%%%%%%%%%%%%%%%%%%%%%%%%%

%%%%%%%%%%%%%%%%%%%%%%%%%%%%%%%%%%%%%%%%%%%%%%%%%%%%%%%%%%%%%%%%%%%%%%%%%%%%%%%%%%%%%%%%%%%%

% \backsection[Supplementary data]{\label{SupMat}Supplementary material and movies are available at \\https://doi.org/10.1017/jfm.2019...}

\backsection[Acknowledgements]{A.V.S.N. thanks the Prime Minister’s Research Fellows (PMRF) scheme, Ministry of Education, Government of India. A.R. acknowledge SERB project SPR/2021/000536 for funding. A.R. and A.V.S.N. acknowledge the support of the Centre for Atmospheric and Climate Sciences (CACS) at IIT Madras. A.R. and A.V.S.N. also acknowledge the support of the Geophysical Flows Lab (GFL) at IIT Madras.}
%Complex Systems and Dynamics Group at IIT Madras}

% {Acknowledgements may be included at the end of the paper, before the References section or any appendices. Several anonymous individuals are thanked for contributions to these instructions.}

% \backsection[Funding]{Please provide details of the sources of financial support for all authors, including grant numbers. Where no specific funding has been provided for research, please provide the following statement: "This research received no specific grant from any funding agency, commercial or not-for-profit sectors." }

% \backsection[Declaration of interests]{A Competing Interests statement is now mandatory in the manuscript PDF. Please note that if there are no conflicts of interest, the declaration in your PDF should read as follows: {\bf Declaration of Interests}. The authors report no conflict of interest.}

% \backsection[Data availability statement]{The data that support the findings of this study are openly available in [repository name] at http://doi.org/[doi], reference number [reference number]. See JFM's \href{https://www.cambridge.org/core/journals/journal-of-fluid-mechanics/information/journal-policies/research-transparency}{research transparency policy} for more information}

\backsection[Author ORCIDs]{A. V. S. Nath, https://orcid.org/0000-0003-2144-2978; A. Roy, https://orcid.org/0000-0002-0049-2653}

% {Authors may include the ORCID identifers as follows.  F. Smith, https://orcid.org/0000-0001-2345-6789; B. Jones, https://orcid.org/0000-0009-8765-4321}

% \backsection[Author contributions]{Authors may include details of the contributions made by each author to the manuscript'}

\appendix
\section{\label{appA} Governing equations in the co-rotating Cartesian reference frame}
The dynamic equations (\ref{eqn2p2}) governing the transport of heavy inertial particles in the Kirchhoff vortex are obtained from the Maxey-Riley equation equation (\ref{eqn2p1}) by writing it in elliptic coordinates ($\xi, \eta$), which is helpful in analytical exercises. However, for numerical simulations, it is convenient to use these equations in the co-rotating Cartesian reference frame ($x,y$), which are
\begin{subequations}
\label{eqnA1}
\begin{eqnarray}
    \dot{v}_x = \frac{u_x-v_x}{St}+x \, \Omega^{2}+2\, \Omega\, v_y~, \quad \dot{v}_x = v_x~, \\
    \dot{v}_y = \frac{u_y-v_y}{St}+y \, \Omega^{2}-2\, \Omega\, v_x~, \quad \dot{v}_y = v_y ~.
\end{eqnarray}
\end{subequations}
 The flow velocity field in the stationary reference frame can be obtained from the stream function as $u_x' = \frac{\partial \psi'}{\partial y'}$ and $u_y' = -\frac{\partial \psi'}{\partial x'}$, which is unsteady. The velocity field in the co-rotating frame can be obtained from the following transformation 
\begin{subequations}
\label{eqnA2}
\begin{eqnarray}
    u_x = u_x' \, \cos \theta + u_y' \, \sin \theta + \Omega \, y~,\\
    u_y = -u_x' \, \sin \theta + u_y' \, \cos \theta - \Omega \, x~.
\end{eqnarray}
\end{subequations}
For Kirchhoff vortex, we get
%%%%%%%%%%%%%%%%%%%%%%%%%%%%%%%%%%%%%%
\begin{subequations}
\begin{eqnarray}
    u_x &=& \left\{
\begin{array}{ll}
      k^{-1}\, \sin \eta\, \{-\cosh \xi+(1+k^2\, \Omega)\, \sinh \xi\}, & \quad \quad \tanh \xi >  r \vspace{0.25cm}\\
      -k\, \Omega\, r^{-1}\, \sinh \xi\, \sin \eta, & \quad \quad \tanh \xi <  r \\
\end{array} 
\right. \\
u_y &=& \left\{
\begin{array}{ll}
      -k^{-1}\, \cos \eta\, \{\sinh \xi+(-1+k^2\, \Omega)\, \cosh \xi\}, & \quad  \, \, \tanh \xi > r \vspace{0.25cm}\\
       k\, \Omega\,r\, \cosh \xi\, \cos \eta, & \quad \, \, \, \tanh \xi <  r~, \\
\end{array} 
\right. 
\end{eqnarray}
\label{eqnA3}
\end{subequations}
where the variables in both the co-ordinate system are related as $x = k\, \cosh \xi \, \cos \eta$ and $y = k\, \sinh \xi\, \sin \eta$. The dynamical system remains four-dimensional; however, now in the phase space of variables $x, y, v_x$ and $v_y$. The fixed points of the system can be obtained by solving for $(\dot{x},\dot{y},\dot{v}_x,\dot{v}_y) = (0,0,0,0)$. Trivially it implies that $\overline{v}_x =\overline{v}_y = 0$ and the fixed point positions $\overline{\textbf{x}} = (\overline{x},\overline{y})$ are solutions of the vector equation $\textbf{u}(\textbf{x})+St\, \Omega^2\, \textbf{x} = 0$. Using numerical solvers like `fsolve' in MATLAB or Newton-Raphson method, one can obtain the fixed points and verify that they exactly match the analytical expressions (\ref{eqn2p6}). For the Cartesian equations (\ref{eqnA1}), the entries in the stability matrix will be much simpler than equation (\ref{eqn2p8}) and are
\begin{eqnarray}
\mathsfbi{J} &=& 
\begin{pmatrix}
0 & 0 & 1 & 0\\
0 & 0 & 0 & 1\\
\Omega^2+\frac{1}{St}\, \frac{\partial u_x}{\partial x} & \frac{1}{St}\, \frac{\partial u_x}{\partial y} & -\frac{1}{St} & 2\, \Omega\\
\frac{1}{St}\, \frac{\partial u_y}{\partial x} & \Omega^2+\frac{1}{St}\, \frac{\partial u_y}{\partial y} & -2\, \Omega & -\frac{1}{St}
\end{pmatrix} \nonumber \\
&=&
\begin{pmatrix}
0 & 0 & 1 & 0\\
0 & 0 & 0 & 1\\
 \Omega^2+\frac{h^2\, \sin 2\eta}{2\, k^2\, St}& \frac{\Omega}{St}+\frac{2-h^2\, \sinh 2\xi}{2\, k^2\, St} & -\frac{1}{St} & 2\, \Omega\\
 -\frac{\Omega}{St}+\frac{2-h^2\, \sinh 2\xi}{2\, k^2\, St} & \Omega^2-\frac{h^2\, \sin 2\eta}{2\, k^2\, St}   & -2\, \Omega & -\frac{1}{St}
\end{pmatrix}~.
\label{eqnA4}
\end{eqnarray}
Note that the eigenvalues of the Jacobian matrix in equation (\ref{eqnA4}) and equation (\ref{eqn2p8}) will be the same since they are invariants of the same dynamical system. However, the components of the eigenvectors differ by the coordinate transformation.
%%%%%%%%%%%%%%%%%%%%%%%%%%%%%%%%%%%%%%%%%%%%%%%
\section{\label{appB} Perturbation analysis for the limit cycles for a Kida vortex - entries of $\mathsfbi{K}$, $\mathsfbi{L}$, $\mathsfbi{M}$ and $N$}
The strain perturbation to the system affects the fixed points ($\overline{\xi},\overline{\eta}$) of a Kirchhoff vortex. Using the method of \cite{ijzermans2006accumulation}, we assume the fixed points are perturbed to a limit cycle and solve for its coordinates ($ \overline{\xi}+s\, \xi'(t) + \mathcal{O}(s^2),  \overline{\eta}+s\, \eta'(t) + \mathcal{O}(s^2)$) using equation \ref{eqn3p5}. The entries in the coefficient matrices depend on the fixed points of the Kirchhof vortex and are listed below:  
\begin{equation}
\setlength{\arraycolsep}{3pt}
\renewcommand{\arraystretch}{1.3}
\mathsfbi{L} = \left[
\begin{array}{cc}
    1   &  -2\, St\, \Omega_0  \\
    2\, St\, \Omega_0   &  1  \\
\end{array}  \right],
\quad 
\setlength{\arraycolsep}{3pt}
\renewcommand{\arraystretch}{1.3}
\mathsfbi{K} = \left[
\begin{array}{cc}
    K_{11}   &  K_{12} \\
    K_{21}   &  K_{22}  \\
\end{array}  \right], \quad \textrm{and} \, \, \, 
\setlength{\arraycolsep}{3pt}
\renewcommand{\arraystretch}{1.3}
\mathsfbi{M} = \left[
\begin{array}{cc}
    M_{11}   &  M_{12} \\
    M_{21}   &  M_{22}  \\
\end{array}  \right] .
\label{eqn3p6}
\end{equation}
Outside the ellipse ($\tanh \xi >r$), corresponding to the fixed points A or B or C or D, denoted as $(\overline{\xi},\overline{\eta})$, the elements of $\mathsfbi{K}$, $\mathsfbi{M}$ and $N$ are
\begin{subequations}
\begin{eqnarray}
K_{11}&=& \overline{h}^2 \, \Omega_0 \left(\sin (2 \, \overline{\eta})-e^{2 \, \overline{\xi}} \, St \, \Omega_0\right)~, \\
K_{12}&=& \, \overline{h}^2 \, \cos (2 \, \overline{\eta}) \left(\, \Omega_0-k_0^{-2}\,e^{-2 \, \overline{\xi}}\right)~,\\
  K_{21}&=&  \overline{h}^2 \left( \Omega_0 \, \cosh (2 \, \overline{\xi})-k_0^{-2}\,e^{-2 \, \overline{\xi}} \, \cos (2 \, \overline{\eta})\right)~,\\
   K_{22}&=&  \overline{h}^2 \, \Omega_0 \left( St \, \Omega_0 \,\cos (2 \, \overline{\eta})-l_1^{-1}\,e^{-2 \, \overline{\xi}} \sin (2 \,\overline{\eta})\right)~,\\
M_{11}&=& \frac{\overline{h}^2}{4} \,\left\{\sin (2 \, \overline{\eta})\, ( l_1-\cosh (2 \, \overline{\xi}))-\, l_3 \, St \, \Omega_0 \sinh (2
   \, \overline{\xi})\right\}~,\\
   M_{12}&=& \frac{\overline{h}^2}{4} \, \left\{\sinh (2 \, \overline{\xi}) (\Lambda_0-\cos (2 \, \overline{\eta}))-2 \, l_1 \, St \, \Omega_0 \sin
   (2 \, \overline{\eta})\right\}~,\\
   M_{21}&=& \frac{\overline{h}^2}{4} \,  \left\{\sinh (2 \, \overline{\xi}) ( l_1-\cos (2 \, \overline{\eta}))+ l_3 \, St \, \Omega_0 \sin (2
   \, \overline{\eta})\right\}~,\\
   M_{22}&=& -\frac{\overline{h}^2}{4}\, \left\{\sin (2 \, \overline{\eta}) (\Lambda_0-\cosh (2 \, \overline{\xi}))+2 \, l_1 \, St \, \Omega_0 \sinh
   (2 \, \overline{\xi})\right\}~,\\
   N_1&=& \frac{\overline{h}^2}{4} \, \cos (2 \, \theta_0) \left(2 \, k_0^{-2}\,\sin (2 \, \overline{\eta})+ l_3 \, St \, \Omega_0
   \sinh (2 \, \overline{\xi})\right)~,\\
   N_2&=& \frac{\overline{h}^2}{4} \,  \cos (2 \, \theta_0) \left(2 \, k_0^{-2}\,\sinh (2 \, \overline{\xi})- l_3 \, St \, \Omega_0
   \sin (2 \, \overline{\eta})\right)~,
    \end{eqnarray}
    \label{eqnB1}
\end{subequations}
where $\overline{h} = (\cosh^2 \overline{\xi} - \cos^2 \overline{\eta})^{-1/2}$, $l_1 = (1-r_0)/(1+r_0)$ and $l_3 = (1-4\, r_0+r_0^2)/(1-r_0^2) = 2\, l_1-\Lambda_0$. Inside the ellipse ($\tanh \xi <r$), corresponding to the fixed point at the origin O, the corresponding elements are
\begin{subequations}
\begin{eqnarray}
K_{11}&=& r_0\, \Omega_0~, \\
K_{12} &=& K_{21}= -St\, \Omega_0^2~,\\
K_{22}&=&-\Omega_0/r_0~,\\
M_{11} &=& M_{12} = M_{21} = M_{22} = N_1 = N_2 = 0~.
\end{eqnarray}
\label{eqnB2}
\end{subequations}
%%%%%%%%%%%%%%%%%%%%%%%%%%%%%%%%%%%%%%%%%%%%%%%%%%%%%%%%%%%%%%%%%%%%%%%%%%%%%%%%%%%%%%%%%%%%%%
\section{\label{appC} Terms required for the evaluation of the Melnikov function for weakly inertial particles in a Kida vortex}
Outside the ellipse $\tanh \xi > r_0+s\, r_1+\textit{O}(s^2)$,
\begin{subequations}
    \begin{eqnarray}
    \hat{f}_1 &=& \frac{h^2}{2}\, \left(-\Omega_0+k_0^{-2}\, e^{-2\xi}\right)\, \sin 2\eta~,\\
    \hat{f}_2 &=& -\frac{h^2}{2}\, \left(\Omega_0\, \sinh 2\xi+k_0^{-2}\, \left[-1+e^{-2\xi}\, \cos 2\eta \right] \right)~,\\
    \hat{g}_1 &=& \frac{h^2}{4}\, \left\{\mathcal{F}(\xi,\eta;r_0,\Omega_0\, t+\theta_0)- 2\, \frac{\Omega_0}{r_0}\, r_1\,\left( l_1-l_2\,e^{-2\xi}\right)\, \sin 2\eta \right\}~,\\
        \hat{g}_2 &=& \frac{h^2}{4}\, \left\{\mathcal{G}(\xi,\eta;r_0,\Omega_0\, t+\theta_0)- 2\, \frac{\Omega_0}{r_0}\, r_1\,\left( l_1\, \sinh 2\xi+l_2\,[-1+e^{-2\xi}\, \cos 2\eta]\right)\right\}~,\\
    \hat{\phi}_1 &=& h^2\,k_0^{-2}\,e^{-2\xi}\, \left( k_0^{-2}-\Omega_0\, \cos 2\eta\right) \\
        \hat{\phi}_2 &=&-h^2\,k_0^{-2}\,e^{-2\xi}\,\Omega_0\,\sin 2\eta~. 
\end{eqnarray}
\label{eqnC1}
\end{subequations}
Inside the elliptic region $\tanh \xi < r_0+s\, r_1+\textit{O}(s^2)$,
\begin{subequations}
    \begin{eqnarray}
    \hat{f}_1 &=& \frac{h^2}{2}\, \left(-\Omega_0+\frac{1}{2}\,\left\{ 1-l_1\,\cosh 2\xi\right\}\right)\, \sin 2\eta~,\\
    \hat{f}_2 &=& \frac{h^2}{2}\, \left(-\Omega_0+\frac{1}{2}\,\left\{ 1-l_1\,\cos 2\eta\right\}\right)\, \sinh 2\xi~,\\
    \hat{g}_1 &=& \frac{h^2}{4}\, \left\{\mathcal{F}(\xi,\eta;r_0,\Omega_0\, t+\theta_0)- 2\,\frac{\Omega_0}{r_0}\, r_1\,\left( l_1-\cosh 2\xi\right)\, \sin 2\eta \right\}~,\\
        \hat{g}_2 &=& \frac{h^2}{4}\, \left\{\mathcal{G}(\xi,\eta;r_0,\Omega_0\, t+\theta_0)- 2\,\frac{\Omega_0}{r_0}\, r_1\,\left( l_1-\cos 2\eta\right)\, \sinh 2\xi \right\} ~,\\
    \hat{\phi}_1 &=& \frac{h^2\,l_1}{2}\,\sinh 2\xi\, \left( k_0^{-2}-\Omega_0\, \cos 2\eta\right)~, \\
        \hat{\phi}_2 &=&-\frac{h^2\, l_1}{2}\,\sin 2\eta \,\left( k_0^{-2}-\Omega_0\, \cosh 2\xi\right)~,  
\end{eqnarray}
 \label{eqnC2}
\end{subequations}
where $l_2 = (1+r_0^2)/(1-r_0)^2$. The functions $\mathcal{F}$ and $\mathcal{G}$ are defined as
\begin{subequations}
    \begin{eqnarray}
                    \mathcal{F}(\xi,\eta;r,\theta) &=& \cosh 2\xi\, \sin 2\eta\, \cos 2\theta+\sinh 2\xi\, \cos 2\eta\, \sin 2\theta \nonumber \\
                    &-&\Lambda\, \left( \sin 2\eta\, \cos 2\theta+\sinh 2\xi\, \sin 2\theta\right)~,\\
        \mathcal{G}(\xi,\eta;r,\theta) &=& \sinh 2\xi\, \cos 2\eta\, \cos 2\theta-\cosh 2\xi\, \sin 2\eta\, \sin 2\theta\nonumber \\
 &+&\Lambda\, \left( \sin 2\eta\, \sin 2\theta-\sinh 2\xi\, \cos 2\theta\right)~.
    \end{eqnarray}
     \label{eqnC3}
\end{subequations}
Expressions for $\hat{f}_1$ to $\hat{g}_2$ for a Kida vortex of $\theta_0 = \pi/4$ can also be found in  \cite{kawakami1999chaotic}, but with a sign mistake in $l_1$ and $l_2$.
%%%%%%%%%%%%%%%%%%%%%%%%%%%%%%%%%%%%%%%%%%%%%%%%
\section{\label{appD} Evaluation of the Melnikov function in the far-field of a Kida vortex}
The streamfunction for a Kida vortex in the stationary reference frame   ($\psi' = \psi'_v+\psi'_e$) can be written in far-field (far from the central elliptic vortex) using polar coordinates as 
%\citep[see][]{kawakami1999chaotic}
\begin{equation}
\label{eqnD1}
    \psi' = -\frac{1}{2}\, \log R-\frac{s\, R^2}{4}\, \cos 2\Theta -\frac{k^2}{16\, R^2}\, \cos (2\, \Theta-2\, \theta)+\textit{O}(R^{-4})~,
\end{equation}
where the Cartesian coordinates in stationary frame $(x',y')$ are related to the polar coordinates in stationary frame $(R,\Theta)$ as $x' = R\, \cos \Theta$ and $y' = R\, \sin \Theta$. The dominant point vortex field balances with external straining to create an integrable system far-field. However, the central rotating ellipse induces a time-periodic perturbation to this system. We re-scale radial and time coordinates with strain rate such that the balance between the point vortex and extensional flow is apparent and the perturbation due to the elliptic vortex is clearly visible. Following \citet{kawakami1999chaotic}, the appropriate scaling is $\Tilde{R} = R\, \sqrt{s}$, $\Tilde{t} = t\, s$, which will reduce equation (\ref{eqnD1}) to the stream function in scaled variables as
\begin{equation}
\label{eqnD2}
    \psi' = -\frac{1}{2}\, \log \tilde{R}-\frac{ \tilde{R}^2}{4}\, \cos 2\Theta -\frac{s\, k^2}{16\, \tilde{R}^2}\, \cos (2\, \Theta-2\, \theta)+\textit{O}(s^2)~.
\end{equation}

\begin{figure}
    \centering
    \includegraphics[width=0.4\linewidth]{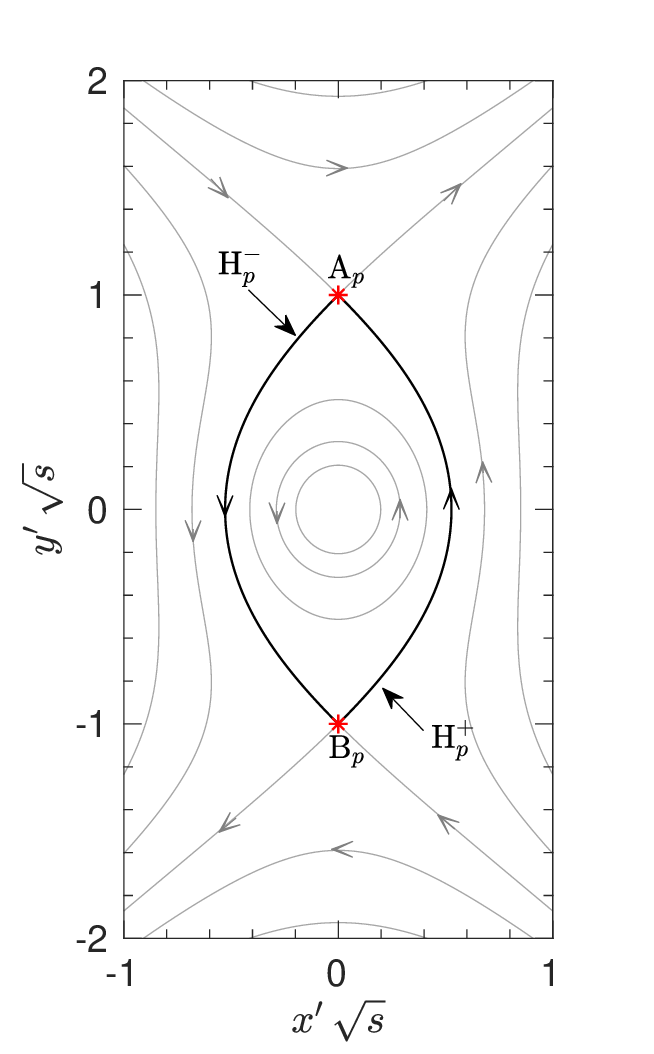}
    \caption{The streamlines of the flow-field result from the balance between the irrotational (point) vortex and external extensional flow. The flow pattern mimics the far-field flow around a Kida vortex (c.f figure \ref{Fig019}). The hyperbolic fixed points and connecting heteroclinic orbits are marked. }
    \label{Fig023}
\end{figure}

\begin{figure}
    \centering
    \includegraphics[width=0.75\linewidth]{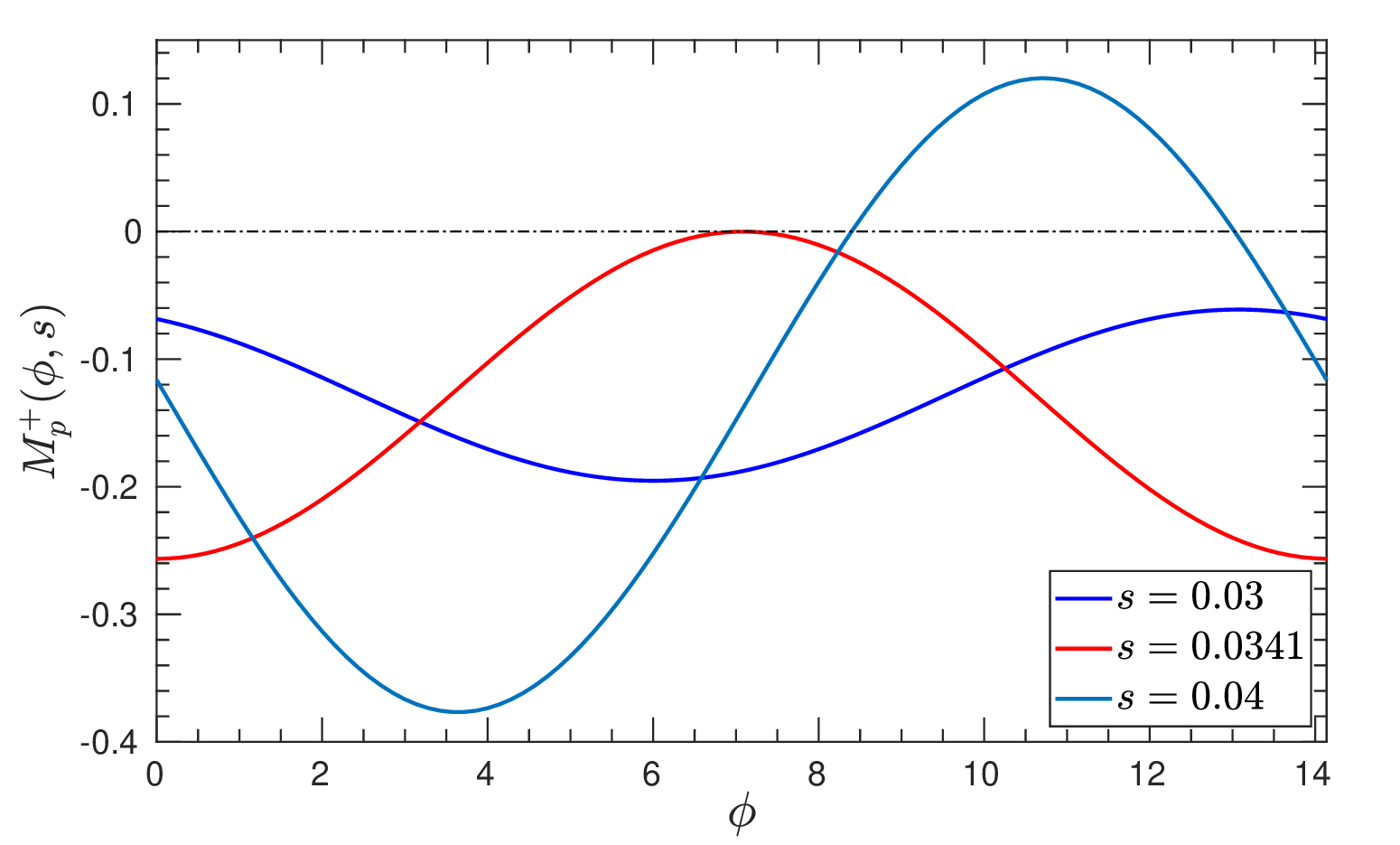}
    \caption{The Melnikov function $M_p^+$ corresponding to a Kida vortex of $r_0 = 0.5$ and $\theta_0 = 0$ for particles of $St = 0.1$ is plotted against $\phi$ for different values of strain rate $s$. The abscissa is shown as a dashed-dotted line to identify the zeros of the Melnikov function. %(b) Curves in the $s - St$ plane demarcate the regions where  $M_p^+$ has odd zeros, and it has not, for various $r_0$ values. 
    }
    \label{Fig024}
\end{figure}

The first two terms on the right-hand side indicate the integrable system formed by the point vortex and the extensional flow. The streamlines of this Hamiltonian system are shown in figure \ref{Fig023}, along with the heteroclinic orbits $H_p^{\pm}$ and far-field fixed points $\textrm{A}_p$ and $\textrm{B}_p$ as marked. Observe the similarity of the far-field streamlines to the snapshots in figure \ref{Fig019}. The third term on the right-hand side is the perturbation due to the central rotating elliptic vortex patch. Using the slow manifold equations for inertial particles in the stationary reference frame
\begin{subequations}
\label{eqnD3}
    \begin{eqnarray}
    \dot{x}' = u_x'-St\, \left[\frac{\partial u_x'}{\partial t}+u_x'\, \frac{\partial u_x'}{\partial x'}+u_y'\, \frac{\partial u_x'}{\partial y'} \right]+\textit{O}(St^2)~,\\
        \dot{y}' = u_y'-St\, \left[\frac{\partial u_y'}{\partial t}+u_x'\, \frac{\partial u_y'}{\partial x'}+u_y'\, \frac{\partial u_y'}{\partial y'} \right]+\textit{O}(St^2)~,
\end{eqnarray}
\end{subequations}
%%%%%%%%%%%%%%%%%%%%%%%%%%%%%%%%%%%%%%%%%%%%%%%%%%%%%%%%%%%%%%%%%%%%%%%%%%5
%farfield_streamlines
%%%%%%%%%%%%%%%%%%%%%%%%%%%%%%%%%%%%%%
%%%%%%%%%%%%%%%%%%%%%%%%%%%%%%%%%%%%%%%%%%%%
we may write the modified kinematic equations for a particle in polar coordinates as
\begin{subequations}
\label{eqnD4}
    \begin{eqnarray}
    \dot{R} &=& \frac{1}{R}\, \frac{\partial \psi'}{\partial \Theta}+\frac{St}{R}\, \left\{ -\frac{\partial^2 \psi'}{   \partial \Theta\, \partial t}+\left(\frac{\partial \psi'}{\partial R}\right)^2+\frac{1}{R}\, \left[\frac{\partial^2 \psi'}{\partial \Theta^2}\, \frac{\partial \psi'}{\partial R}-\frac{\partial^2 \psi'}{\partial R\, \partial \Theta} \, \frac{\partial \psi'}{\partial \Theta}\right]+\frac{1}{R^2}\, \left(\frac{\partial \psi' }{\partial \Theta}\right)^2\right\} \nonumber \\
    &+&\textit{O}(St^2)~,\\
 \dot{\Theta} &=& -\frac{1}{R}\, \frac{\partial \psi'}{\partial R}+\frac{St}{R}\, \left\{ \frac{\partial^2 \psi'}{   \partial R\, \partial t}+\frac{1}{R}\, \left[\frac{\partial^2 \psi'}{\partial R^2}\, \frac{\partial \psi'}{\partial \Theta}-\frac{\partial^2 \psi'}{\partial R\, \partial \Theta} \, \frac{\partial \psi'}{\partial R}\right]+\frac{1}{R^2}\, \frac{\partial \psi' }{\partial \Theta}\, \frac{\partial \psi' }{\partial R}\right\} \nonumber \\
 &+&\textit{O}(St^2)~.  
\end{eqnarray}
\end{subequations}
After substituting the small $s$ expansion of $k$ and $\theta$ (see Section \ref{sec3p1p1}) in equation (\ref{eqnD4}), evolution equations for the scaled coordinates can be obtained as
\begin{subequations}
\label{eqnD5}
    \begin{eqnarray}
        \frac{d \tilde{R}}{d \tilde{t}} &=& \tilde{f}_1 + s\, \left\{\tilde{g}_1+ St\, \tilde{\phi}_1 \right\}+\textit{O}(s^2, St^2\, s, s^2\, St)~,\\
        \frac{d \Theta}{d \tilde{t}} &=& \tilde{f}_2 + s\, \left\{\tilde{g}_2+ St\, \tilde{\phi}_2 \right\}+\textit{O}(s^2, St^2\, s, s^2\, St)~,
    \end{eqnarray}
\end{subequations}
where 
\begin{subequations}
\label{eqnD6}
    \begin{eqnarray}
    \tilde{f}_1 &=& \frac{\tilde{R}}{2}\, \sin 2\Theta~,\\
    \tilde{f}_2 &=& \frac{1}{2\, \tilde{R}^2}+\frac{1}{2}\, \cos 2\Theta~,\\
    \tilde{g}_1 &=& \, \frac{ k_0^2}{8\, \Tilde{R}^3}\, \sin(2\, \Theta-2\, \theta_0-2\, \Omega_0\, \tilde{t}/s)~,\\
    \tilde{g}_2 &=& - \frac{ k_0^2}{8\, \Tilde{R}^4}\, \cos(2\, \Theta-2\, \theta_0-2\, \Omega_0\, \tilde{t}/s)~,\\
    \tilde{\phi}_1 &=& \frac{1}{4\, \tilde{R}^3}\, \left\{1-\tilde{R}^4+k_0^2\, \Omega_0\, \cos(2\, \Theta-2\, \theta_0-2\, \Omega_0\, \tilde{t}/s)\right\}~,\\
    \tilde{\phi}_2 &=& \frac{1}{2\, \tilde{R}^2}\, \left\{\sin 2\Theta+\frac{k_0^2\, \Omega_0}{2\, \tilde{R}^2}\, \sin(2\, \Theta-2\, \theta_0-2\, \Omega_0\, \tilde{t}/s)\right\}~.
\end{eqnarray}
\end{subequations}
Note that the above system of equations reduces to the form given in \citet{kawakami1999chaotic} for the case of $St = 0$ and $\theta_0 = 0$ (with a sign mistake in $\tilde{g}_1$ and $\tilde{g}_2$). For $s \ll 1$ and $St \ll 1$, the Melnikov function associated with the far-field hyperbolic fixed points $\textrm{A}_p$ and $\textrm{B}_p$ can be evaluated as
\begin{equation}
\label{eqnD7}
\begin{split}
        M_p^{\pm}(\phi,s) = \int_{-\infty}^{\infty}\bigg[\tilde{f}_1(\tilde{R}_0^{\pm}(\tilde{t}),\Theta_0^{\pm}(\tilde{t}))\, \left\{\tilde{g}_2\left(\tilde{R}_0^{\pm}(\tilde{t}),\Theta_0^{\pm}(\tilde{t});\frac{\tilde{t}}{s}+\phi\right)+St\, \tilde{\phi}_2\left(\tilde{R}_0^{\pm}(\tilde{t}),\Theta_0^{\pm}(\tilde{t});\frac{\tilde{t}}{s}+\phi\right)\right\} \\
    -\tilde{f}_2(\tilde{R}_0^{\pm}(\tilde{t}),\Theta_0^{\pm}(\tilde{t}))\, \left\{\tilde{g}_1\left(\tilde{R}_0^{\pm}(\tilde{t}),\Theta_0^{\pm}(\tilde{t});\frac{\tilde{t}}{s}+\phi\right)+St\, \tilde{\phi}_1\left(\tilde{R}_0^{\pm}(\tilde{t}),\Theta_0^{\pm}(\tilde{t});\frac{\tilde{t}}{s}+\phi\right)\right\}\bigg]\, e^{-\int_0^{\tilde{t}} \textrm{tr}(D\tilde{f})\,dt'}\, d\tilde{t}~,
\end{split}
\end{equation}
where the integration is performed along the heteroclinic orbits $H_p^{\pm}$ of the unperturbed integrable system parametrised by $(\tilde{R}_0^{\pm}(\tilde{t}),\Theta_0^{\pm}(\tilde{t}))$. Here
\begin{equation}
\label{eqnD8}
    \textrm{tr}(D\tilde{f}) = \frac{\partial \tilde{f}_1}{\partial \Tilde{R}}+\frac{\partial \tilde{f}_2}{\partial \Theta} = -\frac{1}{2}\, \sin 2\Theta 
\end{equation}
is the trace of the Jacobian matrix of the unperturbed system, which is not zero in general. %\citet{kawakami1999chaotic} has not accounted for this trace part in calculating the Melnikov function for fluid tracers, as mentioned in Section \ref{sec3p1p2}. %Though, as mentioned in Section \ref{sec3p1p2}, it did not affect their result qualitatively. 
%In our study, the competition of inertia ($St$) and straining ($s$) exists, and the trace term is crucial in detecting the critical values of these parameters for which the Melnikov function can have zeros.
From equation (\ref{eqnD7}),  one can infer that $M_p^{\pm}$ is periodic in $\phi$ with period $\pi/\Omega_0$. Unlike $M_1^{\pm}$ and $M_2^{\pm}$, $M_p^{\pm}$ has explicit dependancy on the strain rate $s$. However, it does not affect the validity of the Melnikov analysis and related theorems as stated by \citet{kawakami1999chaotic}. The variation of the Melnikov function $M_p^{\pm}$ with $\phi$ for various $s$ values for $St=0.1$ particles in a Kida vortex is shown in figure \ref{Fig024}. Since $M_p^{\pm}$ depends on $s$ implicitly, it will result in a nonlinear relation between the critical values of $s$ and $St$, as can be seen in the figure \ref{Fig018}(b). 
%%%%%%%%%%%%%%%%%%%%%%%%%%%%%%%%%%%%%%%%%%%%%%%%%%%%%%%%%%%%%%%%%%%%%%%%%%5
%melnikovFarfield_r=0p5_St=0p1
%%%%%%%%%%%%%%%%%%%%%%%%%%%%%%%%%%%%%%
%%%%%%%%%%%%%%%%%%%%%%%%%%%%%%%%%%%%%%%%%%%%
%%%%%%%%%%%%%%%%%%%%%%%%%%%%%%%%%%%%%%%%%%%%%%%%%%%%%%%%%%%
\section{\label{appE} Scaling law for large-time dispersion with time}
The inertial particles centrifuged away to infinity by the Kirchhoff vortex spiral outward and form a ring-like structure, as we see in the simulation result (figure \ref{Fig002}). At large times, these particles reach far away from the central ellipse. Thus, the background flow they perceive can be approximated to that created by an irrotational vortex (point vortex). %\citep[see][]{kawakami1999chaotic}. 
Using the polar coordinates $(R,\Theta)$ about the origin, the flow-field for the point vortex can be represented in stationary frame as
\begin{equation}
        U_R = 0~, \quad U_{\Theta} = \frac{1}{2 R}~.
        \label{eqnE1}
\end{equation}
Then, the dynamics of a heavy inertial particle in the point vortex is given by the Maxey-Riley equations in polar coordinates as \citep[see][]{ravichandran2015caustics}
\begin{subequations}
\label{eqnE2}
\begin{eqnarray}
    \dot{v}_R +\frac{v_R}{St} =\frac{v_{\Theta}^2}{R}~, \quad \dot{R} = v_R, \label{eqnE2a}\\
    \dot{(R\, v_{\Theta})} + \frac{R\, v_{\Theta}}{St}=\frac{R\, U_{\Theta}}{St}~, \quad \dot{\Theta} = \frac{v_{\Theta}}{R} ~. \label{eqnE2b}
\end{eqnarray}
\end{subequations}
Solving the first equation among equations (\ref{eqnE2b}) along with equation (\ref{eqnE1}) yields the azimuthal velocity of particle as
\begin{equation}
    v_{\Theta} = \frac{1}{2\, R}\left(1+C_1\, e^{-t/St}\right)~,
    \label{eqnE3}
\end{equation}
where $C_1$ is an integration constant. The form of the solution says that the azimuthal velocity of the particle approaches that of fluid tracers at a large time, i.e. $v_{\Theta} = U_{\Theta}$ as $t \rightarrow \infty$. Substituting this result in equations (\ref{eqnE2a}) and solve for $R$ asymptotically ($\dot{v}_R \ll v_R$) yields, at large time $R(t) \sim (t\, St)^{1/4}$. I.e., in terms of the dispersion parameter MSD, $\sigma^2(t) = \langle R(t)^2  \rangle \sim \sqrt{t\, St}$ at large time.

The same scaling law can also be obtained from an analysis using the method of characteristics for the evolution of the number density field of particles. The single fluid model is an Eulerian model for particle-laden flows, where the suspended particles are treated as a field, characterised by their number density. The model is valid if the particles are only weakly inertial and their concentration is just right. The model yields the evolution equation for the axisymmetric number density field $n(t,R)$ as
\begin{equation}
\label{eqnE4}
    \frac{\partial n}{\partial t}+\frac{St}{R}\, \frac{\partial}{\partial R}(U_{\Theta}^2\, n) = 0~.
\end{equation}
The solution for this partial differential equation in a point vortex field equations (\ref{eqnE1}) can be obtained using the method of characteristics as, 
\begin{equation}
\label{eqnE5}
    n(t,R) = \frac{R^2}{\sqrt{R^4-t\, St}}\, n_0\left(\left\{R^4-t\, St\right\}^{1/4}\right)~,
\end{equation}
where $n(0,R)=n_0(R)$ is the initial number density distribution. Not that the scaling $R^4 \sim t\, St$ automatically appears in this solution.
%%%%%%%%%%%%%%%%%%%%%%%%%%%%%%%%%%%%%%%
%%%%%%%%%%%%%%%%%%%%%%%%%%%%%%%%%%%%%%%%%%%%%%%%%%%%%%%%%%%%%%%%%%%%%%%%%%%%%%%%%%%%%%%%%%%%%%%%%%%%%%%%%%%%
\bibliographystyle{jfm}
\bibliography{jfm}

\end{document}